\documentclass[aps,prb,twocolumn,superscriptaddress,floatfix,longbibliography]{revtex4-2}

\usepackage{amsmath,amssymb} 
\usepackage{bm} 
\usepackage{graphicx} 
\usepackage{comment} 
\usepackage{textcomp} 
\usepackage{float}
\usepackage{enumitem}
\setlist{noitemsep,leftmargin=*,topsep=0pt,parsep=0pt}

\usepackage{xcolor} 
\definecolor{lightgray}{gray}{0.6}
\definecolor{medgray}{gray}{0.4}
\usepackage[colorinlistoftodos]{todonotes}
\usepackage [autostyle, english = american]{csquotes}
\MakeOuterQuote{"}

\usepackage{hyperref}
\hypersetup{
	colorlinks=true,
	urlcolor= blue,
	citecolor=blue,
	linkcolor= blue,
}

\newif\ifptitle
\newif\ifpnumber
\newcounter{para}
\newcommand\ptitle[1]{\par~\refstepcounter{para}
	{\ifpnumber{\noindent\textcolor{lightgray}{\textbf{\thepara}}\indent}\fi}
	{\ifptitle{\textbf{[{#1}]}}\fi}}

\newcommand{\mytitle}{Effect of topological length on Bound states signatures in a Topological nanowire}
\begin{document}
	
	\title{\mytitle}
	
	\author{Dibyajyoti Sahu}
	\email[]{dibyajyoti20@iiserb.ac.in}
	\author{Vipin Khade}
	\author{Suhas Gangadharaiah}
	\email[]{suhasg@iiserb.ac.in}
	\affiliation{Department of Physics, Indian Institute of Science Education and Research, Bhopal, India}
	
	\date{\today}
	\begin{abstract}
		Majorana bound states
		(MBS) at the end of  nanowires 
		have been proposed as  one of the most important candidate for the topological qubits. 
		However, similar tunneling conductance features for both the MBS and 
		Andreev bound states (ABS) have turned out to be 
		a major obstacle in the verification of the presence of MBS in semiconductor-superconductor heterostructures. In this article, we use a  protocol to probe properties specific to the MBS and use it to distinguish the topological 
		zero-bias peak (ZBP) from a trivial one.  For a scenario involving   quantized ZBP  in the nanowire, we propose a scheme
		wherein the length of the  topological region in the wire is altered. The tunneling conductance signatures can then be utilized to gauge the impact 
		on the energy of the low-energy states. We show that the topological and trivial ZBP behave differently under our protocol, in particular,  the topological ZBP remains robust at zero bias throughout the protocol, while the trivial ZBP splits into two peaks at finite bias. This protocol probes the protection of near zero energy states due to their separable nature, allowing us to distinguish between topological and trivial ZBP.
	\end{abstract}

	\maketitle
	\section{\label{sec:intro}Introduction}
	\ptitle{motivation, MBS}
	{\color{red}  
		
		
		
		
		
	}
	The   Kitaev's one-dimensional topological superconducting  model~\cite{Kitaev} predicts   
	Majorana modes  at the ends of the 1D chain. These well separated states form robust  nonlocal fermionic states and in addition under braiding the Majorana states  obey
	non-abelian statistics,  
	thus potentially providing an ideal base for  fault-tolerant topological quantum computation~\cite{Chetan_Das_sharma_Non_Abelian_anyons_2008, alicea_New_directions_in_the_pursuit_of_MF, Das_Sharma_Chetan_Majorana_zero_modes_TQC_2015}. Since then  number of theoretical proposals  have been made for potentially realizing systems that could  host the Majoranas. Some of the earliest proposals  include,  fractional quantum Hall
	states at filling $\nu=5/2$~\cite{Moore_Nonabelions_FQH_1991},  spinless
	topological $p_x + ip_y$ superconductors 
	with the cores hosting Majoranas~\cite{rice1995sr2ruo4}. Others include,   the cores of superconducting vortices present on the surface of a 3D topological insulator proximitized to s-wave superconductor~\cite{Fu_Superconducting_Proximity_Effect_2008}, 
	a semiconductor film with   spin-orbit interaction and  proximity coupled  to an s-wave superconductor and 
	magnetic field~\cite{Sau_Das_sharma_Generic_new_platform_2010}, in a quantum-wire with strong spin-orbit interaction proximity coupled to  superconductor in the presence of magnetic-field~\cite{Oreg_Von_Helical_Liquids_2010, Stanescu_Das_sharma_Proximity_effect_2010}
	which drives the system from a trivial phase with a finite gap to a topological phase containing a zero-energy MBS in the gap.
	Intensive experimental efforts have been made on realizing the above setups, in particular the last setup involving InAs, InSb,  etc., quantum wires have been quite promising~\cite{Albrecht_Exponential_protection_2016,Churchill_SC_nanowire_2013,Das_ZBP_InAs_2012,mourik_signatures_MBS_InSb_2012}

	

	\ptitle{ABS, Reference Research} 
	Of all the signatures of MBS, the overwhelming emphasis has been on   experimentally detecting the  zero-bias tunneling conductance  and their corresponding  interpretation~\cite{Das_Sharma_Sengupta_Midgap_edge_state_2001, Das_sharma_Setiawan_ZBP_Electron_temp_and_tunnel_coupling_dependence, Yu_Forlov_non_MBS_nearly_quantized_2021, Pan_Das_sharma_Physical_mechanism_ZBP_2020, Vuik_Akmerov_Reproducing_topological_2019, Pan_Das_Sharma_Generic_quantized_ZBP_2020, Chen_Akmerov_Ubiquitous_non_Majorana_ZBP_2019, Woods_Stanescu_Zero_energy_pinning_trivial_BS_2019, Vernek_Subtle_leakage_of_Majorana_2014, Grivnin_MZM_2019, Moore_Tiwari_Quantized_ZBP_trivial_2018, Huang_Sau_Das_sharma_Metamorphosis_ABS_to_MBS_2018, Setiawan_Das_Sharma_Electron_temp_dependance_ZBP_2017, Nichele_Flensberg_Scaling_Majorana_ZBP_2017, Kammhuber_Wimmer_Conductance_halical_states_2017, Reeg_Zero_energy_ABS_2018, Liu_Sau_Das_Sharma_ABS_vs_MBS_ZBP_2017, Cayao_SNS_2015, Zhang_Balastic_SC_SM_nanowire_2017, higginbotham_parity_2015, Dmytruk_Pinning_2020, Liu_Sau_Das_sharma_Role_of_dissipation_MBS_2017, Prada_Transport_2012, Lee_ZBA_2012, prada_andreev_2020, san_majorana_2016, Deng_Flensberg_MBS_in_QD_hybrid_nanowire_2016,Song_Large_ZBP_2022, avila_non_hermitian_2019}. From the theoretical perspective, tunneling at the edge of the  topological superconductor (where the MBS are localized) 
	results in a resonant  effect, 
	consequently the  Zero-bias peak (ZBP) conductance is expected to  acquire the quantized value of $2e^2/h$~\cite{Das_Sharma_Sengupta_Midgap_edge_state_2001}. 
	Previous studies have shown that the ZBP can also appear due to the Andreev bound states (ABS), with the peak height close to the quantization values~\cite{Moore_Tiwari_Quantized_ZBP_trivial_2018,Huang_Sau_Das_sharma_Metamorphosis_ABS_to_MBS_2018,Pan_Das_sharma_Physical_mechanism_ZBP_2020}. The origin of ABS  is often due to  inhomogeneous profiles and quantum dots at the wire's end and many a times they mimic MBS's tunneling signatures. 
	
	The typical ABS consists of overlapping Majorana components, making them very sensitive to parameter changes. Therefore, the ZBP due to the ABS 
	can generally be attributed to the fine tuning of the parameters and hence not robust~\cite{Tiwari_Two-terminal_2018}.
	Recent theoretical works have shown that the trivial in-gap states can actually be pinned close to  zero energy for an extended range of parameter space for certain inhomogeneous profiles, making the ZBP robust. These ABS have their  Majorana components  partially  separated (termed as the  ps-ABS) from each other~\cite{Tiwari_Robust_low-energy_ABS_2019}.
	On the other hand, for magnetic field strengths greater than the critical value, the wire will be in the topological regime wherein the Majorana components are completely 
	separated and localized at the two ends of the wire. 
	It turns out that the ps-ABS appear just before the critical Zeeman term, thus  
	it is   challenging to unambiguously pin the origin of the ZBP as arising from either  the topological or the trivial states. 
	
	To distinguish them, use of two leads each placed at the opposite ends of the wire have been proposed as they can measure  the correlations between the conductance from the two local leads. In addition, this setup can be used to probe non-local conductances which arise 
	due to the bulk states and can produce the band gap closing signature at the topological phase transition point beyond  which MBS appears~\cite{Tiwari_Two-terminal_2018, Das_sharma_Three_terminal_ZBP_2021}.
	
	However, as discussed in Refs.\cite{Loss_Local_and_nonlocal_2021},  the presence of  inhomogeneities like quantum dots at both the ends can yield correlated trivial ZBP in both the local conductances.
	It turns out that the non-local conductance signatures in case of  the heterostructures are not strong enough to unambiguously depict the band gap closing. 
	Besides the above, dephasing, leakage dynamics of Majorana~\cite{Mishmash_Dephasing_leakage_2020} and recent studies focused  on the 
	robustness of quantization of ZBP with respect to tunneling barrier height and temperature~\cite{Das_sharma_Quality_ZBP_2021} have been suggested as potential schemes  to distinguish the topological Majorana modes from the Andreev states.

	\ptitle{Our idea, moving bound states}
	
	As mentioned in the preceding paragraph the presence of quantized ZBP is insufficient to conclude the existence of MBS as trivial ABS can also produce quantized ZBP. To distinguish topological states from normal zero-modes, one needs to go beyond the presence of Quantized ZBP and probe properties specific to MBS only. 
	Consider that the ends of a wire are moved into trivial parameter space by changing the topological length of the wire; in such a case, one produces a trivial-topological-trivial structure with the MBS still present at the edge of the topological region. This property specific to the MBS has been proposed before for Majorana braiding protocols~\cite{alicea_non_Abelian_2011, Das_Sharma_MZM_QC_2015, Alicea_Antipov_Dynamic_of_Majorana_qubit_2018, Zolle_MBS_noisy_kitaev_2015}. Motivated by this, we define the "moving protocol" to systematically decrease the topological length by fixing the wire's ends in a parameter regime  below the topological range by either reducing the Zeeman term or the external potential. 
	We employ "Zeeman moving protocol" (ZMP) by applying Zeeman term profile and "Potential moving protocol" (PMP) with external potential in our numerical simulation.
	Since local conductance measures the local property of the wire close to the lead, the effect of change in the topological region should be apparent in the local conductance. We apply both the protocol to diverse scenarios and investigate their impact on the trivial and the topological ZBP. We show that the topological ZBP from MBS remains robust under the moving protocol. At the same time, the trivial ZBP from ps-ABS has entirely different behaviour, splitting into two separate peaks at finite bias. This difference in behavior arises due to the fundamental difference between the MBS and ps-ABS in the overlap of their Majorana components. Our "moving protocol" indirectly probes this separable nature of the Majorana components. 
	We also show that the PMP is more effective compared to ZMP as far as distinguishing trivial and topological ZBPs are concerned, moreover, the practical implementation of this scheme is also relatively more feasible.

	\ptitle{Outline}
	The organization of the paper is as follows. In section~\ref{sec:model}, we describe the model and profile of the semiconductor-superconductor nanowire used to produce the different heterostructures. In this section, we also define our moving protocols, under which we will study tunneling conductance signatures. In section~\ref{sec:Result}, we present our results. First, we look into a homogeneous nanowire and its tunneling conductance signatures. We also look at the inhomogeneous system, which can have both trivial and topological ZBP. We show the effect of ZMP on the trivial and topological ZBP to demonstrate the different ways they behave. In subsection~\ref{subsec:mu_moving}, we present the behaviour of topological and trivial ZBP under PMP. In subsection~\ref{subsec:SSS}, we consider another system, S${}^\prime$SS${}^\prime$ (where S${}^\prime$ and S are superconducting region of the wire with different chemical potential) system to showcase the shortcomings of the ZMP and how we go around it by using the PMP to distinguish topological ZBP from trivial ZBP. Finally, we present our conclusions in section~\ref{sec:Conclusion}.
	
	\section{\label{sec:model}Model}
	We study a 1D semiconductor nanowire
	with spin-orbit coupling having a superconducting gap induced via
	the proximity effect. In addition, we take into consideration magnetic field  applied parallel to the wire. 
	The corresponding Hamiltonian which incorporates all of the above terms  is given by 
	\begin{eqnarray}
		&& H =-  \sum_{n}\Bigg\{ \Big[ \sum_{\sigma, \sigma'} c_{n+1,\sigma}^{\dagger}(t_n \delta_{\sigma,\sigma'}-i \alpha_n \sigma_{\sigma, \sigma'}^y)c_{n,\sigma'}  + {}\nonumber\\
		&c_{n,\sigma}^{\dagger}&\frac{(-\mu_n\delta_{\sigma,\sigma'}+\Gamma \sigma_{\sigma, \sigma'}^x)}{2}c_{n,\sigma'} \Big] + \Delta_n c_{n,\uparrow}^{\dagger}c_{n,\downarrow}^{\dagger} + h.c.\Bigg\},
		\label{eqn:Hamiltonian}
	\end{eqnarray}
	where $c^\dagger_{n,\sigma}$($c_{n,\sigma}$) represents  creation (annihilation) operators of fermions with spin $\sigma$ at site $n$. The parameters $t_n$, $\alpha_n$, $\mu_n$, $\Gamma$ and $\Delta$ represent the tunneling, spin-orbit, chemical potential, Zeeman energy and the superconducting terms, respectively at the site $n$.  This will be   our base model on top of which different profiles of  $\mu_n$ and $\Delta_n$ are added to create different heterostructures. 
	
	We will focus our attention to the following two scenarios.  The first
	one is the homogeneous case in which all the parameters in 
	the Hamiltonian are taken to be position independent. We 
	reproduce the result of this well studied parameter regime to validate 
	our numerical scheme and to interpret the  effect of our 
	protocol on this system.  In the second scenario we will consider the presence of smooth variation in the  $\Delta_n$ and $\mu_n$ profiles.  The latter differentiates the different chemical potenital in the normal (N) regions and  the superconductor (S) regions, while the former creates regions of N and S resulting in   NS and NSN type of heterostructures. 
	For certain parameter regimes these heterostructures can host ABS or quasi-MBS which can stay close to zero energy over large range of magnetic fields. The smooth variation of  the spatial  profile is modeled by the function \cite{Loss_Local_and_nonlocal_2021,Das_sharma_Three_terminal_ZBP_2021,Tiwari_Two-terminal_2018}
	\begin{equation}
		\Omega_{n_0,s}(n) = \frac{1}{2}\left[1+ \tanh\left(\frac{n-n_0}{s}\right)\right]
	\end{equation}
	\noindent where $n_0$ represents the point around which the profile changes, while $s$ is a measure of  smoothness in the variation (which has been fix to $s = 20$ for all profiles).
	\begin{figure}[ht]
		\includegraphics[clip=true,width=\columnwidth]{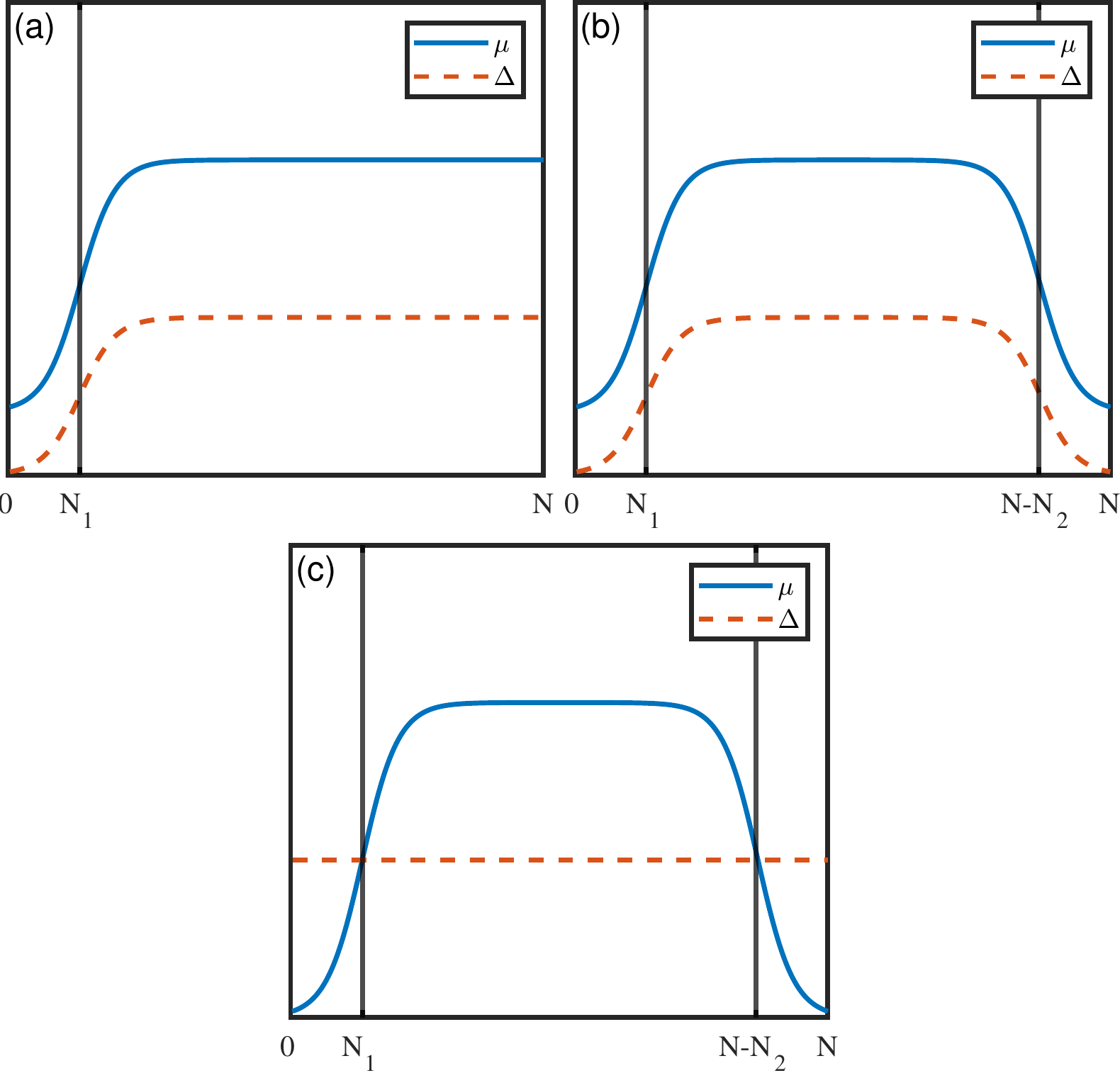}
		\caption{Plots of inhomegeneous profile (a) NS , (b) NSN and (c) S${}^\prime$SS${}^\prime$ system.}
		\label{fig:Inhomo_profile}
	\end{figure}
	
	For example, an NS profile  with smoothly varying chemical potential  can be represented via the following set of parameters
	\begin{align}
		\Delta_n &= \Delta_0 \Omega_{N_1,s}(n), \nonumber\\
		\mu_n &= \mu_N + (\mu_S - \mu_N) \Omega_{N_1, s}(n),
		\label{eqn:NS_profile}
	\end{align}
	where $N_1$ is the length of the Normal region in the heterostructure, $\mu_S$ and $\mu_N$ are the chemical potential for the S and N regions respectively, and  $\Delta_0$ is the effective superconducting coupling strength in the S region. The  NSN heterostructure can be modeled by the following set of 
	parameters with the profile given by
	\begin{align}
		\Delta_n =& \Delta_0\left[ \Omega_{N_1, s}(n) - \Omega_{N_1+N_S+1, s}(n)\right],\nonumber \\
		\mu_n =& \mu_{N_1} + (\mu_S - \mu_{N_1}) \Omega_{N_1, s}(n) \nonumber\\
		&+ (\mu_{N_2}-\mu_S ) \Omega_{N_1+N_S+1, s}(n)
		\label{eqn:NSN_profile}
	\end{align}
	where $\mu_{N_1}$ and $\mu_{N_2}$ are the chemical potentials for the Normal regions. we note that $N_1$, $N_2$ and $N_S$, denote the  length of left and right normal regions and the superconducting region respectively.
	Together with these systems we also consider S${}^\prime$SS${}^\prime$ which has a homogeneous $\Delta$ but inhomogenity in $\mu$, which distinguish the two superconducting region S${}^\prime$ and S.
	\begin{align}
		\Delta_n =& \Delta_0 \nonumber \\
		\mu_n =&  \mu_S  \left[\Omega_{N_1, s}(n) - \Omega_{N_1+N_S+1, s}(n)\right]
		\label{eqn:SSS_profile}
	\end{align}
	Plots of the NS, NSN and S${}^\prime$SS${}^\prime$ systems $\mu$ and $\Delta$ profile are shown in Fig.~\ref{fig:Inhomo_profile}.
	\begin{figure}[ht]
		\includegraphics[clip=true,width=\columnwidth]{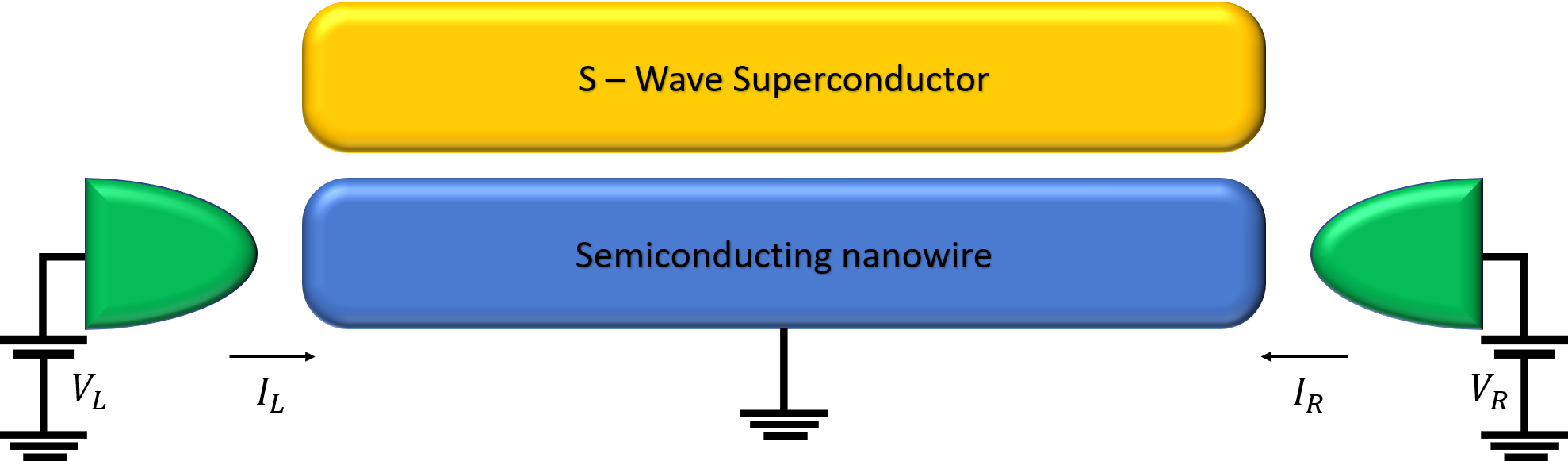}
		\caption{Tunneling conductance setup with two lead.}
		\label{fig:Tunnel_setup}
	\end{figure}
	
	As a simple diagnostic tool to distinguish the presence of MBS from the ABS in the various parameter regimes we calculate the boundary topological invariant (TI) using the scattering matrix $S$.  The scattering matrix  relates the incoming and the outgoing wave amplitudes 
	\begin{equation}
		\label{eqn:scattering matrix}
		S = \begin{pmatrix}
			R & T' \\
			T & R'
		\end{pmatrix}.
	\end{equation}
	Note that the reflection subblocks $R,R'$ and the transmission subblocks $T,T'$ connects the two ends of the chains and are obtained in terms of the Fermi level wave amplitudes.   The TI $=\text{sgn Det} (R)=\text{sgn Det} (R')$, obtains a value $  +1$ when the wire is in the trivial phase and $-1$ for the topological phase~\cite{Nori}, further details are provided in Appendix~\ref{app:TI}. 
	Along with the topological invariant we also plot the wavefunction of the Majorana components to characterise the topological nature of the bound states. The Hamiltonian has particle-hole symmetry, giving rise to symmetric spectrum with two fermionic wave functions $\Psi_{\epsilon}(x)$ and $\Psi_{-\epsilon}(x)$ for every eigenvalue $\epsilon$. 
	Thus one constructs the following symmetric and anti-symmetric  wave functions from the  low energy eigenstates 
	\begin{align}
		\gamma_1(x) = \frac{1}{\sqrt{2}}\left[\Psi_{\epsilon}(x)+\Psi_{-\epsilon}(x)\right], \nonumber \\
		\gamma_2(x) = \frac{i}{\sqrt{2}}\left[\Psi_{\epsilon}(x)-\Psi_{-\epsilon}(x)\right].
	\end{align}
	The wire can thus be characterised as topological (with MBS) if $\gamma_1(x)$ and $\gamma_2(x)$ are spatially separated, while the overlapping and partially overlapping states characterises ABS and ps-ABS, respectively~\cite{Tiwari_Two-terminal_2018,Das_sharma_Quality_ZBP_2021}.
	
\begin{figure}[ht]
		\includegraphics[clip=true,width=\columnwidth]{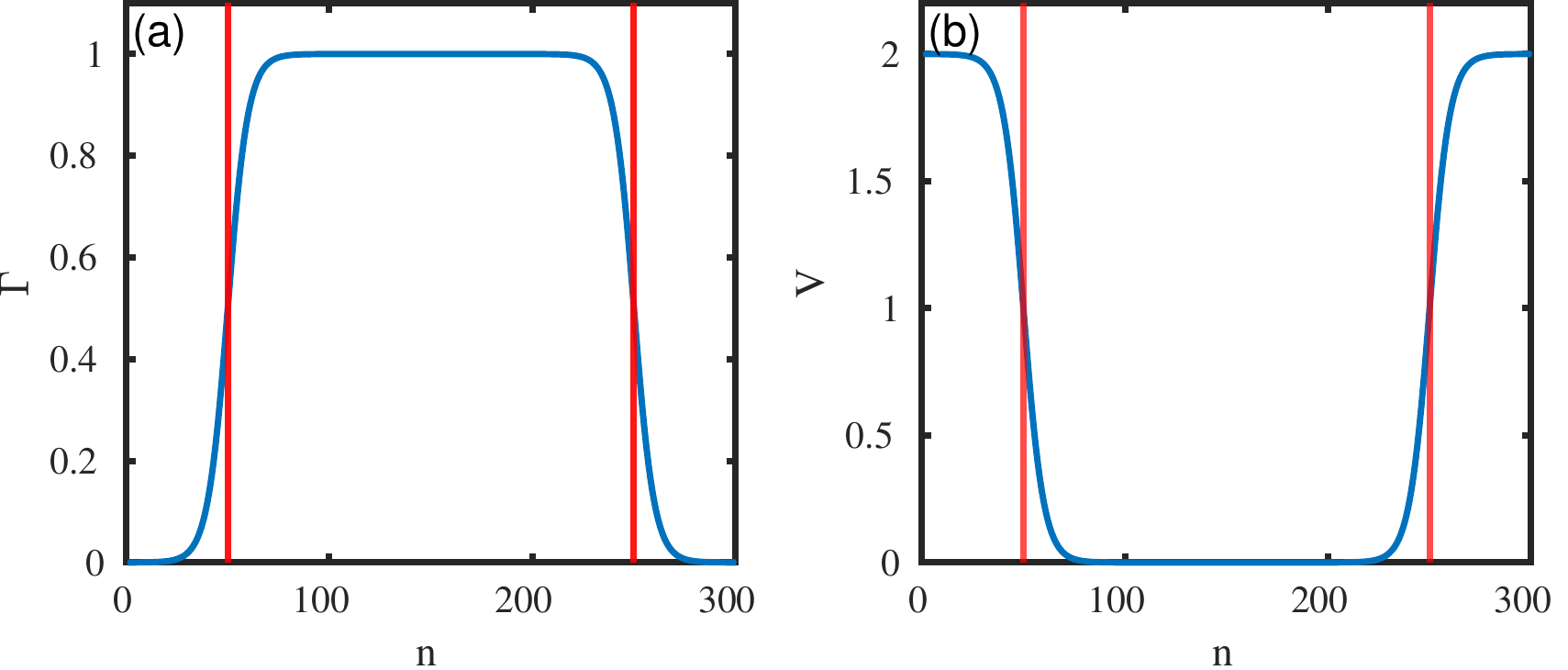}
		\caption{Moving protocol profile for $N = 300$ and $n_0 = 50$ (a) with Zeeman profile (ZMP) for $\Gamma_0 =  1$ and (b)with External potential (PMP) profile for $V_0 =  2$}
		\label{fig:moving_profile}
	\end{figure}

	\ptitle{Tunneling Conductance}
	We  perform  tunneling conductance calculations for a setup involving 
	two normal leads attached to either side of the wire to create a three terminal device (see Fig.~\ref{fig:Tunnel_setup})\cite{Das_sharma_Three_terminal_ZBP_2021}.
	The conductance matrix  has the following form
	\begin{equation}
		G = 
		\begin{pmatrix}
			G_{LL} & G_{LR}\\
			G_{RL} & G_{RR}
		\end{pmatrix},
	\end{equation}
	where the elements of the matrix are given by
	\begin{eqnarray}
		G_{ij} = \left.\frac{dI_i}{dV_j}\right|_{V_i = 0}.
	\end{eqnarray}
	The local and the non-local conductance are calculated using the Green's function method. Details related to the  calculations are   provided in the Appendix~\ref{app:Tunneling}.
	We focus our attention on the effect the topological length of the wire has  on the tunneling conductance signatures. By appropriately choosing site-dependent Zeeman or the chemical potential terms, different regions of the 1D wire can be tuned to be in the topological or the trivial phase~\cite{alicea_non_Abelian_2011,Das_Sharma_MZM_QC_2015}.
	Consider the following spatial dependent Zeeman strength
	\begin{align}
		\Gamma_n = \Gamma_0 \text{F}(n,n_0,N,s'),\nonumber
	\end{align}
	where
	\begin{eqnarray}
		\text{F}(n,n_0,N,s') =  n_F\left[2\left(\frac{n_0-n}{s'}\right)\right]n_F\left[2\left( \frac{n-N+n_0}{s'}\right)\right],\nonumber
	\end{eqnarray}
	and $n_F(x)=1/(1+e^x),$
	here $N$ represents the full length/site of the nanowire and $s'$ is smoothness parameter (fixed to $s' = 10$). A sample plot of the profile is shown in Fig.~\ref{fig:moving_profile}(a) for $n_0 = 50$. For the   protocol that we consider,    $\Gamma_0 $ is kept unchanged while $n_0$ is increased from negative to  positive values.  If $\Gamma_0$  is above the critical field,  the wire will be  topological for approximately $n_0< n < N-n_0$ regions.  Thus, increasing $n_0$ decreases the length of the topological region due to the Zeeman strength being below the critical values in those regions. 
	For $\Gamma_0 > \Gamma_c $  this profile creates a domain wall structure of trivial-topological-trivial superconductor, changing $n_0$ thus  results in the change of the topological length. We denote this protocol of changing the topological length as the "Zeeman Moving protocol" (ZMP) and we will be exploring the 
	effect it has on the ZBP due to the presence of either the ps-ABS or the  MBS.
	In addition to the Zeeman term, one could also use tunable local gates to add an external potential which increases the effective chemical potential in the ends of the wire to put them in a trivial state, thus decreasing the topological length. To consider this we will add external potential of the form Eq.~(\ref{eqn:mu_moving}) to our chemical potential the plot of which is shown in Fig.~\ref{fig:moving_profile}(b). We denote this protocol of changing the topological length as the "Potential Moving protocol" (PMP).
	\begin{equation}
		\label{eqn:mu_moving}
		V_n = V_0\left(1-F(n,n_0,N,s')\right),
	\end{equation}


	\section{\label{sec:Result}Results}
	In this section, we will be considering the effects of the moving-protocol on the ZBP for the  homogeneous and inhomogeneous systems. For either of them, we begin by exploring the phase diagram by calculating the topological boundary invariant. Further, we calculate the Majorana components to classify the in-gap states.  We reproduce some of the earlier results on the ZBP by calculating the tunneling conductance.  The main thrust of our work is to distinguish the ZBP arising due to the presence of topological and trivial bound states via the moving protocol. 
	

	
	\begin{figure}[t]
		\includegraphics[clip=true,width=\columnwidth]{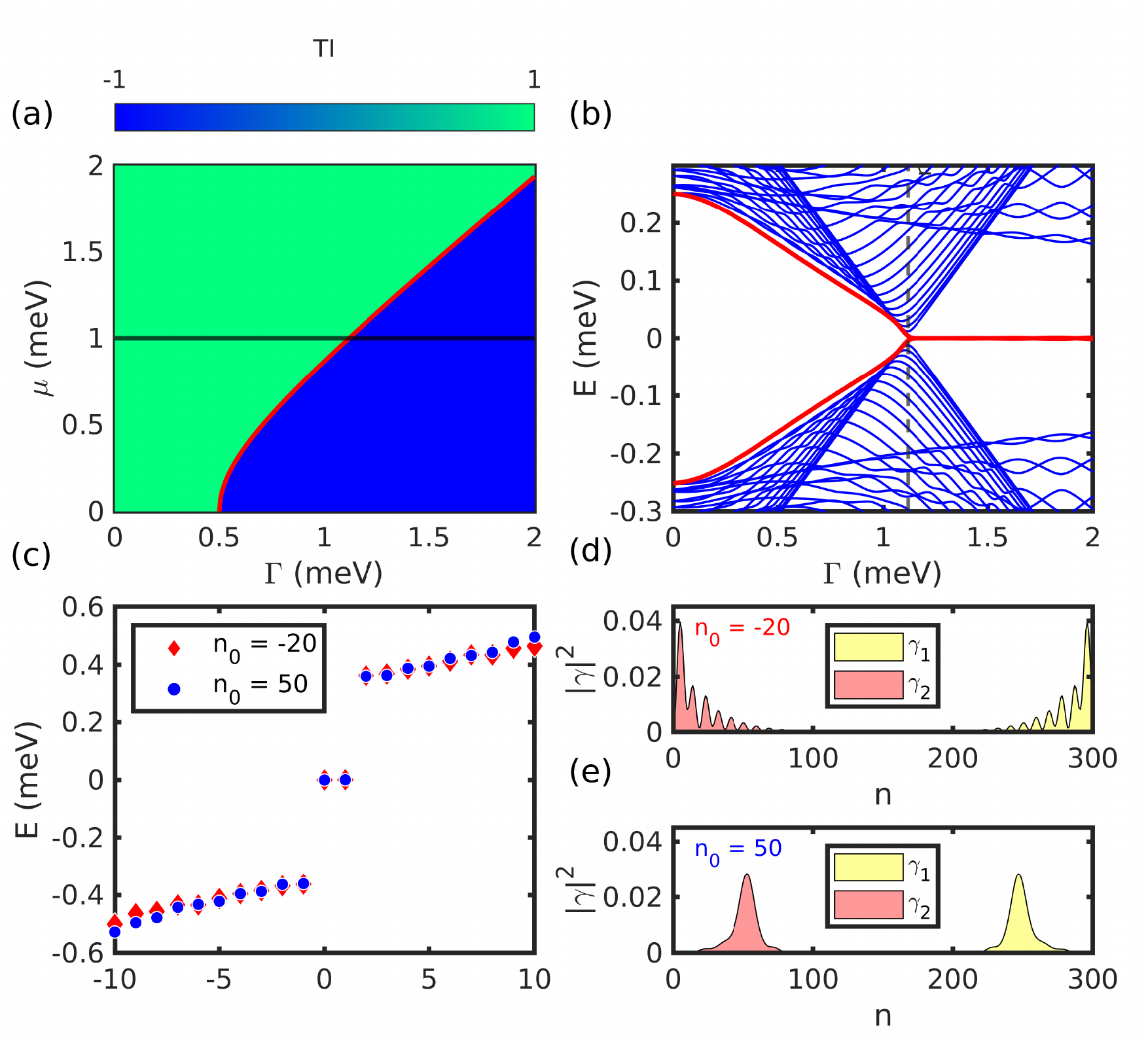}
		\caption{Homogeneous system with parameter $t =  25.4$ meV, $\alpha = 2$ meV and $\Delta  = 0.5$ meV. (a) phase portrait with boundary topological invariant for range of $\mu$ and $\Gamma$ values, the red line denotes the critical zeeman term for each $\mu$ value separating topological (TI = -1) and trivial phase (TI = 1). (b) Energy spectrum for $\mu = 1$ meV showing MBS after $\Gamma_c$ (vertical grey dash line). (c) Energy eigenvalues with moving profile for different $n_0$ value. (d) Majorana components of low energy states for different instances of moving protocol. It shows MBS moving away from the ends.}
		\label{fig:Homogeneous_all}
	\end{figure}

	\subsection{\label{subsec:Homogeneous_1}Homogeneous System}
	We choose the parameter space of InSb-Al nanowire that are often  used in theoretical and experimental studies\cite{Tiwari_Two-terminal_2018,Das_sharma_Three_terminal_ZBP_2021}.  The effective mass considered is $m^* = 0.015 m_e$ with lattice constant $a = 10$ nm and Rashba spin orbit coupling $\alpha_R = 0.4$ eV$\AA$  which yields  $t =  \hbar^2/2 m^* a^2 =   25.4$ meV and  $\alpha =\alpha_R/2a =  2$ meV. The effective superconducting coupling  term  considered is $\Delta  = 0.5$ meV \cite{Das_sharma_Quality_ZBP_2021,Das_sharma_Three_terminal_ZBP_2021}.  The ideal system enters the topological non-trivial region when the  Zeeman strength is greater than the  critical value given by $\Gamma_c \approx \sqrt{\mu^2+\Delta^2}$. Using the scattering matrix approach we have calculated the TI for a range of $\mu$ and $\Gamma$, 
	and as shown in Fig.~\ref{fig:Homogeneous_all}(a) 
	the  numerically obtained  boundary between the topological and the non-topological region coincides with the theoretical prediction of the critical field shown by the red curve.  For concretness, we consider $\mu = 1$ meV as the onsite potential, the corresponding critical Zeeman strength for which the band closing takes place is  $\Gamma_c\approx 1.12$ meV. From the plot of energy spectrum Vs $\Gamma$ shown in  Fig.~\ref{fig:Homogeneous_all}(b), we observe the expected appearance of zero energy modes after the band closing which takes place at the critical Zeeman field.

	\begin{figure}[t]
		\includegraphics[clip=true,width=\columnwidth]{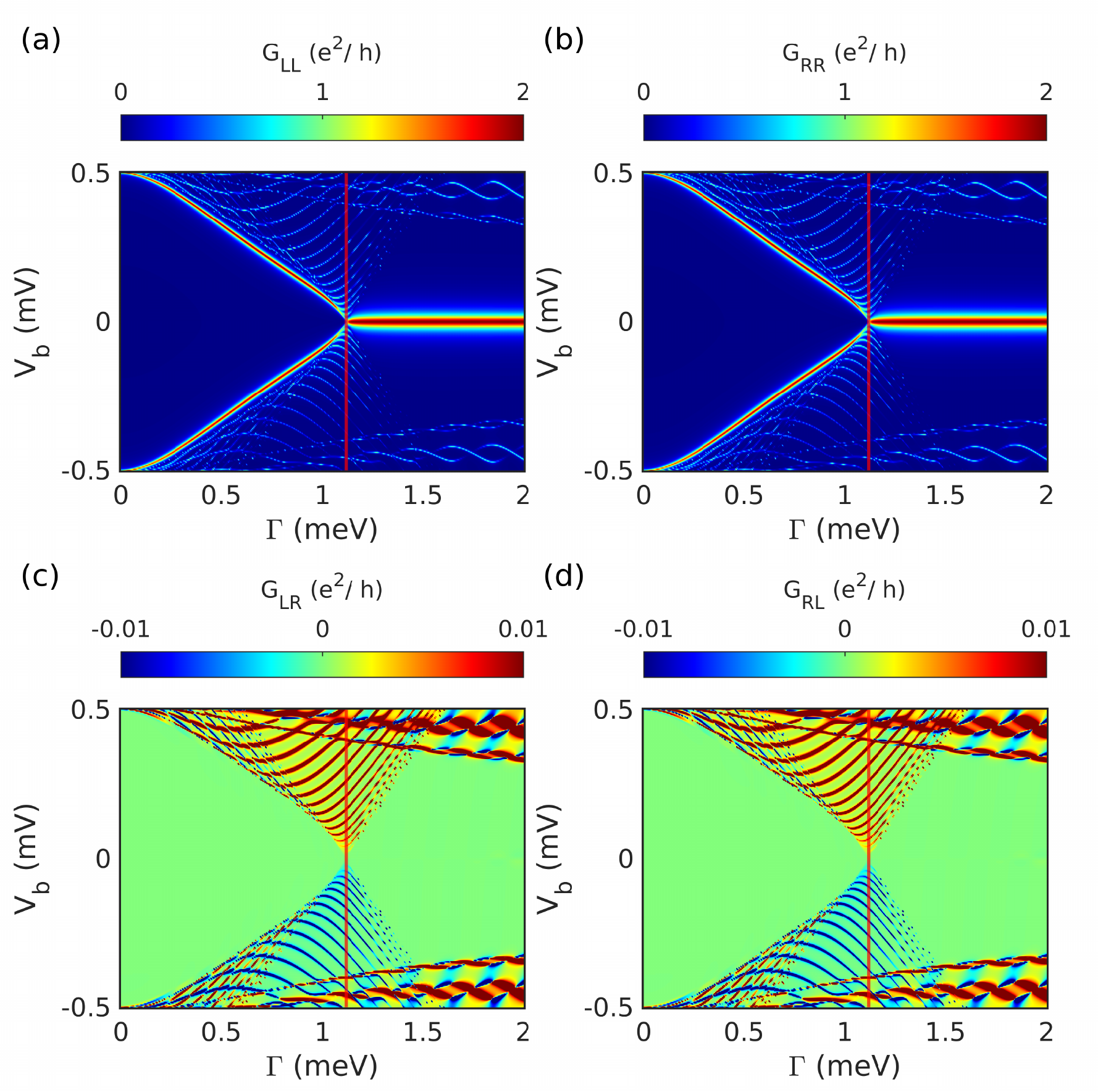}
		\caption{(a-b)Local and (c-d) non-local Tunneling conductance for the homogeneous system. Local conductance shows the presence of ZBP and the non-local conductance captures the band closing signature.}
		\label{fig:Homogeneous_ZBP}
	\end{figure}
	
	The numerically calculated tunneling conductance  for a range of $\Gamma$ is plotted in Fig.~\ref{fig:Homogeneous_ZBP}.  It is clear from the figure that ZBP in the local conductance appears  after $\Gamma_c$.
	The ZBP is quantized at $2e^2/h$ which is also the predicted conductance for the MBS. The left($G_{LL}$) and right($G_{RR}$) local conductances exhibit coherence denoting the presence of MBS on either ends of the nanowire and at the same time the non-local conductance $G_{LR}$ and $G_{RL}$ exhibit the signature of band gap closing. 
	These tunneling conductance signatures arising from a ideal nanowire with MBS at the edges of the pristine nanowire have been well studied in the past  literature \cite{Das_sharma_Quality_ZBP_2021, Das_sharma_Three_terminal_ZBP_2021}.

	
	Before we discuss the effect of the ZMP on the ZBP, we will first explore its effect  on the system's energy levels. Results of spectrum under ZMP for two different values of $n_0$  are shown in Fig.~\ref{fig:Homogeneous_all}(c). From the plot of the energy spectrum it is clear that the MBS  remain fixed at  the zero energy since the protocol only changes the length of the topological region by putting some part of the wire in a Zeeman term below the critical term($\Gamma_c$). As long as  the length of the topological region remains much larger than the  Majorana localization length, the energies of the MBS will be  fixed at zero energy due to the topological protection. Further insight into the effect of the protocol on the state is obtained by plotting  the  Majorana components of the low energy states. 
	For uniform Zeeman term  (which is the case for $n_0 = -20$) 
	the MBS are localized at the edges of the nanowire with the two Majorana components spatially separated (see Fig.~\ref{fig:Homogeneous_all}(d)). However, as the Zeeman term profile is moved so that  $n_0 = 50$, the length of the topological region as well as the   bound state  position  changes but at the same time they remain localized at the edges of the topological region as shown in Fig.~\ref{fig:Homogeneous_all}(e).  The Majorana components are still separated so they stay at zero energy. 
	
	\begin{figure}[t]
		\includegraphics[clip=true,width=\columnwidth]{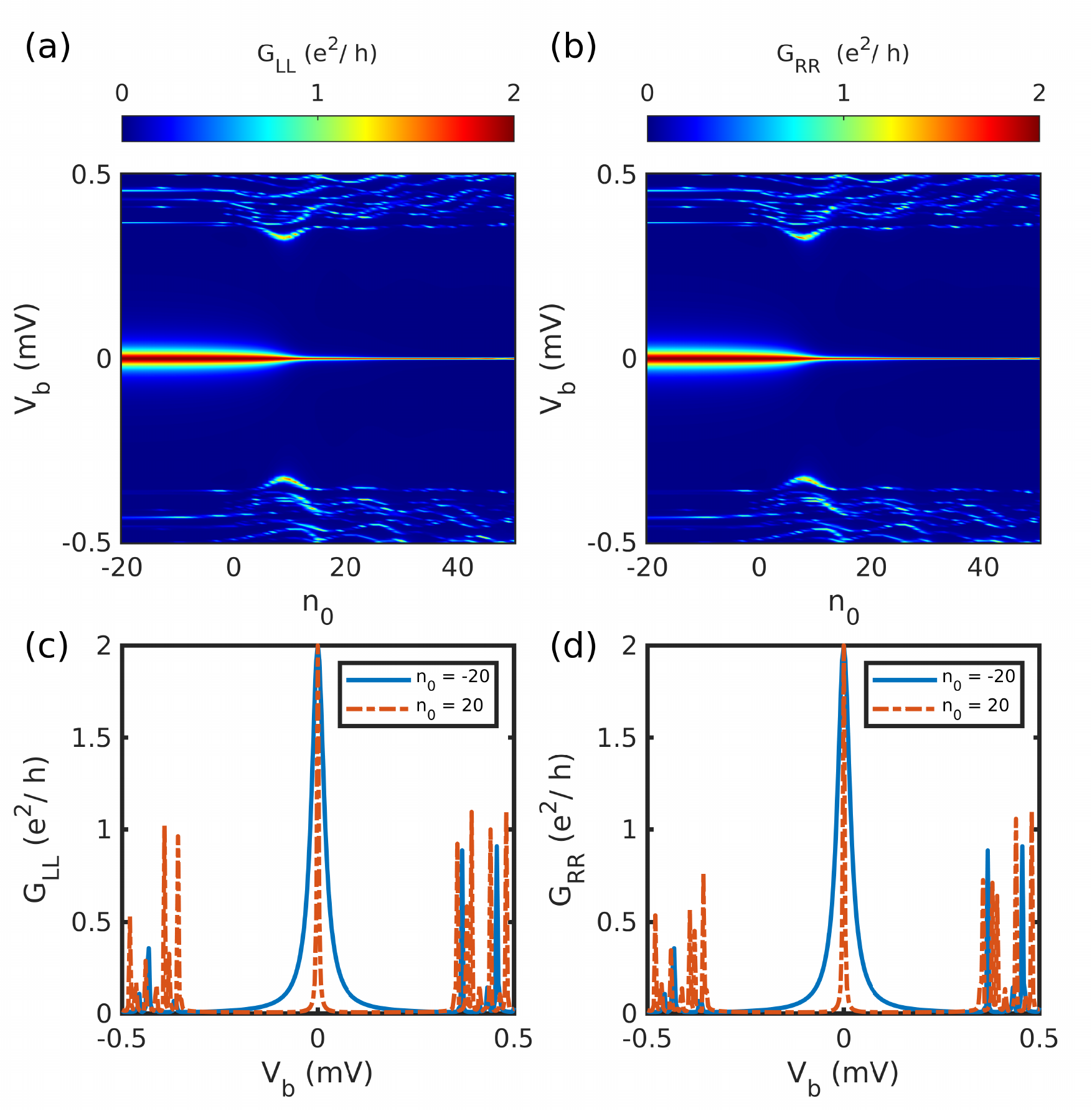}
		\caption{Tunneling conductance signature on the application of ZMP with $\Gamma_0 = 1.5$ meV. (a-b) Local tunneling conductance for range of $n_0$ values and, (c-d) vertical line-cut of conductance for specific $n_0$. Both local conductance shows robustness of topological ZBP under ZMP.}
		\label{fig:Homogeneous_ZBP_moving}
	\end{figure}

	We next focus the effect of ZMP on the ZBP.  For the ZMP, the  calculated local conductance for a range of  $n_0$ can be seen in Figs.~\ref{fig:Homogeneous_ZBP_moving}(a-b).
	As $n_0$ is increased (i.e., the topological region is reduced while the position of the lead remains unchanged), a subsequent decrease in the width of the ZBP becomes apparent, while at the same time the height of the peak remains  quantized to the value $2e^2/\hbar$. We also plot value of the conductance for two values of $n_0$  in Figs.~\ref{fig:Homogeneous_ZBP_moving}(c-d), where the quantization of peak is clearly observed. This feature has a simple explanation due to the   property of the MBS.  Application of the moving protocol changes the topological length of the wire and therefore the  MBS moves further away from the lead while always staying at the edge of the topological region. Since the edge states move further away from the lead the coupling between the lead and the MBS decreases, resulting in the reduced width  of the ZBP. For sufficiently large separation between the lead and the edge mode,
	the ZBP disappears altogether. Through out this protocol the Majorana components remain spatially separated, 
	as a result the MBS remains at the zero energy causing the peak to be at zero bias through out this protocol.

	\subsection{\label{subsubsec:Inhomogeneous_1}Inhomogeneous Wire}
	
	\begin{figure}[t]
		\includegraphics[clip=true,width=\columnwidth]{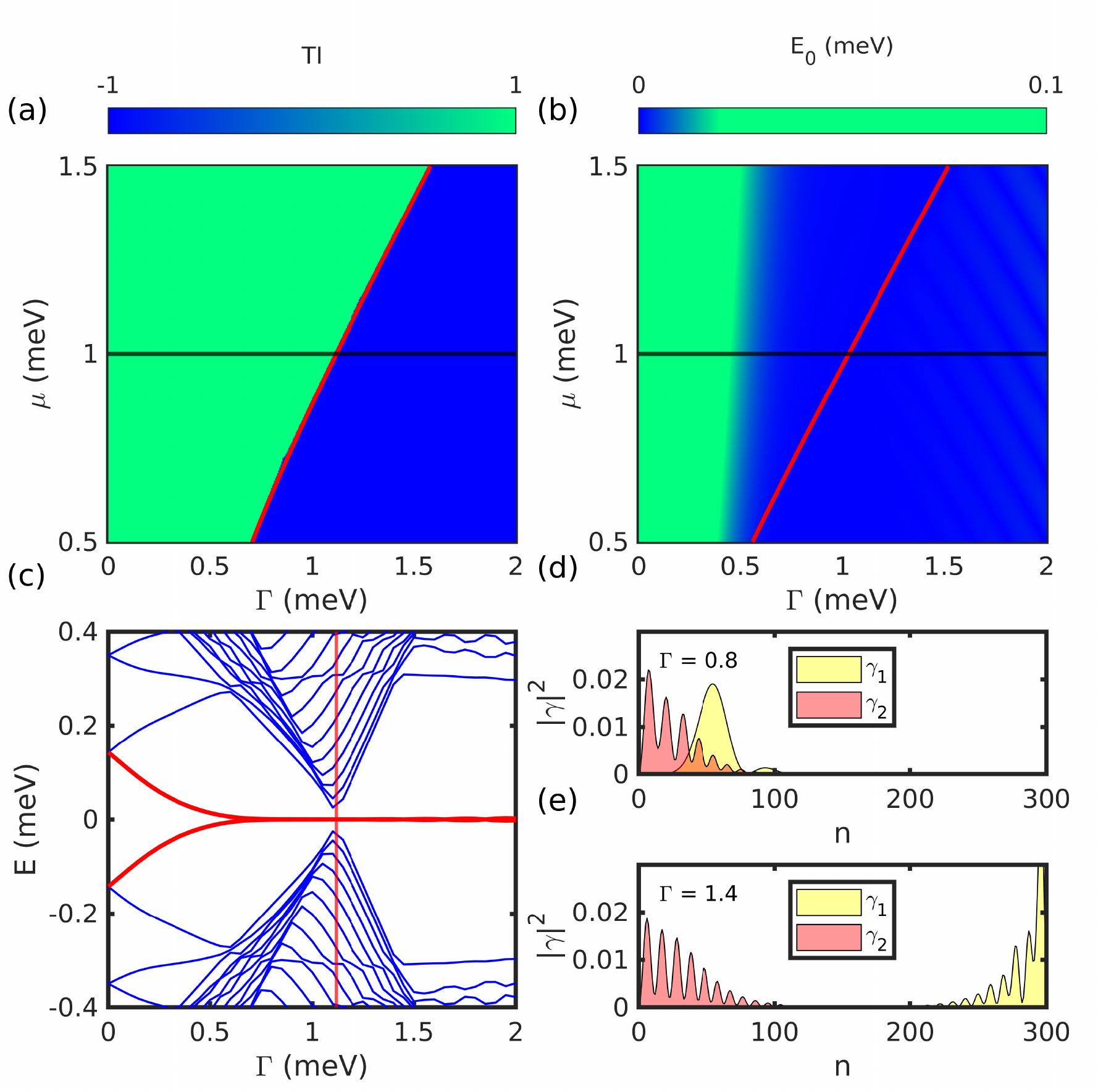}
		\caption{NS system with parameters $t =  25.4$ meV, $\alpha = 2$ meV, $\Delta_0 = 0.5$ meV, $N_1 = 40$, $\mu_N = 0.2$ meV and $N = 300$. Phase portrait for a  range of chemical potential in the S region ($\mu_S$) and Zeeman term ($\Gamma$), (a) depicted as the topological invariant and (b) low energy eigenvalues to show the presence of zero energy states in the trivial region. (c) Energy spectrum of NS system with $\mu_S = 1$ meV, the vertical line denotes critical Zeeman term ($\Gamma_c) = \sqrt{\mu_S+\Delta_0} = 1.118$ meV. (d) The Majorana component in trivial region with ps-ABS and (e) MBS in topological region.}
		\label{fig:NS_data}
	\end{figure}
	
	\ptitle{NS heterostructure}
	In quantum wires  it is difficult to achieve a scenario wherein the parameters are homogeneous through out the length of the wire. Instead,
	the  inhomogeneous profile in the heterostructure better
	captures the realistic scenario. 
	We begin by considering an inhomogeneous NS heterostructure where the chemical potential and the superconducting profile used to model the heterostructure is given by Eq.~(\ref{eqn:NS_profile}) and shown in Fig.~\ref{fig:Inhomo_profile}(a). 
	The numerically calculated topological boundary invariant  
	is shown in Fig.~\ref{fig:NS_data}(a), which shows the presence of topological phase for Zeeman terms greater than $\Gamma_c$ for a given value of the chemical potential (in the $S$ region). This figure is similar to the corresponding Fig.~\ref{fig:Homogeneous_all}(a) for the homogeneous case. However, the difference between the two scenarios can be seen  from the presence of low lying energy states close to zero energy even in the non-topological regime , see Fig.~\ref{fig:NS_data}(b). 
	It turns out that the zero energy modes in the trivial region  persist for large range of Zeeman coupling and chemical potential. The energy spectrum for the system for $\mu = 1$ meV is shown in Fig.~\ref{fig:NS_data}(c), which shows the appearance of ABS before the topological phase transition at $\Gamma_c \approx  1.12$ meV.
	For $\Gamma > \Gamma_c$ the Majorana components obtained from the low-lying states are  localized in non-overlapping regions (at the edges) and are the MBS (see Fig.~\ref{fig:NS_data}(e)), while for $\Gamma < \Gamma_c$ the components have significant overlap with each other and are the partially separated ABS, as shown in Fig.~\ref{fig:NS_data}(d).

	
	
	
	\begin{figure}[t]
		\includegraphics[clip=true,width=\columnwidth]{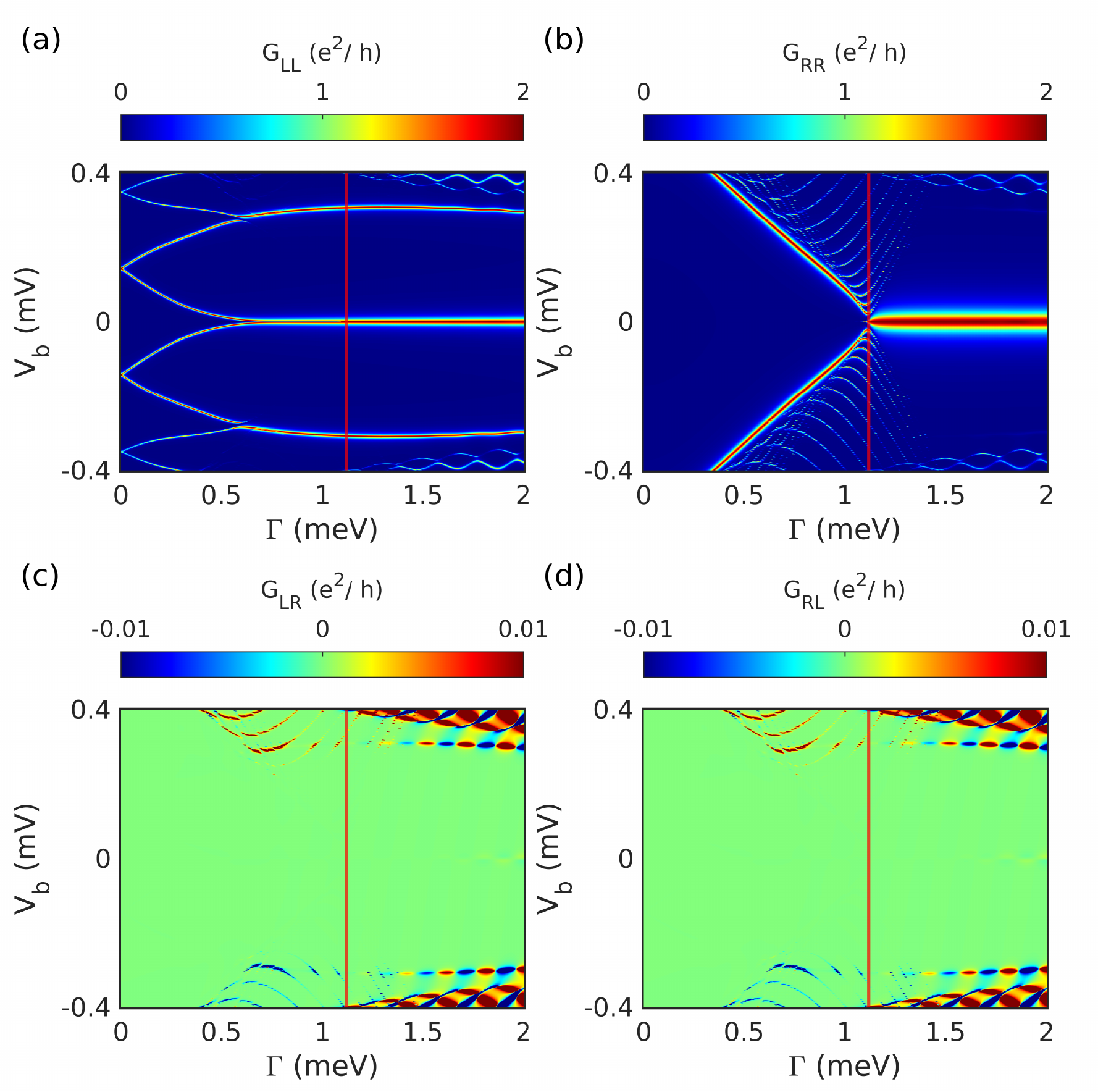}
		\caption{NS system: (a-b) Local and (c-d) non-local conductances. Both conductance shows the presence of ZBP after $\Gamma_c$(red vertical line) but left conductance also shows ZBP for range of $\Gamma < \Gamma_c$.}
		\label{fig:NS_ZBP}
	\end{figure}
	
	Consider next the Fig.~\ref{fig:NS_ZBP} which depicts  the tunneling conductance calculated for the NS heterostructure. 
	For $\Gamma>\Gamma_c$ (denoted by the red vertical line) both the local conductances, $G_{LL}$ and $G_{RR}$, exhibit ZBP with a   peak height of $2e^2/h$ which is consistent with the presence of MBS. 
	It turns out that for the left lead this signature of MBS persists even for extended values of $\Gamma < \Gamma_c$.  
	The trivial ZBP has a splitting at zero energy, however, the
	splitting is negligible making it difficult to resolve even at
	low non-zero temperatures~\cite{Das_sharma_2017_ABS_ZBP_split, akmerov_reproducing_2019}.
	Another aspect which is different from the signatures in the topological regime is the absence of a ZBP on the right tunneling conductance (see Fig.~\ref{fig:NS_ZBP}(b)). The reason for this is the presence of bound states only on the left side of the wire, which  couples  to the corresponding left lead only. The absence of correlation in the local conductance is associated with the signature of ABS~\cite{Tiwari_Two-terminal_2018}. 
	As for the non-local conductance, the  plots in Figs.~\ref{fig:NS_ZBP}(c-d) exhibit weak signature to unambiguously distinguish bulk gap closing at $\Gamma_c$. Similar results for NS heterostructure have been demonstrated previously in the literature\cite{Loss_Local_and_nonlocal_2021, Das_sharma_Three_terminal_ZBP_2021}.

	
	
	\begin{figure}[!th]
		\includegraphics[clip=true,width=\columnwidth]{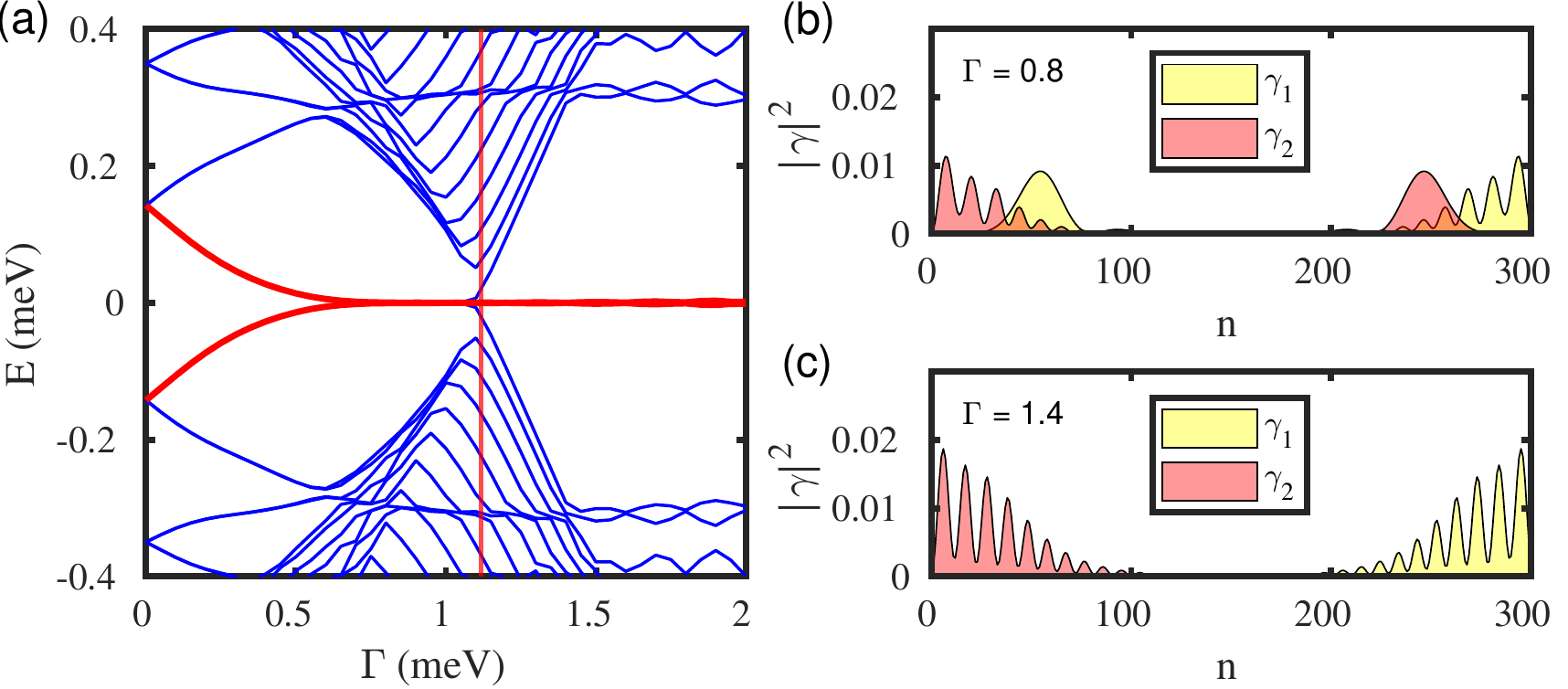}
		\caption{NSN system with parameters $t =  25.4$ meV, $\alpha = 2$ meV, $\Delta_0 = 0.5$ meV, $N_1 = N_2 = 40$, $\mu_{N_1} = \mu_{N_2} = 0.2$ meV, $\mu_S = 1$ meV and $N = 300$. (a) Energy spectrum, (b) the Majorana component in trivial region with ps-ABS  and, (c) MBS in topological region.}
		\label{fig:NSN_data}
	\end{figure}
	
	\ptitle{NSN heterostructure}
	In this scenario where the tunneling setup involves two leads it is more likely that the inhomogeneity will be present on both sides of the nanowire.
	The profile used to reproduce the NSN heterostructure is given in Eq.~(\ref{eqn:NSN_profile}) and shown in Fig.~\ref{fig:Inhomo_profile}(b). We find that for a range of parameters the NSN structure also hosts ABS. 
	The same is verified from the energy spectrum in Fig.~\ref{fig:NSN_data}(a), with ABS appearing much before the critical Zeeman term similar to that  for the NS system. Even if the energy spectrum for both the systems are nearly the same, the difference between the two can be seen from the Majorana component plots for the zero energy modes as shown in Fig.~\ref{fig:NSN_data}(b). For the NSN heterostructure the bound states for $\Gamma < \Gamma_c$ are also localized at both ends just like the MBS. These bound states  are the ps-ABS, with the partially separated Majorana components at both ends of the wire.
	
	\begin{figure}[!t]
		\includegraphics[clip=true,width=\columnwidth]{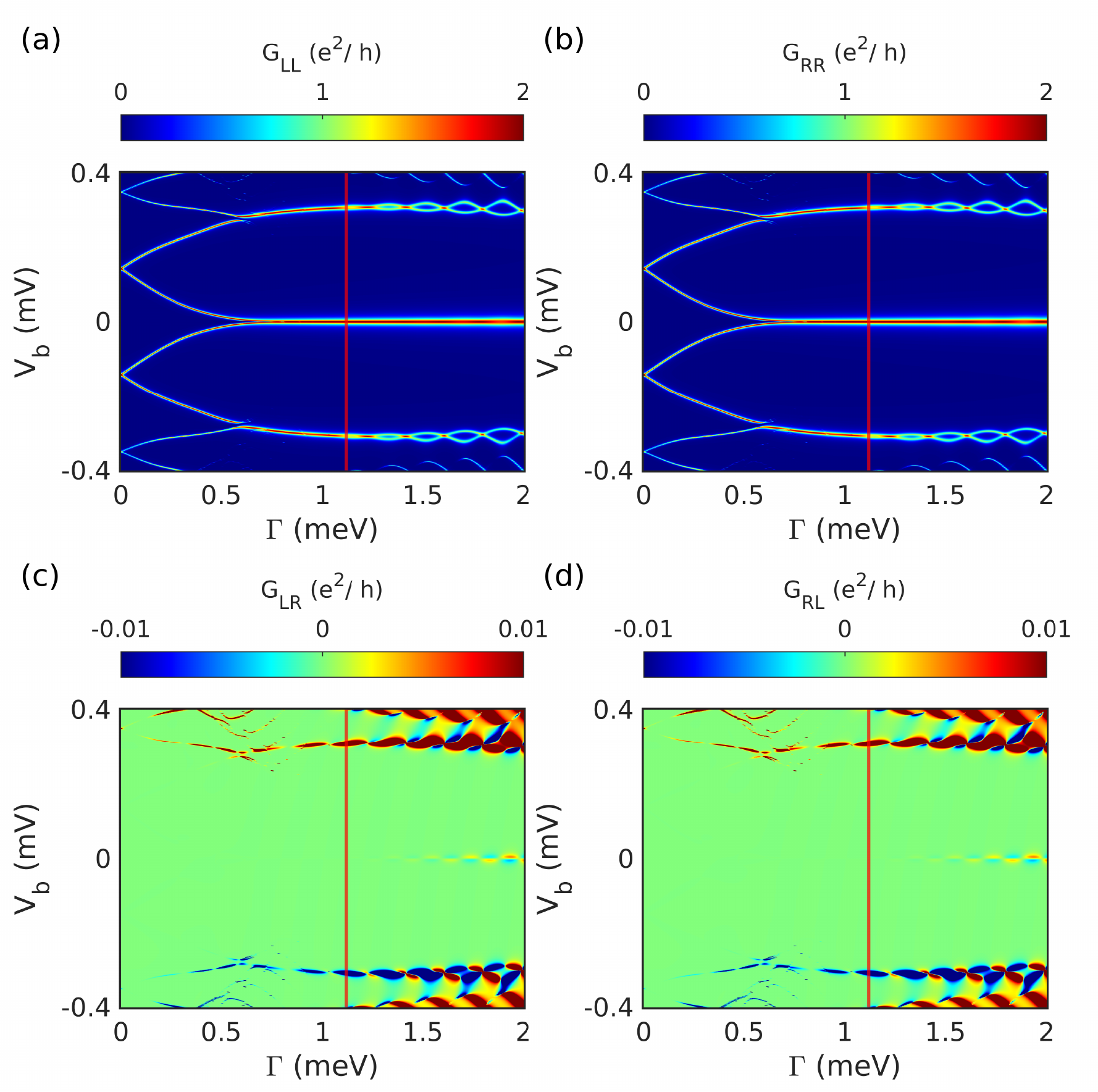}
		\caption{NSN system: (a-b)Local and (c-d) non-local tunneling conductance. Both local conductances shows ZBP much before $\Gamma_c$(vertical red line). Non-local conductance plots do not capture band closing signature.}
		\label{fig:NSN_ZBP}
	\end{figure}
	
	\ptitle{NSN heterostructure ZBP}
	The tunneling conductance of the NSN heterostructure is shown in Fig.~\ref{fig:NSN_ZBP}. Now both the left ($G_{LL}$) and the right ($G_{RR}$) local conductance exhibit quantized ZBP and are also correlated. As before, these signatures are present even for $\Gamma<\Gamma_c$.
	However, even for this system the non-local conductance is 
	unable to capture the band gap closing. This can again be attributed to the bulk states being localized in the superconducting region.
	Thus the correlated quantized ZBP produced by the ps-ABSs  and the absence of band gap signature implies no distinguishing feature between the  ABS and the MBS.
	These conclusions were   reported previously by Hess et. al.~\cite{Loss_Local_and_nonlocal_2021}.
	
	\begin{figure}[t]
		\includegraphics[clip=true,width=\columnwidth]{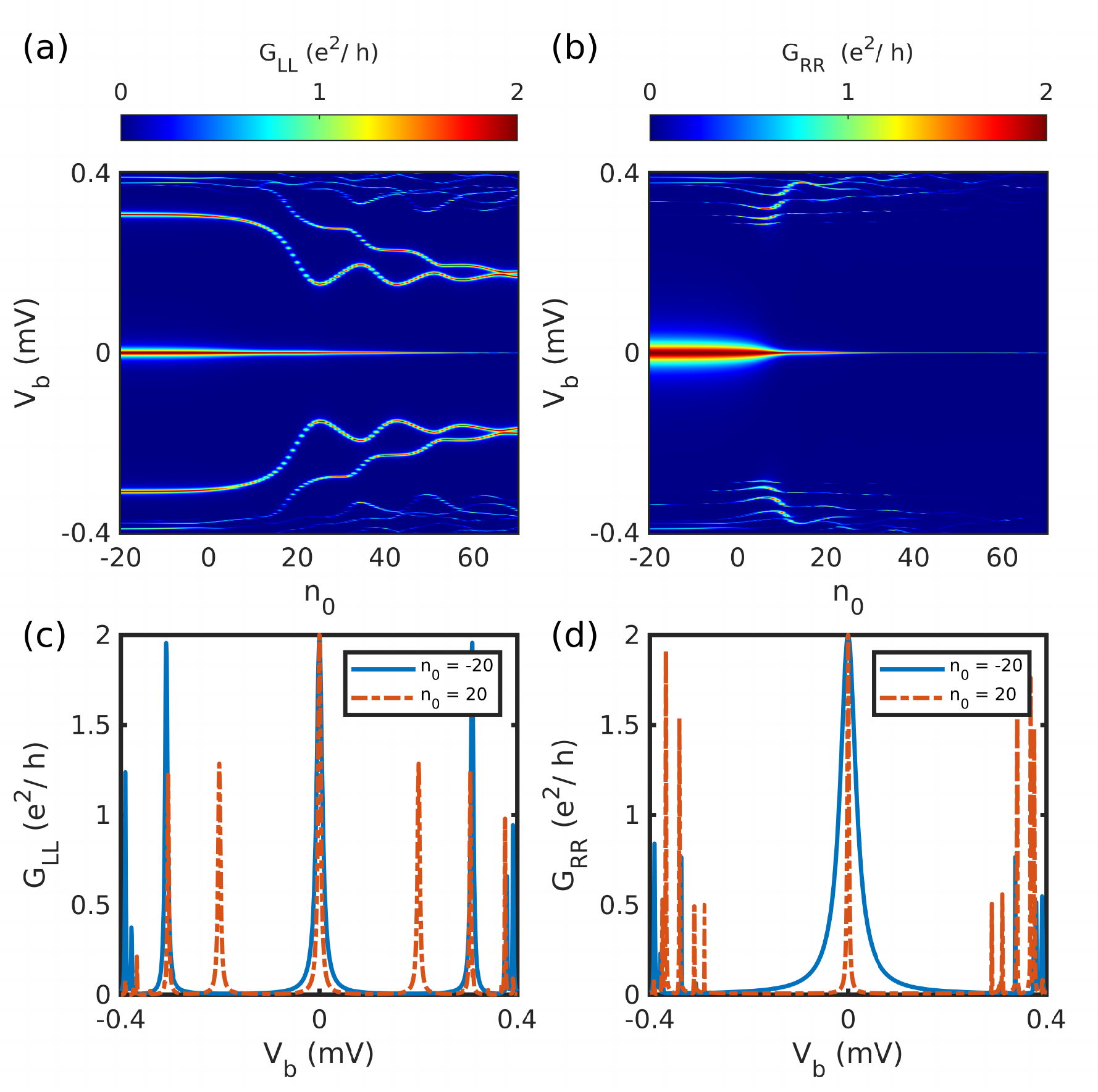}
		\caption{NS system: tunneling conductance signature on the application of ZMP to topological ZBP at $\Gamma > \Gamma_c$  with $\Gamma_0 = 1.4$ meV. (a-b) Local tunneling conductance for range of $n_0$ values and, (c-d) vertical line-cut of conductance for specific $n_0$, showing robustness of topological ZBP.}
		\label{fig:NS_M}
	\end{figure}
	
	\begin{figure}[t]
		\vspace{0.5cm}
		\includegraphics[clip=true,width=\columnwidth]{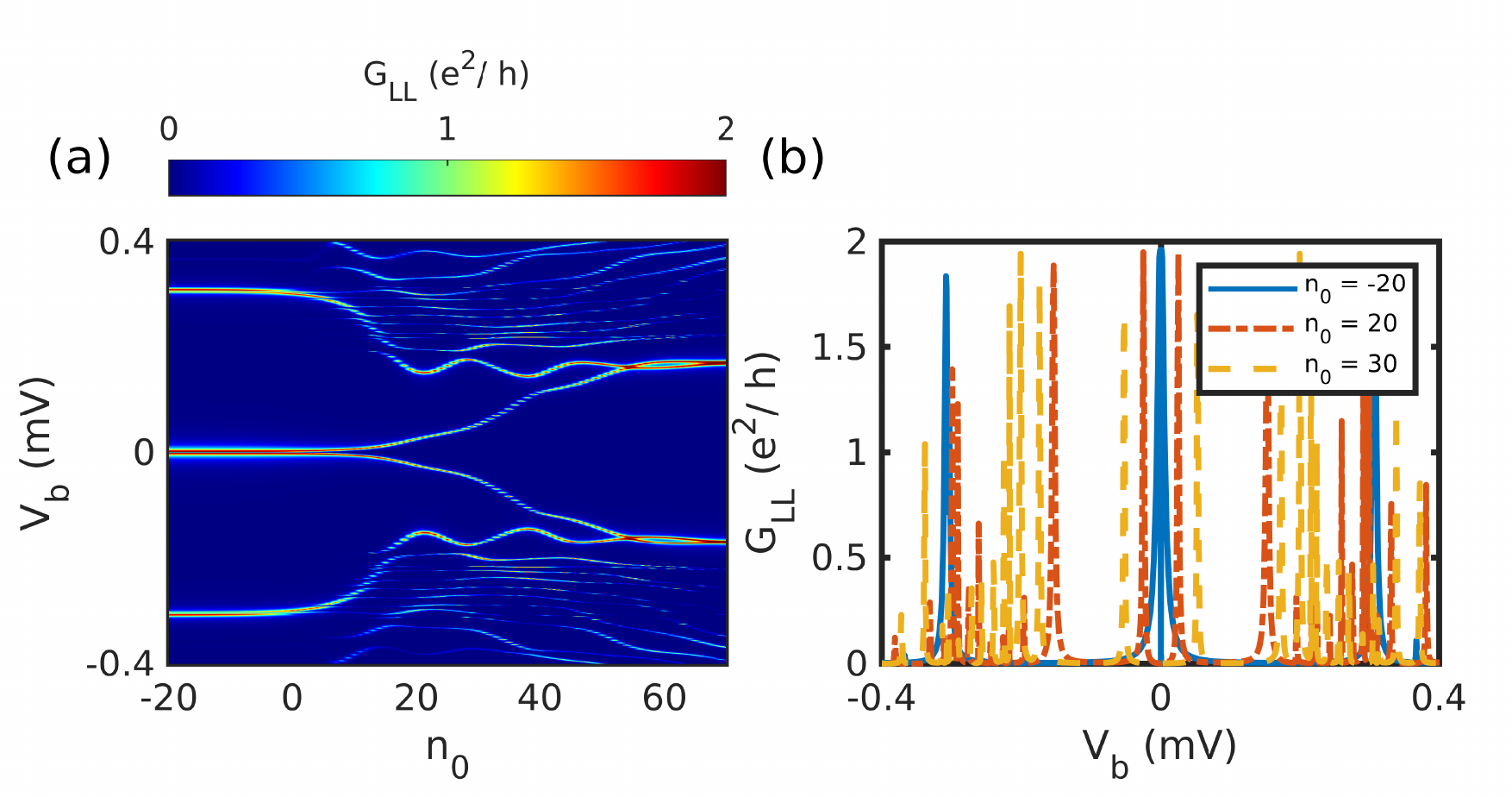}
		\caption{NS system: tunneling conductance signature on the application of ZMP to trivial ZBP at $\Gamma < \Gamma_c$  with $\Gamma_0 = 0.8$ meV. (a) Left local tunneling conductance for range of $n_0$ values and, (b) vertical line-cut of conductance for specific $n_0$ which captures the splitting of trivial ZBP.}
		\label{fig:NS_A}
	\end{figure}
	
	\begin{figure}[t]
		\includegraphics[clip=true,width=\columnwidth]{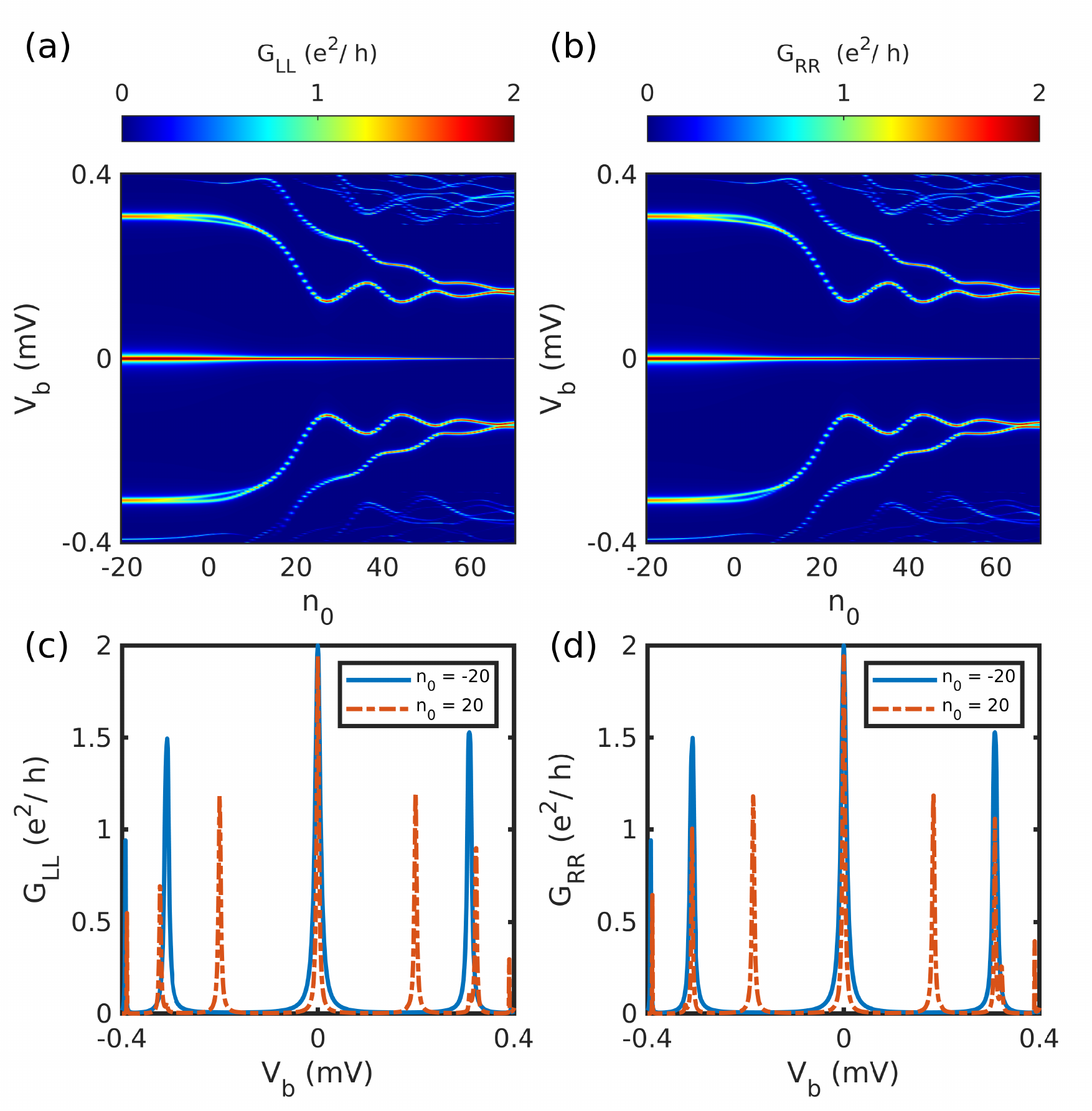}
		\caption{NSN system: tunneling conductance signature on the application of ZMP to topological ZBP at $\Gamma > \Gamma_c$ with $\Gamma_0 = 1.4$ meV. (a-b) Local tunneling conductance for range of $n_0$ values and, (c-d) vertical line-cut of conductance for specific $n_0$, capturing robustness of topological ZBP.}
		\label{fig:NSN_M}
	\end{figure}
	
	\begin{figure}[t]
		\includegraphics[clip=true,width=\columnwidth]{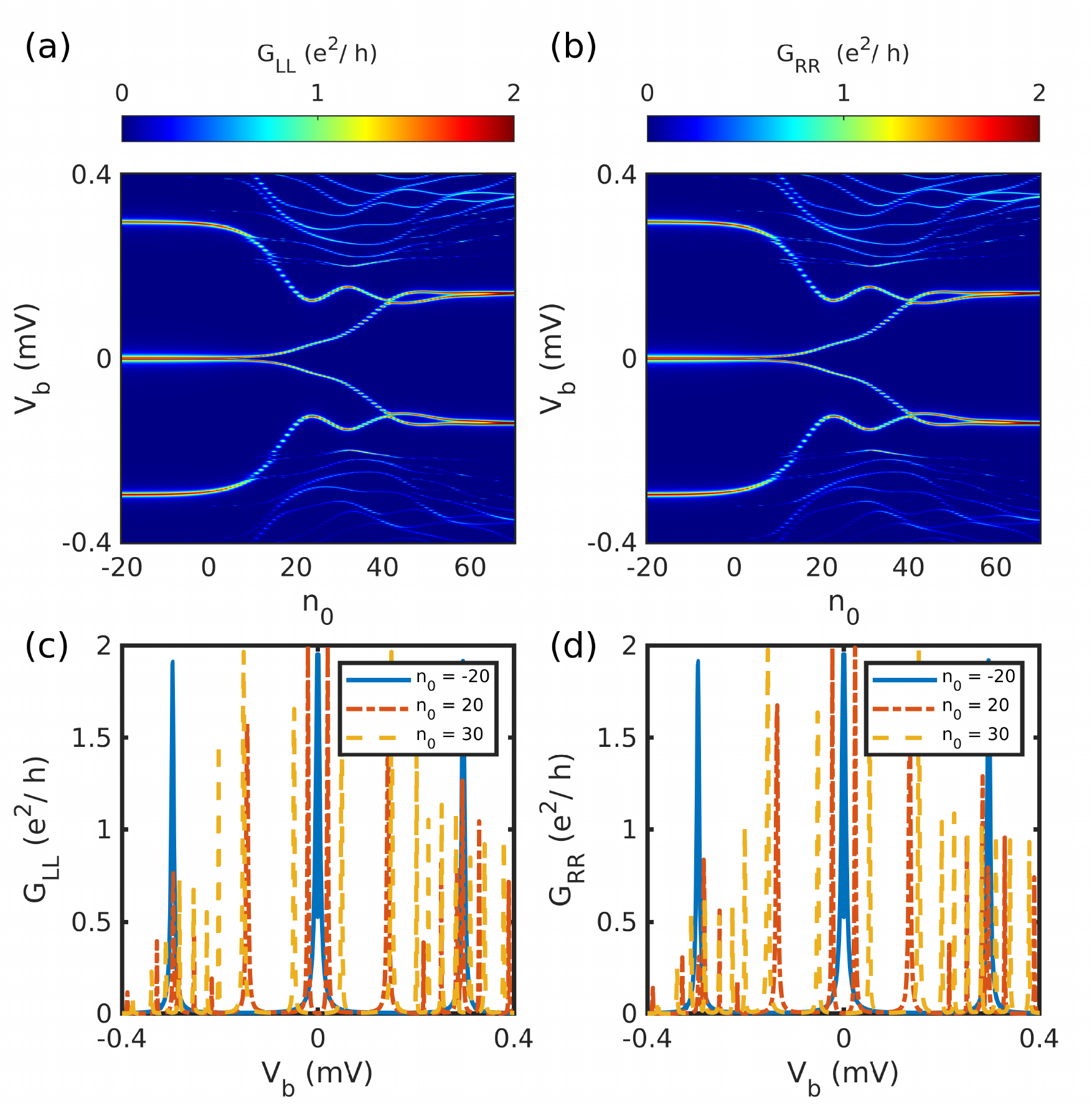}
		\caption{NSN system: tunneling conductance signature under the application of ZMP to trivial ZBP at $\Gamma < \Gamma_c$  with $\Gamma_0 = 0.8$ meV. (a-b) Local tunneling conductance for range of $n_0$ values and, (c-d) vertical line-cut of conductance for specific $n_0$, shows the spliting of trivial ZBP in both local conductance.}
		\label{fig:NSN_A}
	\end{figure}
	
	\ptitle{Moving protocol : NS system}
	
	We will now present the effect of ZMP on the trivial and topological ZBP for the inhomogeneous system. As we saw before, the NS system has both the topological and trivial quantized ZBP.
	The quantized  ZBP for $\Gamma > \Gamma_c$ is present in both the left and the right local conductance, which on the application of ZMP persists for a range of $n_0$ values. 
	This result is similar to the case of homogeneous wire with topological ZBP as discussed in Sec.~\ref{subsec:Homogeneous_1} . Interestingly, effect of ZMP on the trivial ZBP present on the left conductance for the NS system is significantly different (see Fig.~\ref{fig:NS_A}). The ZBP splits into two separate peaks as $n_0$ increases. 
	As discussed before, the topological and trivial peak for the NSN system is the most challenging to distinguish. So we apply ZMP to the tunneling conductance signature to the NSN system. The topological ZBP persists for an extended range of $n_0$, and the peak width decreases with an increase in $n_0$ as is the case for the NS system (see Fig.~\ref{fig:NSN_M}). In contrast, under ZMP, the trivial ZBP present on both local conductances splits into two separate peaks as shown in Fig.~\ref{fig:NSN_A}.
	
\begin{figure}[t!]
		\includegraphics[clip=true,width=\columnwidth]{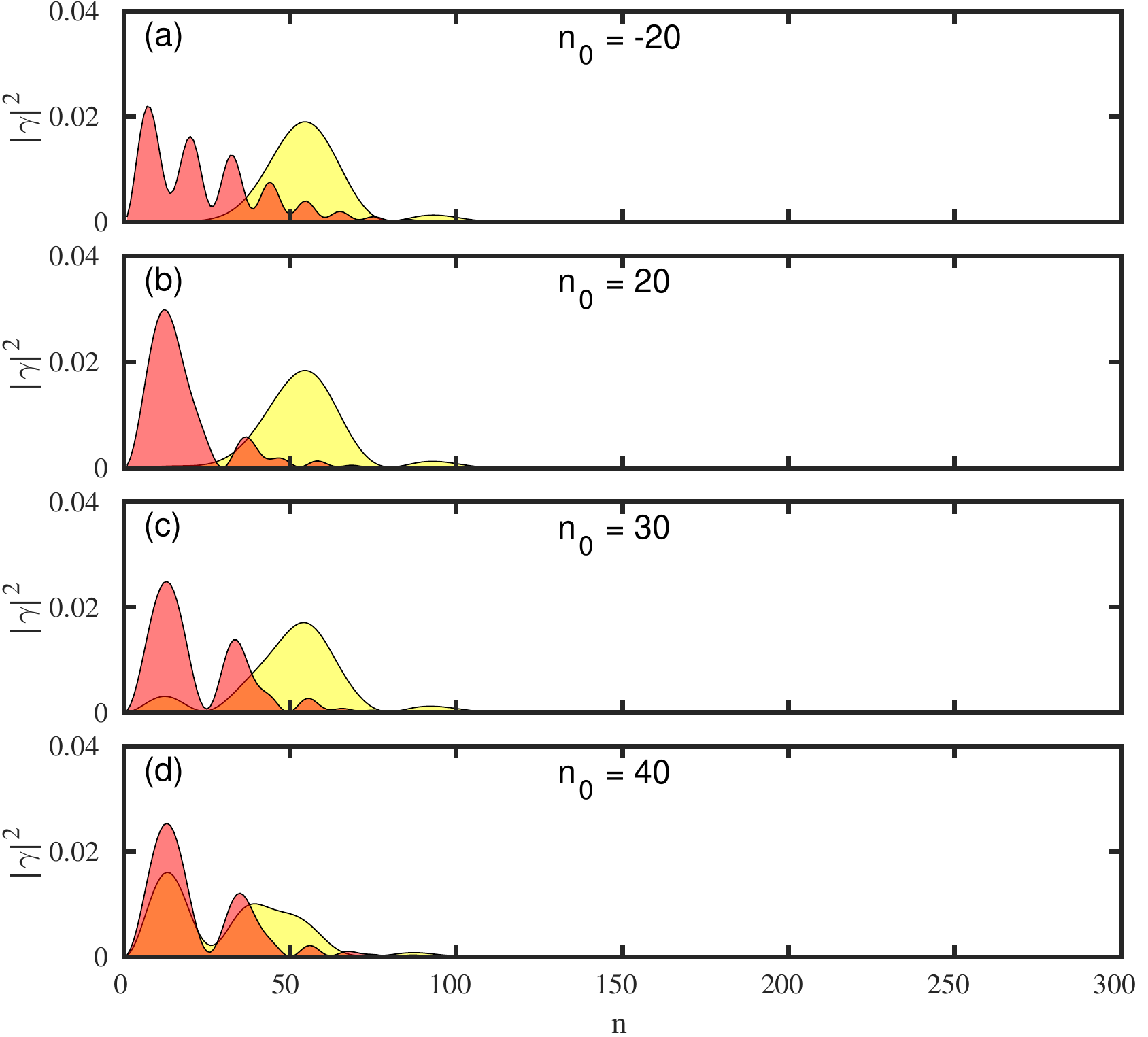}
		\caption{Majorana components of NS system on the application of ZMP to trivial state at $\Gamma < \Gamma_c$ with $\Gamma_0 = 0.8$ meV for different $n_0$ values. The Majorana component overlap increases with $n_0$.}
		\label{fig:NS_overlap_A}
	\end{figure}

\begin{figure}[t!]
		\includegraphics[clip=true,width=\columnwidth]{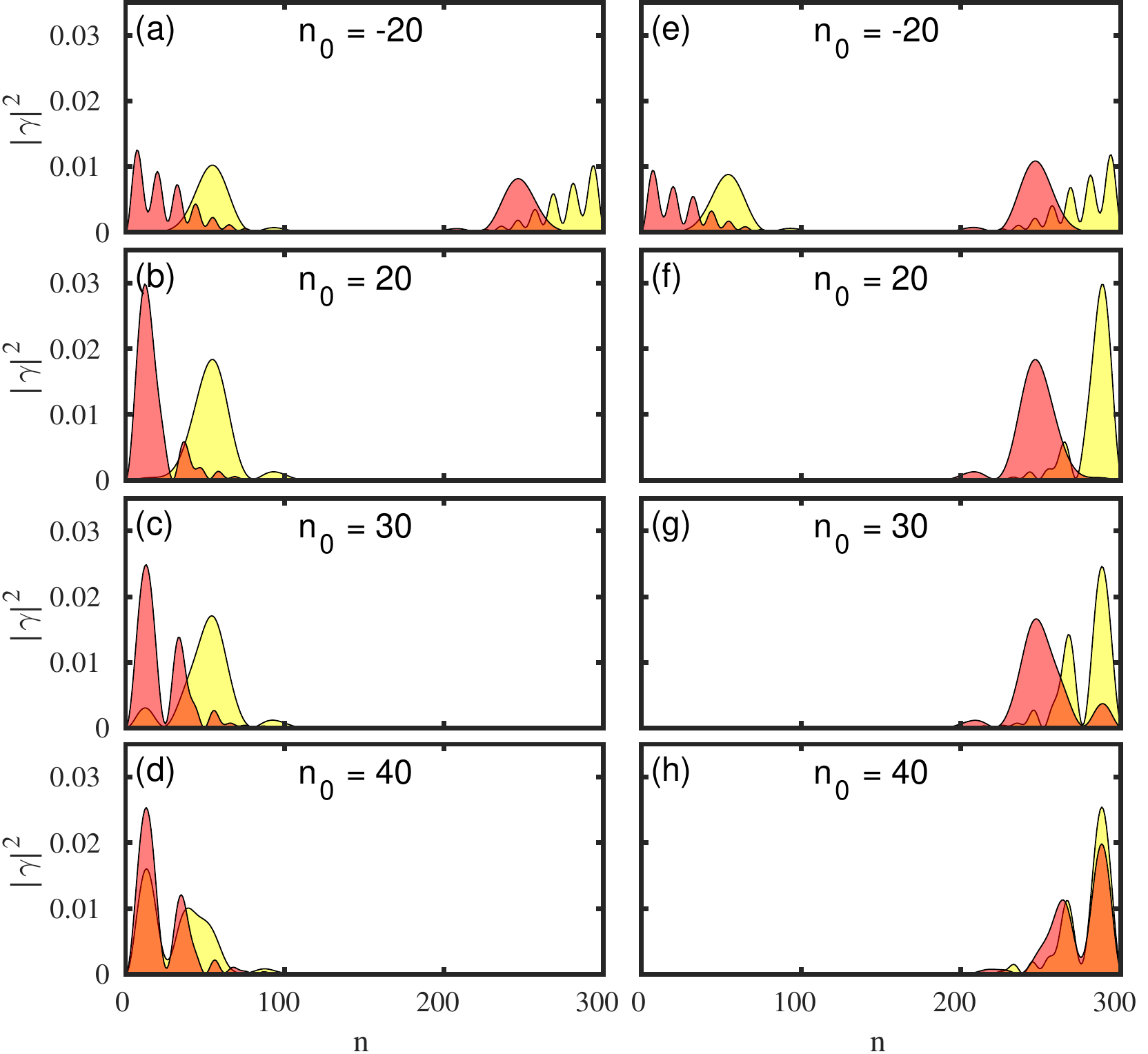}
		\caption{Majorana components of NSN system under ZMP for the trivial state at $\Gamma < \Gamma_c$ with $\Gamma_0 = 0.8$ meV for different $n_0$ values. For both ps-ABS the overlap increases with $n_0$.}
		\label{fig:NSN_overlap_A}
	\end{figure}

	To get an insight into the behavior of topological and trivial peaks under the moving protocol, we focus our attention on the Majorana components of the state close to zero energy. 
	Figure~\ref{fig:NS_overlap_A} shows the effect of ZMP on the Majorana components of the trivial states close to zero energy for the NS system. 
	The trivial ZBP originates due to the ps-ABS which involves partially separated Majorana components. Under the application of ZMP, the overlap between the Majorana components increases, causing them to acquire a finite energy split, resulting in splitting of the ZBP into two peaks away from zero bias. For the case of NSN heterostructure we have two states close to zero energy, both of which are present at either ends of the wire. These states are ps-ABSs with small overlapping Majorana components. On the application of ZMP, the overlap increases for both the states (see Fig.~\ref{fig:NSN_overlap_A}), resulting in a finite energy split and the trivial ZBP on both local conductances splits into two separate peaks at finite bias.
	
	\begin{figure}[t!]
		\includegraphics[clip=true,width=\columnwidth]{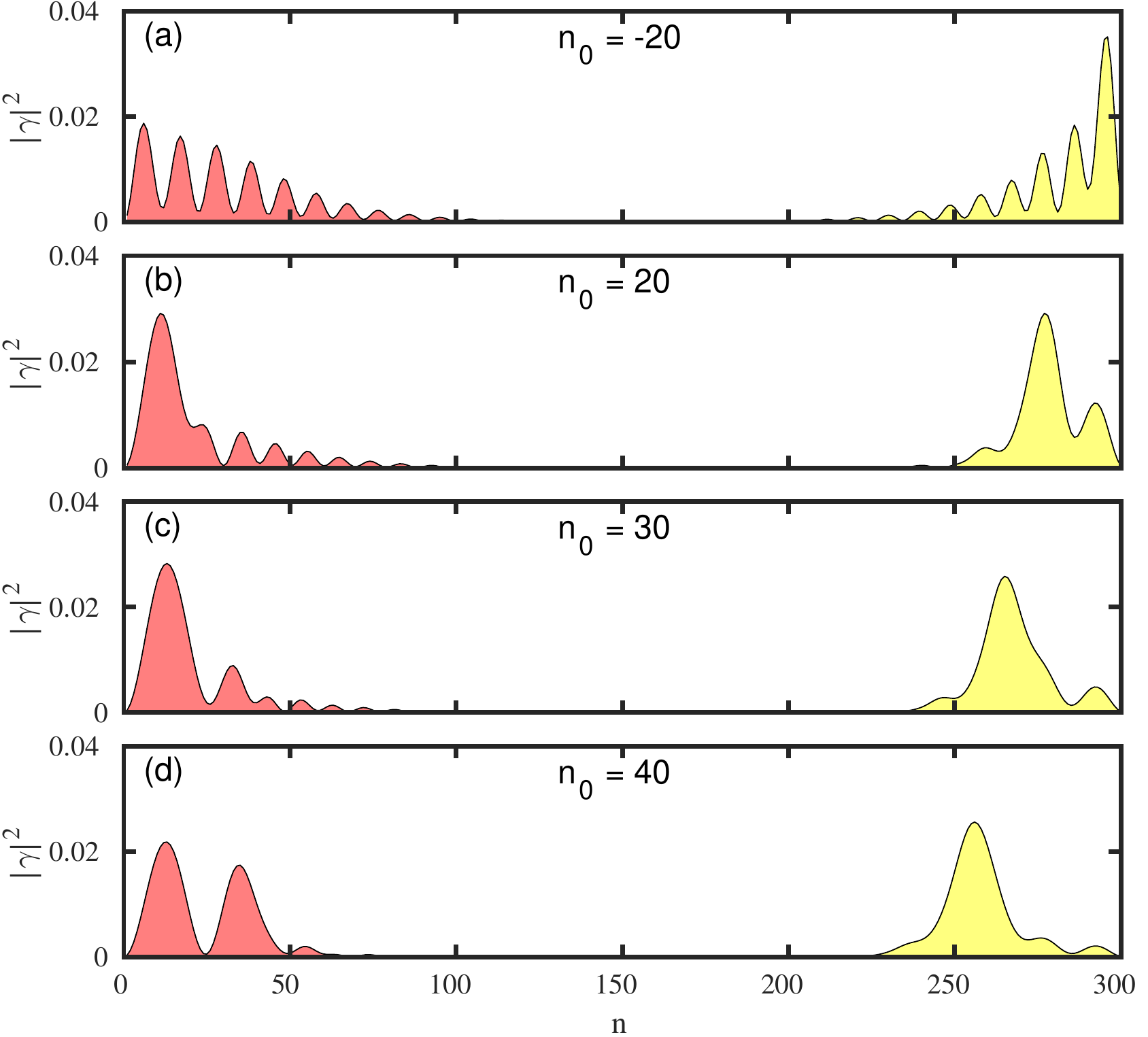}
		\caption{Majorana components of NS system on the application of ZMP to topological state at $\Gamma > \Gamma_c$ with $\Gamma_0 = 1.4$ meV for different $n_0$ values. Majorana components remain separated.}
		\label{fig:NS_overlap_M}
	\end{figure}
	
	\begin{figure}[t!]
		\includegraphics[clip=true,width=\columnwidth]{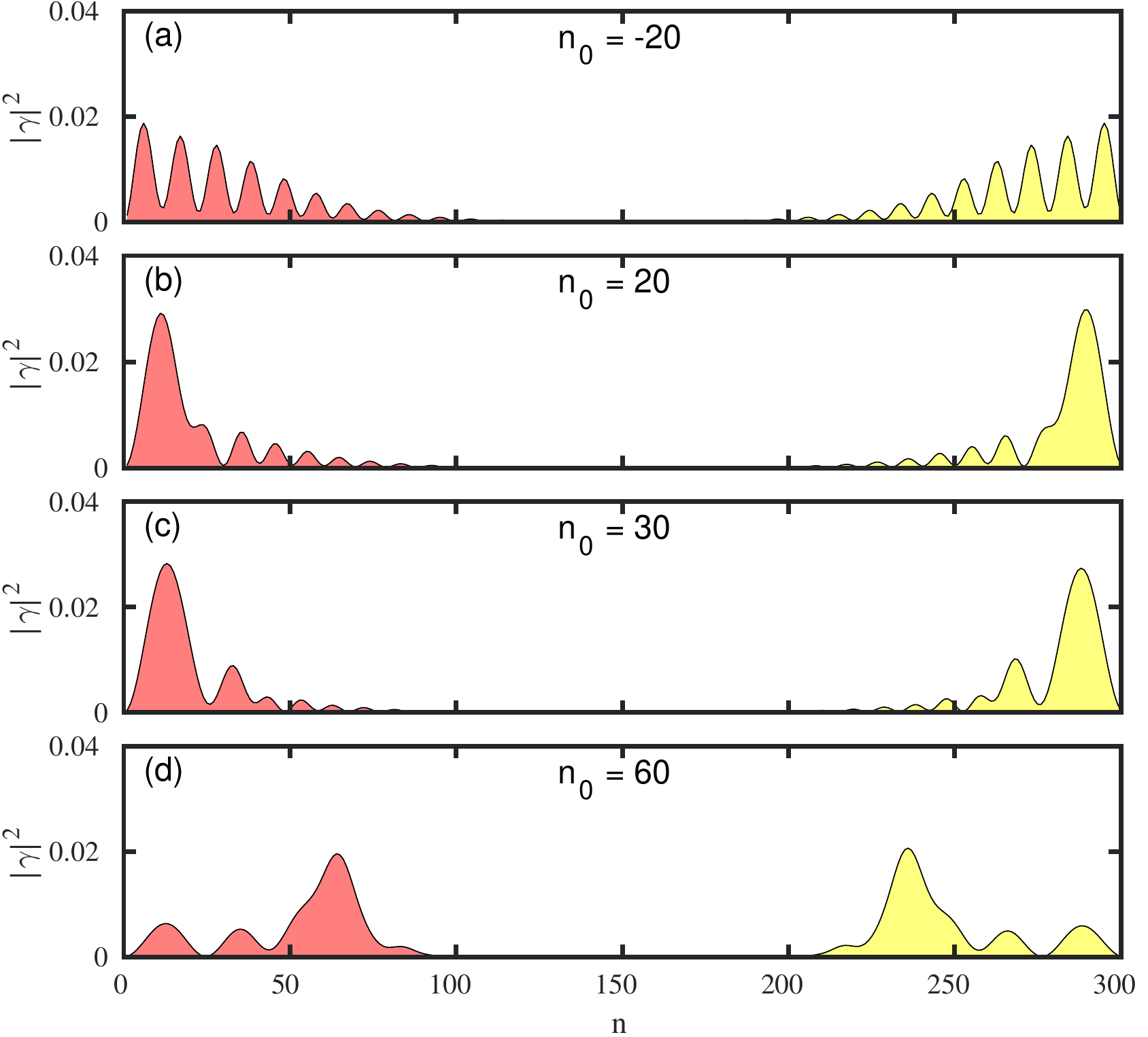}
		\caption{Majorana components of NSN system on the application of ZMP to topological state at $\Gamma > \Gamma_c$ with $\Gamma_0 = 1.4$ meV for different $n_0$ values. The Majorana components remain separated for range of $n_0$ values.}
		\label{fig:NSN_overlap_M}
	\end{figure}
	
	Performing similar analysis for the topological states, we show the effect of ZMP on the Majorana components of MBS in NS and NSN systems in Fig.~\ref{fig:NS_overlap_M} and Fig.~\ref{fig:NSN_overlap_M}, respectively. Fully separated Majorana components represent the MBS, and the components remain separate throughout the protocol for both the NS and NSN systems. As a result, the MBS stays at zero energy resulting in the ZBP staying at zero bias throughout the protocol. Under the protocol, the components move away from the edge (see Fig.~\ref{fig:NSN_overlap_M}), resulting in weak coupling with the leads, leading to decreased peak width. For a large $n_0$ value($ \approx 70$), the states will be far from the edge, making the peak disappear.
	Another thing to notice is that the decrease in peak height on both the local conductances of NS is different under the protocol as shown in Fig.~\ref{fig:NS_M}(a-b). This is mainly due to the presence of the N region on the left side and MBS leaking to this region, see Fig.~\ref{fig:NSN_overlap_M}. 
	
	\begin{figure}[t!]
		\includegraphics[clip=true,width=\columnwidth]{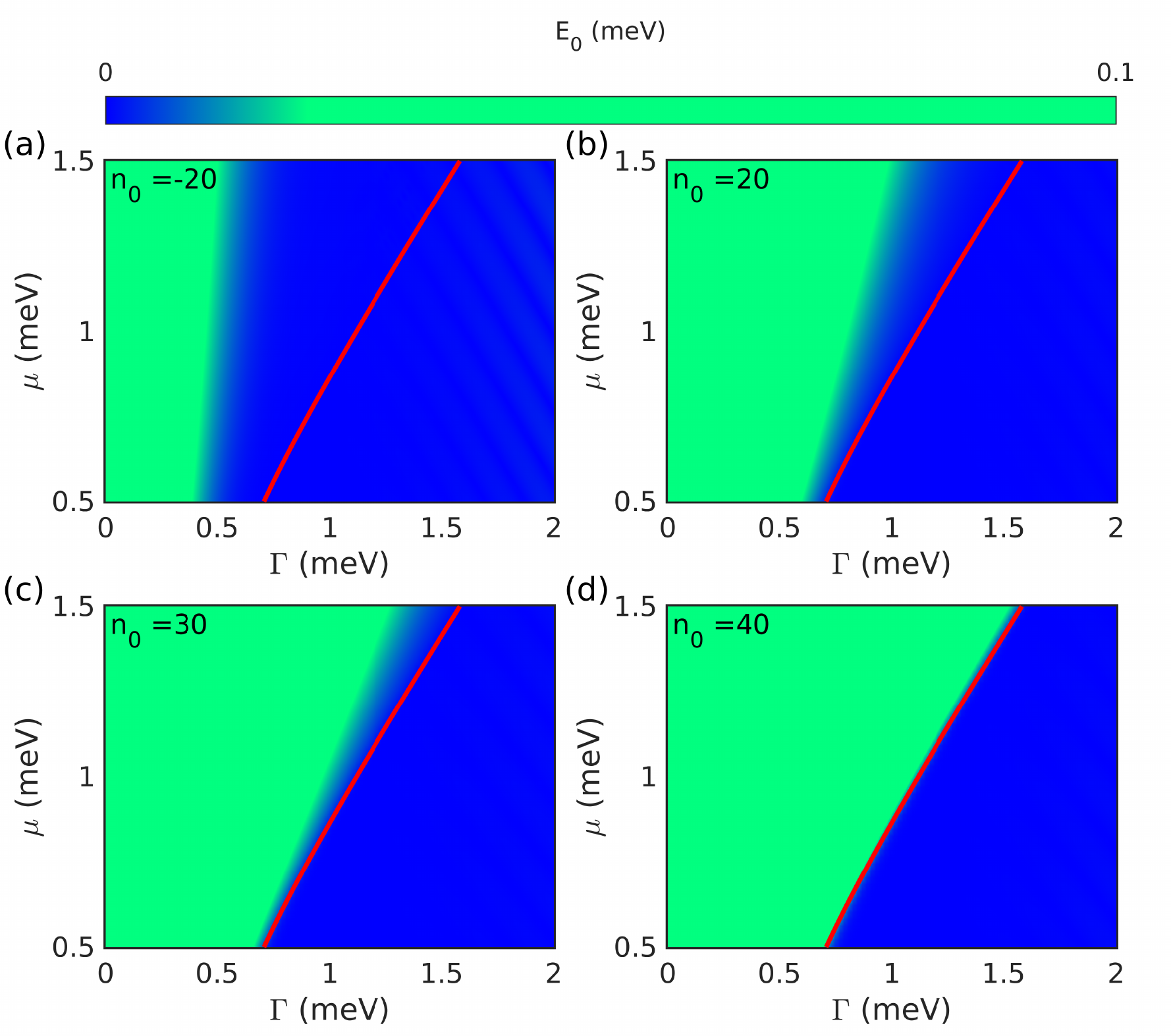}
		\caption{Phase potrait with low energy eigenvalues of NSN system for a range of chemical potential ($\mu$) in S region and Zeeman term ($\Gamma$) with different instances of ZMP. The ABS moves from zero energy to finite energy as $n_0$ is increased from (a) $n_0 = -20$ to (d) $n_0 = 40$, while MBS in topological regime (after red line) continues to be at zero energy.}
		\label{fig:phase_spliting}
	\end{figure}
	
	To showcase that the robustness of MBS energy to ZMP and the splitting is present for all parameter range of ABS, we plot the phase potrait corresponding to the lowest energy eigenvalues for the parameter range with the ZMP for different $n_0$ values.
	As shown in Fig.~\ref{fig:phase_spliting}, with $n_0 = -20$ we have normal NSN system with low energy ABS present before the topological region (separated by red line) with the application of the ZMP these low energy states moves away from zero energy and for $n_0 = 40$ there are no zero energy states outside the topological region. These plots show that the  splitting is not specific to certain parameter values and the ABS move away from zero energy for all the parameter ranges. On the other hand the eigenvalues of the MBS in the topological region remains unaffected.

	\subsection{Moving protocol with external potential\label{subsec:mu_moving}}
	As discussed in Sec.~\ref{sec:model}, external potential can also be used to modify the topological length of a nanowire and produce a trivial-topological-trivial structure. In earlier works, this strategy has been suggested in the context of braiding protocol~\cite{alicea_non_Abelian_2011, Das_Sharma_MZM_QC_2015, Alicea_Antipov_Dynamic_of_Majorana_qubit_2018, Zolle_MBS_noisy_kitaev_2015}. However, a moving protocol with tunneling measurements at the edges has not been considered before for the purpose of distinguishing topological states from trivial ones. In this subsection we will be presenting the effect of the PMP on trivial and topological ZBP.
	
	\subsubsection{Homogeneous wire\label{subsubsec:Homogeneous_1_2}}
	\begin{figure}[t]
		\includegraphics[clip=true,width=\columnwidth]{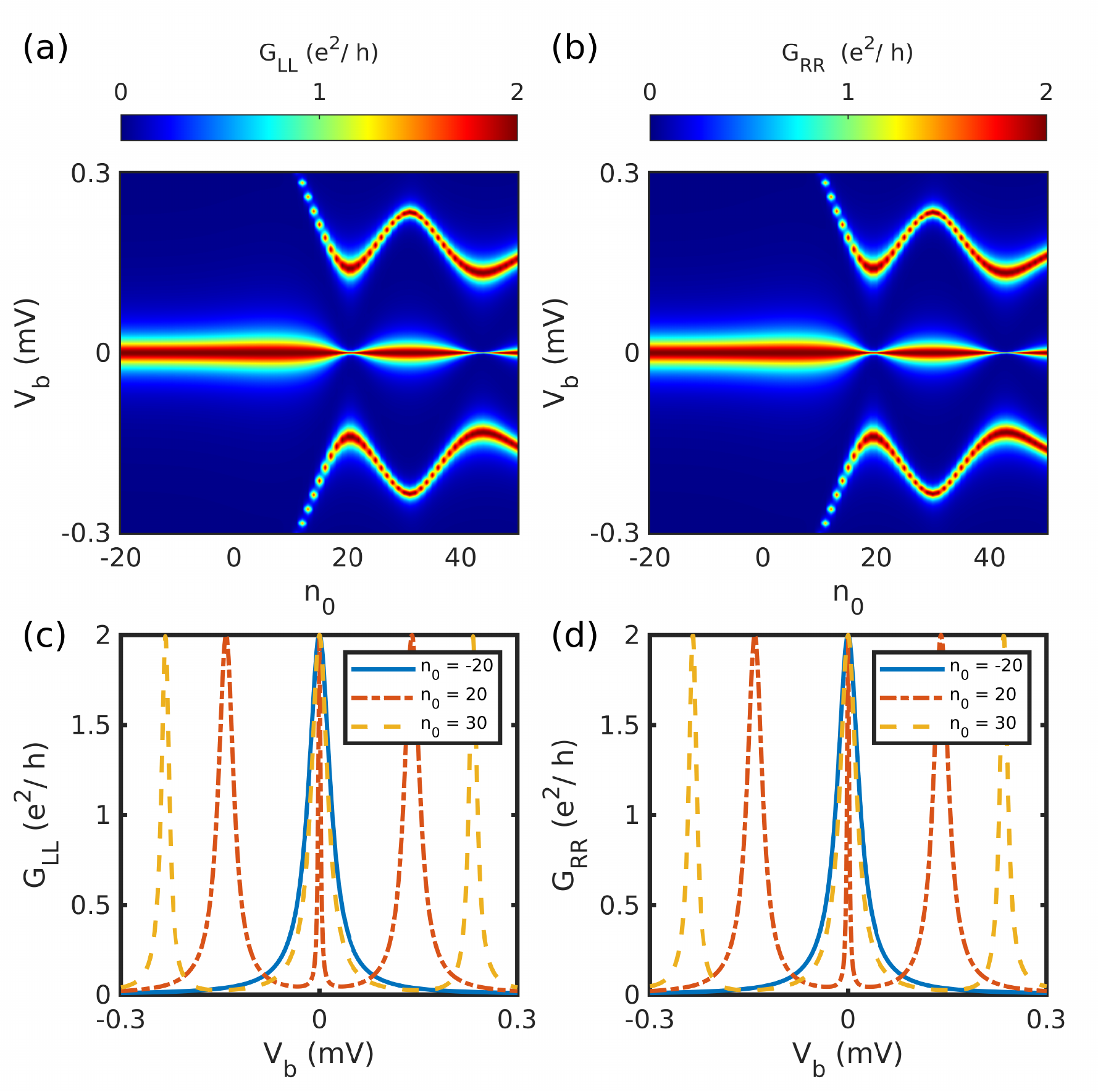}
		\caption{Homogeneous wire tunneling conductance signature on  the application of PMP to topological ZBP at $\Gamma = 1.5 meV > \Gamma_c$ with $V_0 = 2$ meV. (a-b) Local tunneling conductance for range of $n_0$ values and, (c-d) vertical linecut of conductance for specific $n_0$, shows the robustness of topological ZBP.}
		\label{fig:Homogeneous_ZBP_moving_mu}
	\end{figure}
	
	\begin{figure}[t]
		\includegraphics[clip=true,width=\columnwidth]{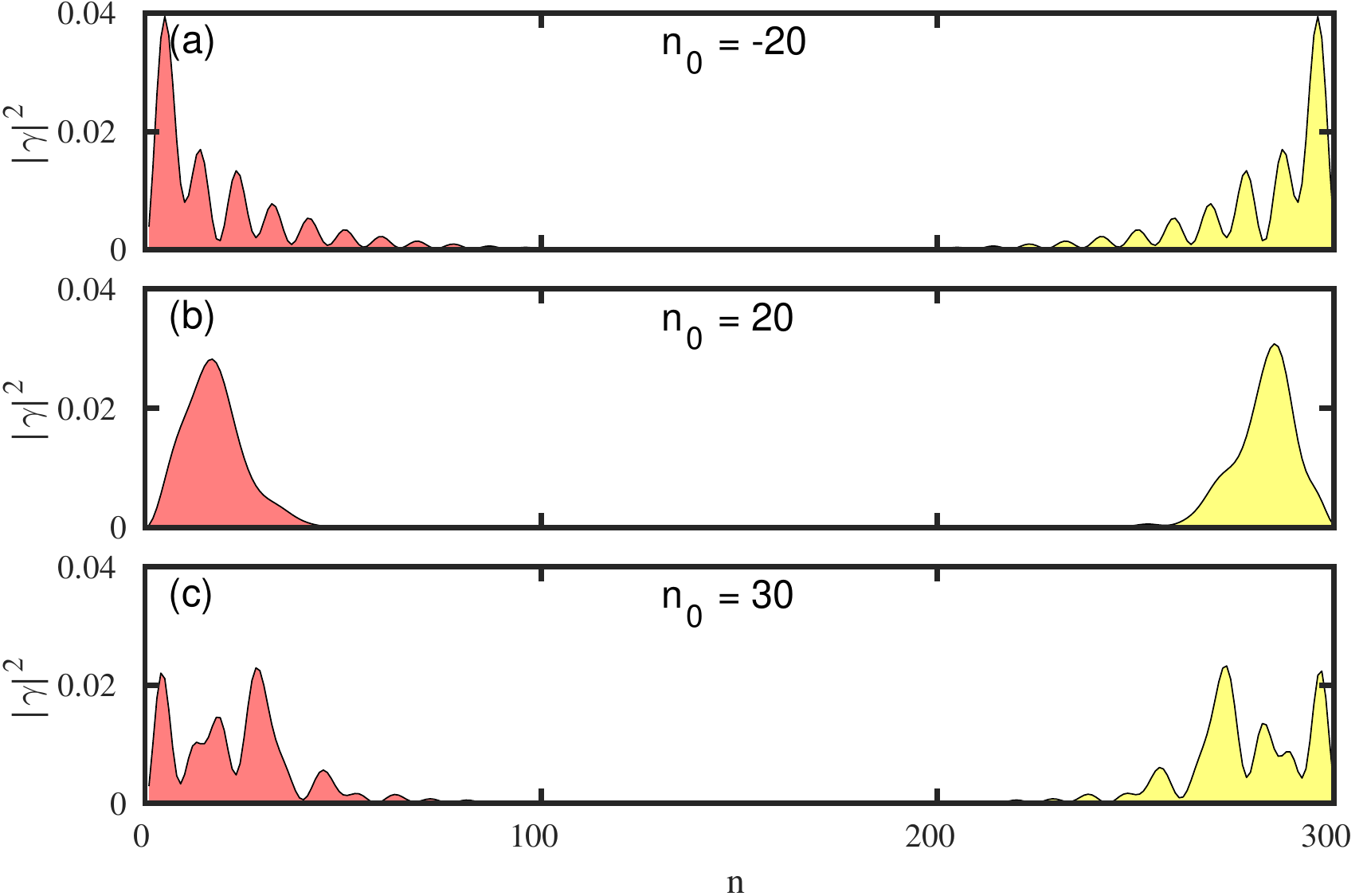}
		\caption{Majorana components of low energy states in Homogeneous wire under the application of PMP to topological states at $\Gamma = 1.5 meV > \Gamma_c$  with $V_0 = 2$ meV. Majorana components remain separated, they also leak towards the ends.}
		\label{fig:Homogeneous_overlap_mu}
	\end{figure}
	
	In Sec.~\ref{subsec:Homogeneous_1}, we presented the result of ZMP on the homogeneous wire. In the case of PMP, the behavior of topological ZBP remains unchanged. For a range of $n_0$ values the topological ZBP remains robust (see Fig.~\ref{fig:Homogeneous_ZBP_moving_mu}). The most important difference between the ZMP and the one with the PMP is the MBS's confinement to the topological region. While for the PMP, the Majorana leaks into the trivial region (see  Fig.~\ref{fig:Homogeneous_overlap_mu}), it is comparatively suppressed for ZMP. The leakage into the ends allow coupling with the lead, restoring the peak width. We find that the peak width exhibits oscillatory dependence on $n_0$. For example, plotted are the ZBP width for $n_0 = 20$ and $30$ in Fig.~\ref{fig:Homogeneous_ZBP_moving_mu}, and the corresponding Majorana components in  Figs.~\ref{fig:Homogeneous_overlap_mu}(b-c). For $n_0 = 20$, the Majorana components are away from the lead resulting in small peak width, while for $n_0 = 30$, the leakage of MBS towards the ends of the wire, results in an increase in peak width. While there are certain differences in the signatures corresponding to the two ways of applying the moving protocol, we would like to draw attention to the main signature corresponding to the MBS which remain unaltered for both the protocols i.e. the peak doesn't move from zero bias throughout the moving protocol.

	\subsubsection{Inhomogeneous wire\label{subsubsec:Inhomogeneous_2}}
	
	\begin{figure}[t!]
		\includegraphics[clip=true,width=.48\textwidth]{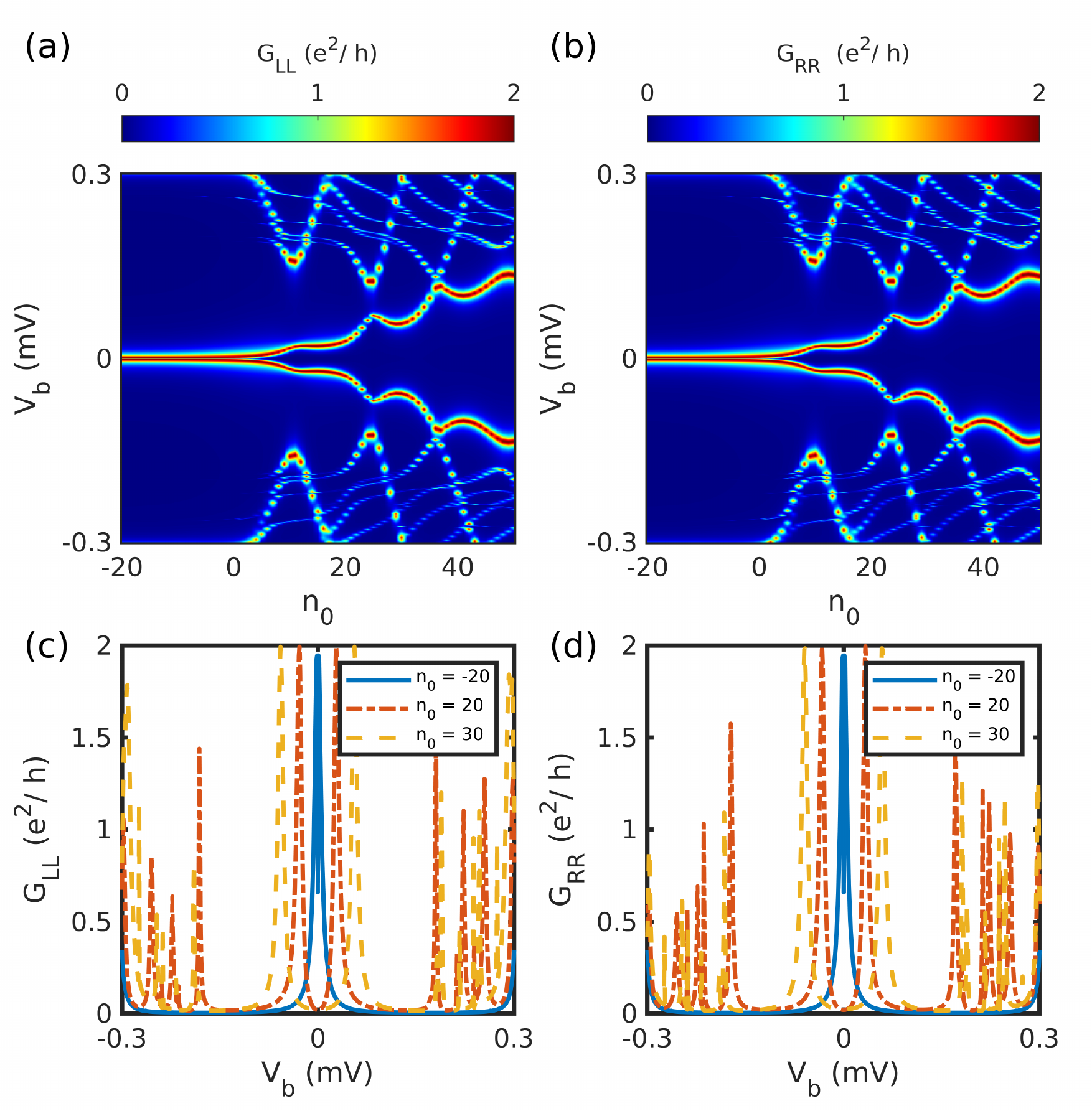}
		\caption{NSN system: tunneling conductance signature under the PMP to topological ZBP at $\Gamma = 0.8 meV > \Gamma_c$ with $V_0 = 2$ meV. (a-b) Local tunneling conductance for range of $n_0$ values and, (c-d) vertical linecut of conductance for specific $n_0$, the trivial ZBP splits under the protocol.}
		\label{fig:NSN_A_mu}
	\end{figure}
	
	\begin{figure}[t!]
		\includegraphics[clip=true,width=.48\textwidth]{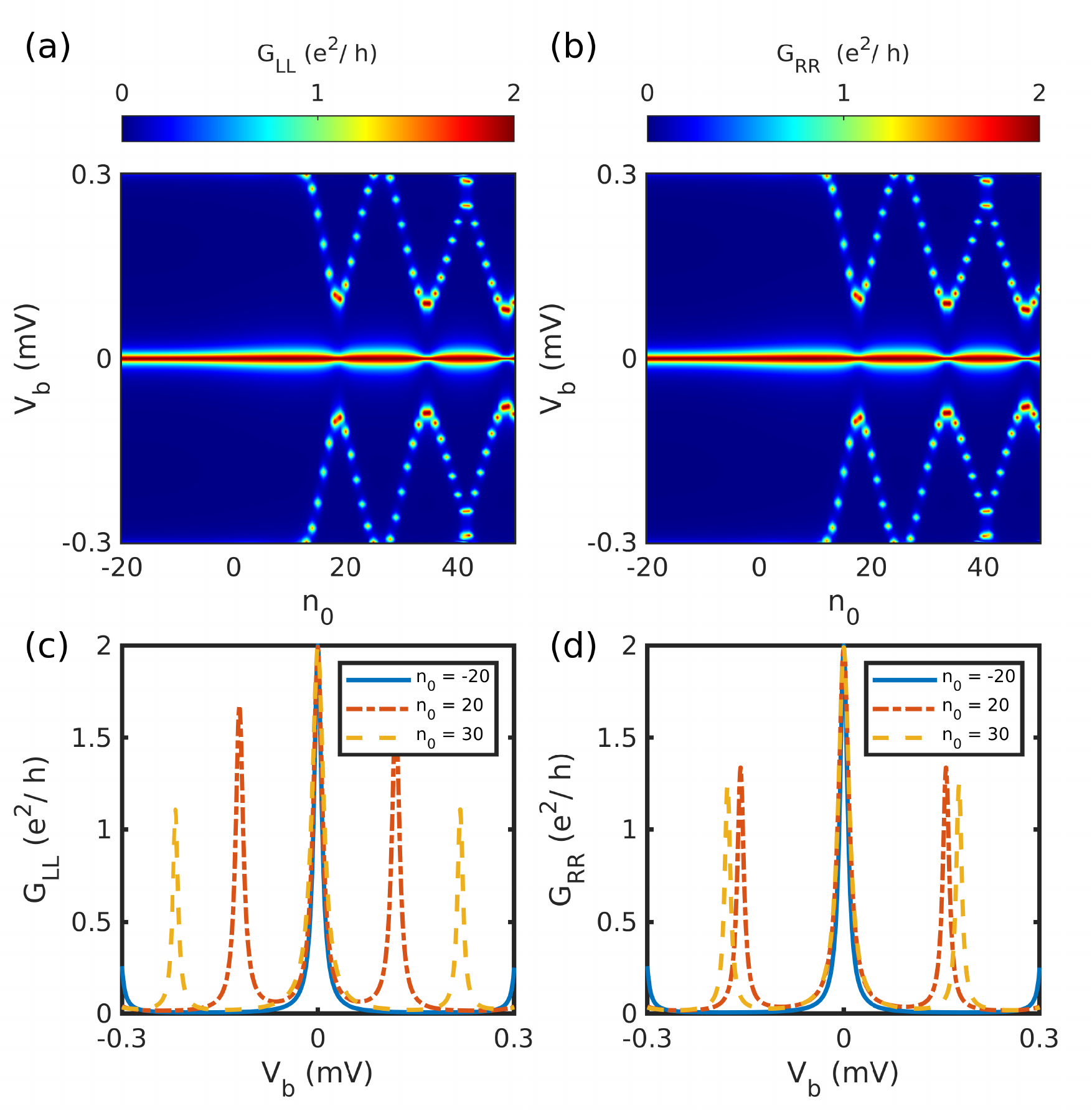}%
		\caption{NSN system: tunneling conductance signature on the application of the PMP to topological ZBP at $\Gamma = 1.4 meV > \Gamma_c$ with $V_0 = 2$ meV. (a-b) Local tunneling conductance for range of $n_0$ values and, (c-d) vertical linecut of conductance for specific $n_0$, showing robustness of topological ZBP.}
		\label{fig:NSN_M_mu}
	\end{figure}

	\begin{figure}[t!]
		\includegraphics[clip=true,width=.48\textwidth]{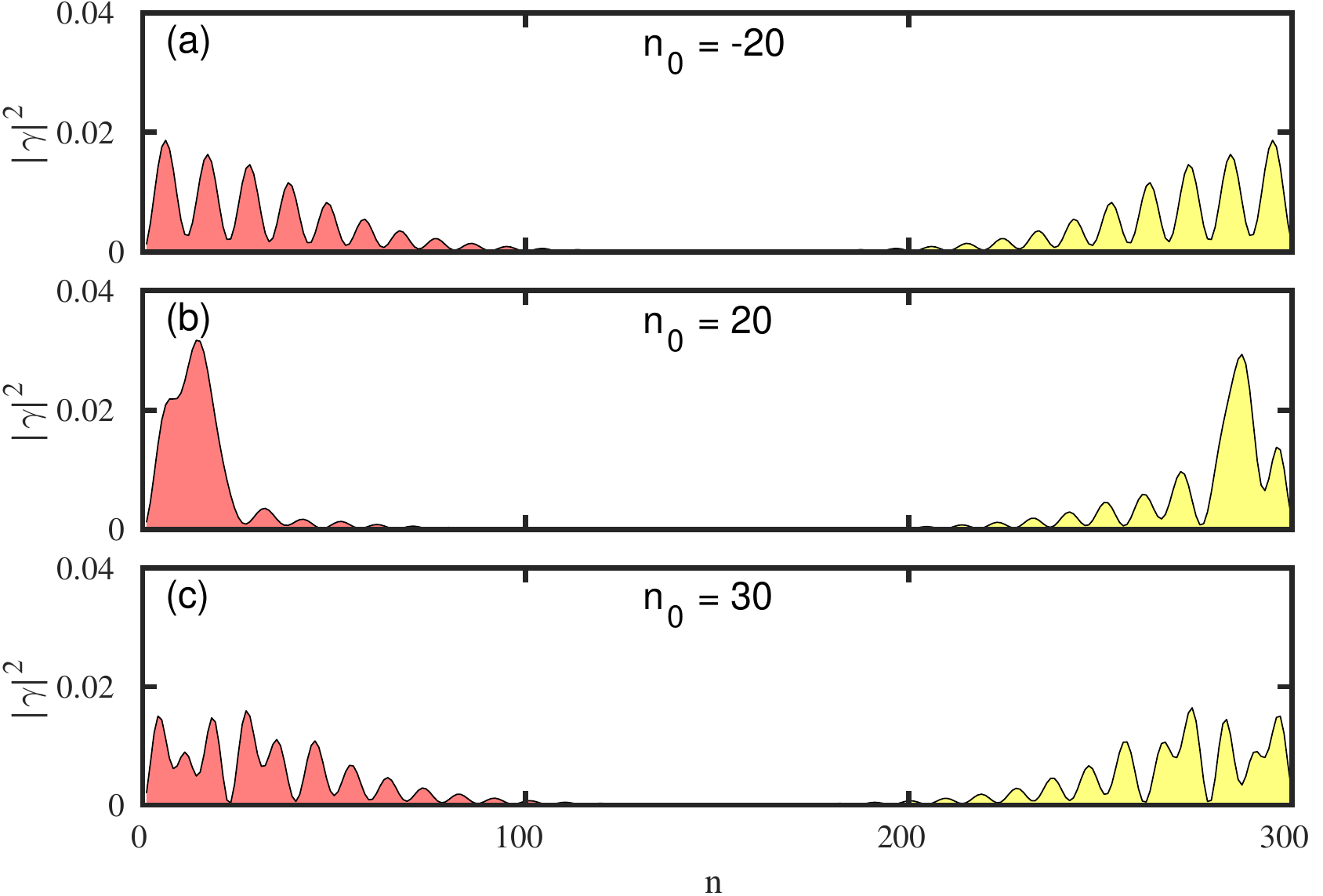}
		\caption{Majorana components of low energy states in NSN system under PMP for $\Gamma = 1.4 meV > \Gamma_c$ with $V_0 = 2$ meV. The components remain separated.}
		\label{fig:NSN_overlap_M_mu}
	\end{figure}
	
	\begin{figure}[t!]
		\includegraphics[clip=true,width=.48\textwidth]{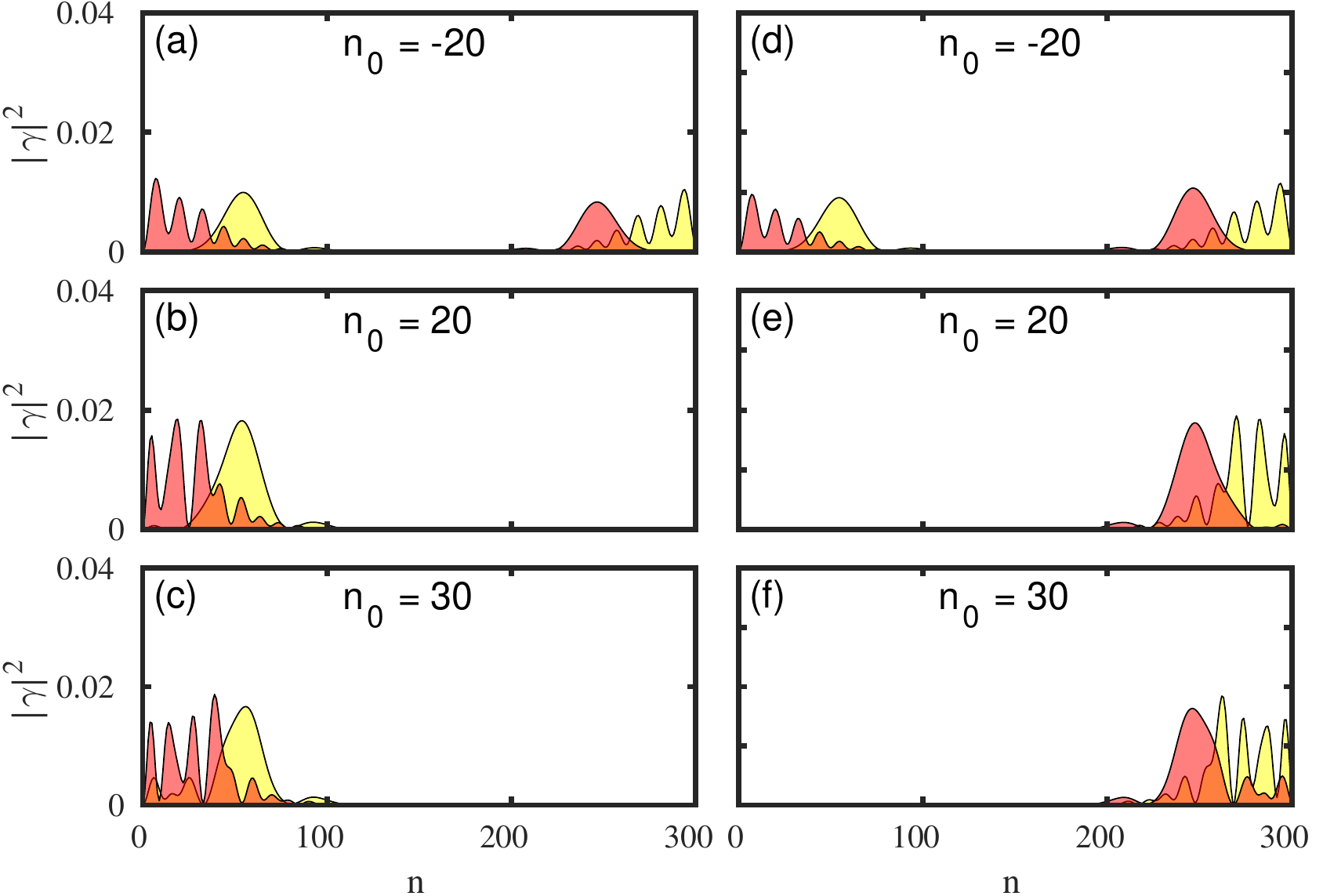}
		\caption{Majorana components of low energy states in NSN system under the PMP for $\Gamma = 0.8 meV < \Gamma_c$ with $V_0 = 2$ meV. The Majorana component overlap increases, also they leak towards the ends.}
		\label{fig:NSN_overlap_A_mu}
	\end{figure}

	We next consider the NSN system and study the effect of the PMP on the trivial and topological ZBP appearing due to the inhomogenity. As shown in Figs.~\ref{fig:NSN_A_mu} and \ref{fig:NSN_M_mu}, the main features regarding the robustness of topological ZBP and the splitting of trivial ZBP remains similar to the ZMP. As for the homogeneous case considered above, oscillations in the topological ZBP width with $n_0$ can be seen. This is again due to the leakage of MBS in the ends of the wire for certain $n_0$ values. For topological zero energy states Majorana components remain separated (see Fig.~\ref{fig:NSN_overlap_M_mu}). On the other hand for trivial zero energy states the Majorana components exhibit an increase in overlap, resulting in the splitting of ZBP (see Fig.~\ref{fig:NSN_overlap_A_mu}). This result suggests the idea of moving protocol is independent of the method used to put certain parts of the wire in a trivial regime. Both ZMP and PMP can be used to carry out the moving protocol, especially to distinguish topological from trivial ZBP, based on the robustness of the ZBP.

	\subsection{S${}^\prime$SS${}^\prime$ System\label{subsec:SSS}}
	
	In this last subsection we will study the tunneling conductance and moving protocol results for the S${}^\prime$SS${}^\prime$ system. We'll utilize this system to highlight some of the shortcomings of the "moving protocol". Nevertheless it turns out that compare to Zeeman moving protocol the Potential moving protocol is partially immune to some of the shortcomings and can clearly differentiate the topological ZBP signature from the trivial ZBP signature.
	
	
	\begin{figure}[th]
		\includegraphics[clip=true,width=.48\textwidth]{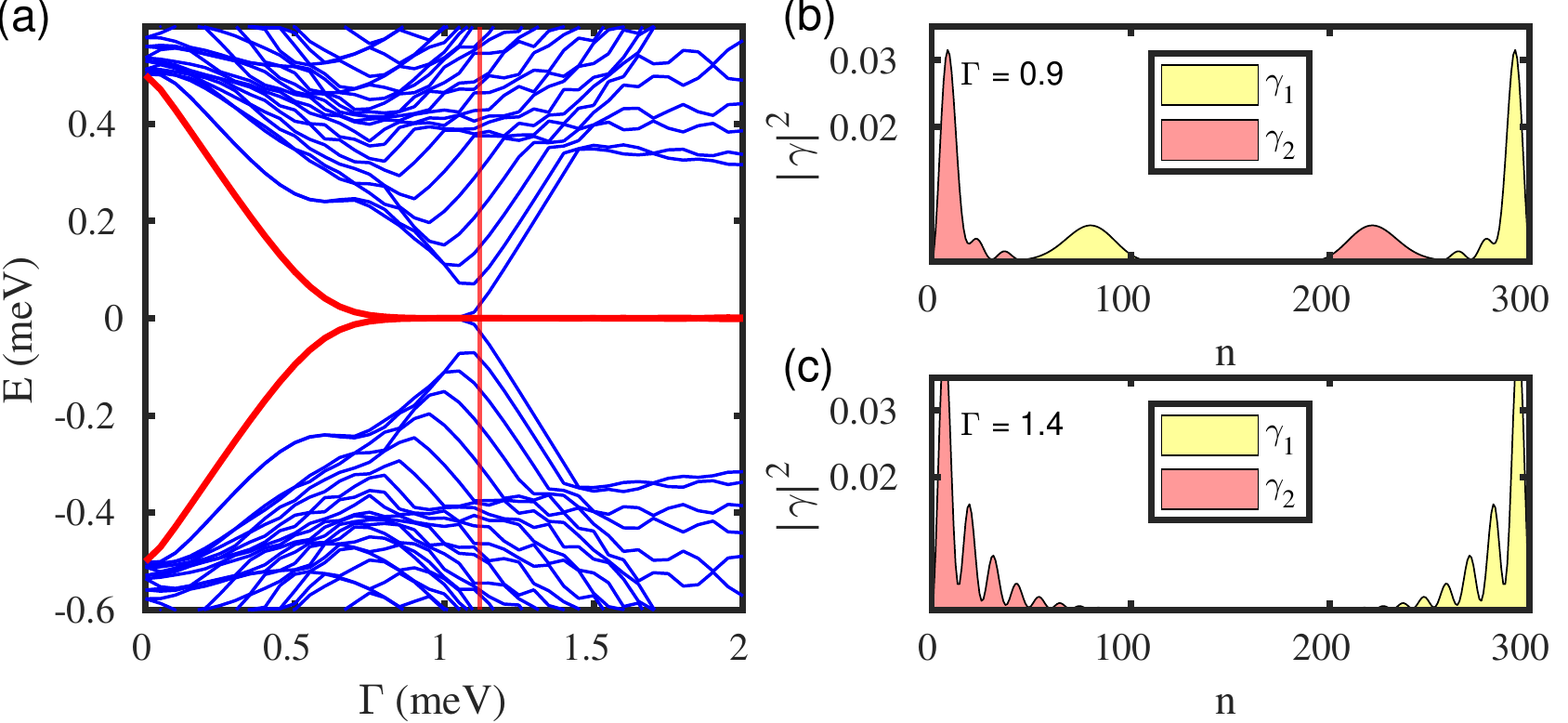}
		\caption{S${}^\prime$SS${}^\prime$ System with parameters $t =  25.4$ meV, $\alpha = 2$ meV, $\Delta_0 = 0.5$ meV, $N_1 = N_2 = 60$, $\mu_S = 1$ meV and $N = 300$. (a) Energy spectrum, (b) Low energy Majorana components for ($\Gamma < \Gamma_c$) and, (c) separated MBS for ($\Gamma > \Gamma_c$)}
		\label{fig:SSS_all}
	\end{figure}
	
	\begin{figure}[th]
		\includegraphics[clip=true,width=.48\textwidth]{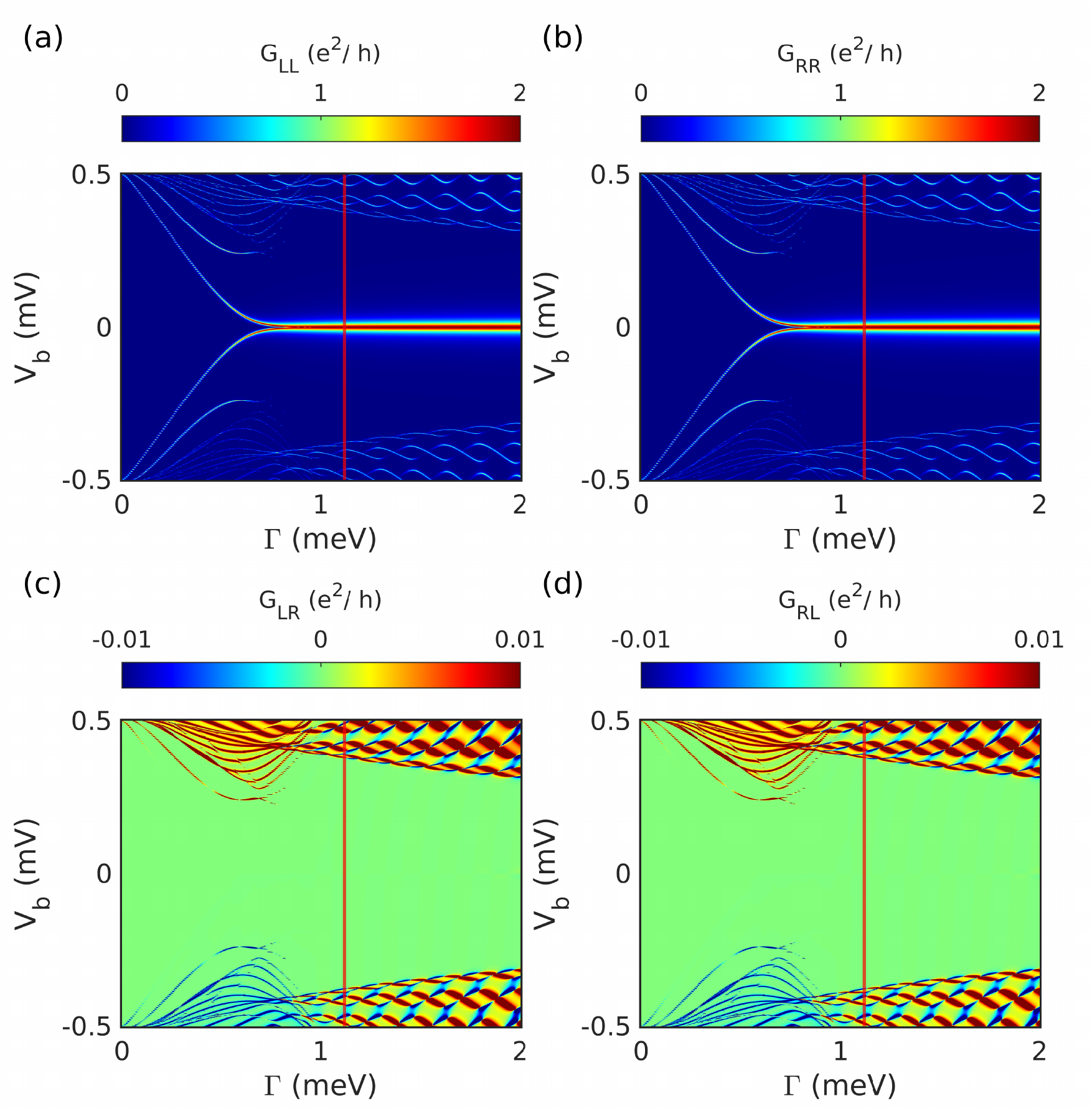}
		\caption{S${}^\prime$SS${}^\prime$ system: (a-b) local conductance exhibit quantized ZBP in trivial and topological region and, (c-d) non-local tunneling conductance unable to capture the band closing signature.}
		\label{fig:SSS_ZBP}
	\end{figure}

\begin{figure}[t!]
		\includegraphics[clip=true,width=\columnwidth]{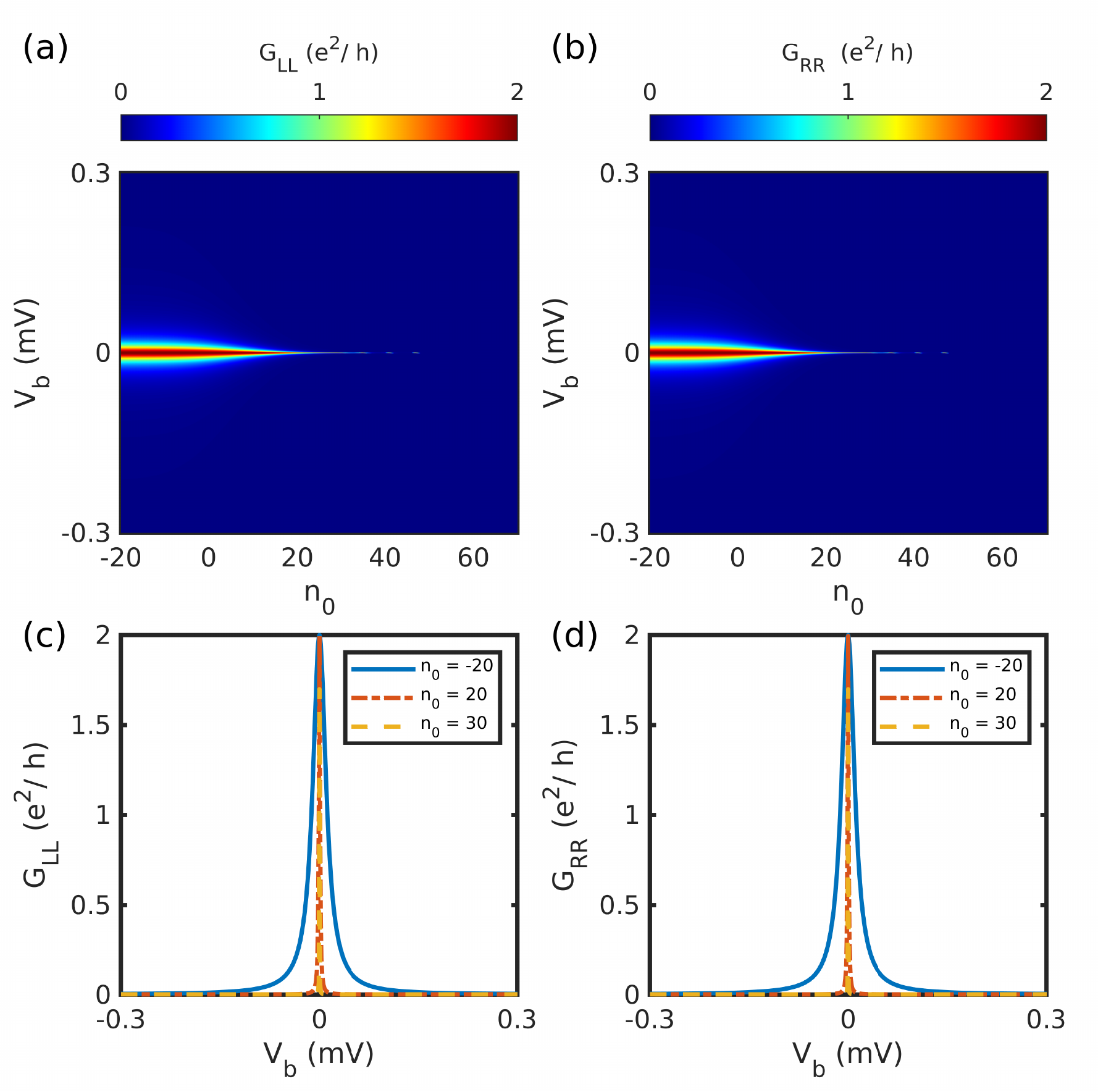}
		\caption{S${}^\prime$SS${}^\prime$ System: tunneling conductance signature on the application of ZMP to trivial ZBP at $\Gamma < \Gamma_c$ with $\Gamma_0 = 0.9$ meV. (a-b) Local tunneling conductance for range of $n_0$ values and, (c-d) vertical line-cut of conductance for specific $n_0$. The peak split is moderately pronounced.}
		\label{fig:SSS_A_all}
	\end{figure}

\begin{figure}[t!]
		\includegraphics[clip=true,width=\columnwidth]{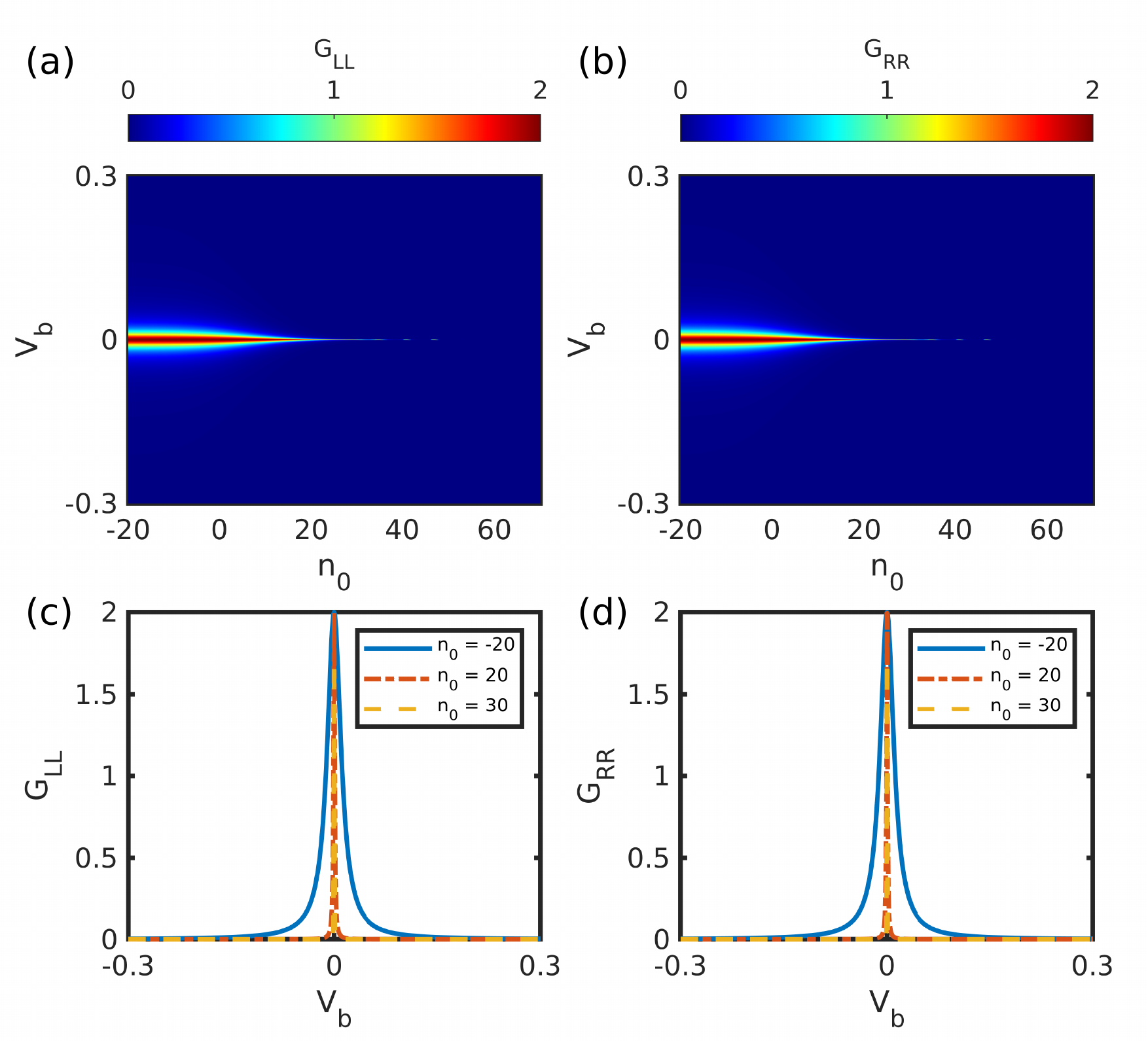}%
		\caption{S${}^\prime$SS${}^\prime$ system: tunneling conductance signature on the application of ZMP to topological ZBP at $\Gamma > \Gamma_c$ with $\Gamma_0 = 1.4$ meV. (a-b) Local tunneling conductance for range of $n_0$ values and, (c-d) vertical line-cut of conductance for specific $n_0$, the peak remain robust before disappearing for higher $n_0$.}
		\label{fig:SSS_M_all}
	\end{figure}

\begin{figure}[t!]
		\includegraphics[clip=true,width=\columnwidth]{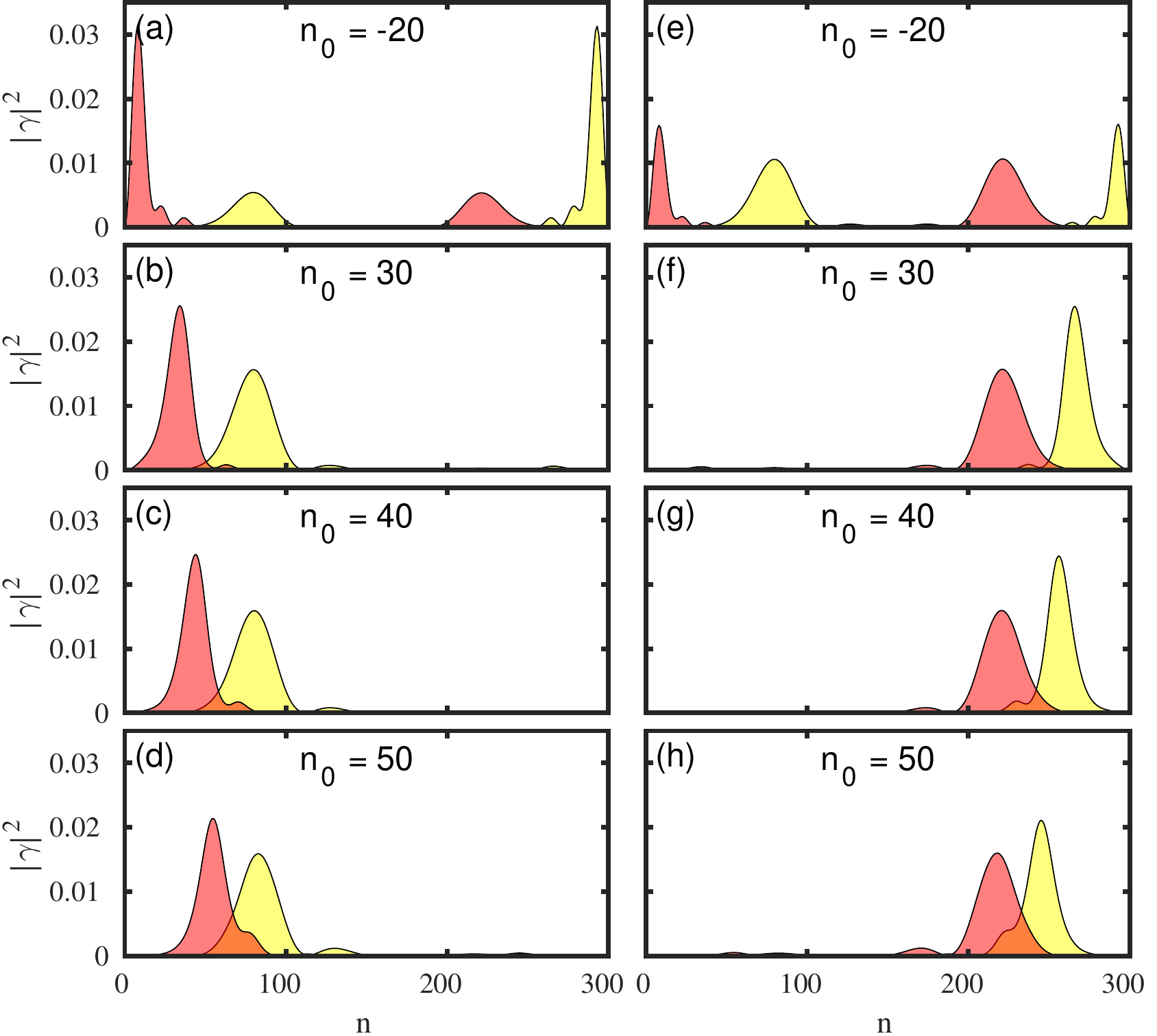}
		\caption{Majorana components of S${}^\prime$SS${}^\prime$ system on the application of ZMP to trivial state at $\Gamma < \Gamma_c$  with $\Gamma_0 = 0.9$ meV for different $n_0$ values, shows increase in the overlap and the components moving away from the ends.}
		\label{fig:SSS_overlap_A}
	\end{figure}

\begin{figure}[t!]
		\includegraphics[clip=true,width=\columnwidth]{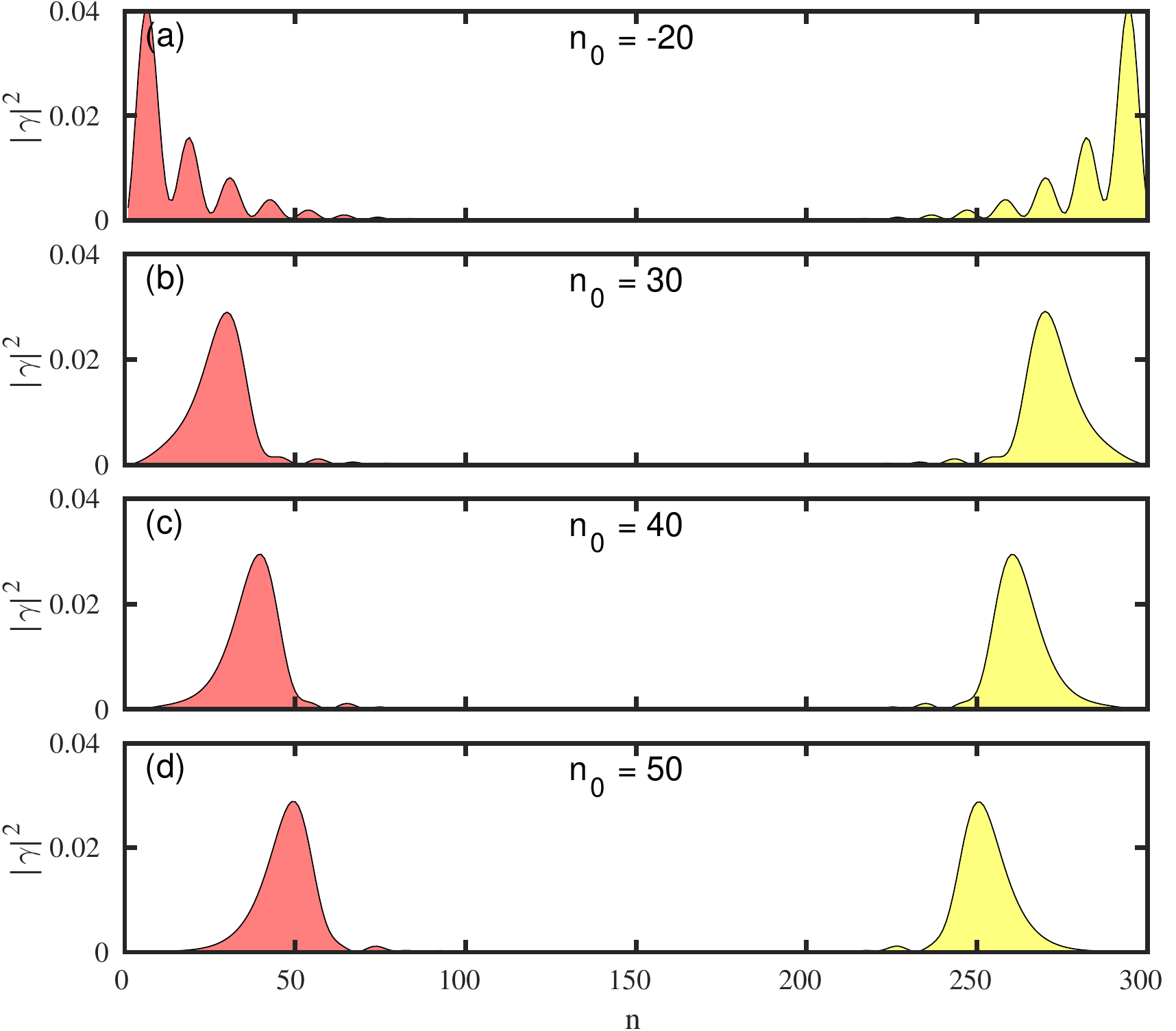}
		\caption{Majorana components of S${}^\prime$SS${}^\prime$ system on the application of ZMP to topological state at $\Gamma > \Gamma_c$ with $\Gamma_0 = 1.4$ meV for different $n_0$ values, shows the components moving away from edge while being separated.}
		\label{fig:SSS_overlap_M}
	\end{figure}

	On the either ends of the wire, this system has zero energy states that appear even before the critical Zeeman term ($\Gamma_c \approx \sqrt{\mu_S^2+\Delta_0^2}$), however, as expected MBS emerges after the critical Zeeman term (see Fig.~\ref{fig:SSS_all}). The zero energy states before the critical zeeman field ($\Gamma_c$) are topological in their own right but have partially separable Majorana components, these states prior to the band gap closing are referred to as the quasi-MBS~\cite{Das_sharma_Three_terminal_ZBP_2021,Das_sharma_Quality_ZBP_2021}. The tunneling conductance of the system is shown in Fig.~\ref{fig:SSS_ZBP}, where the ZBP corresponding to both the local conductances and the non-local conductance fail to detect the band closing signature. We note that this system's tunneling conductance signatures are similar to that of the NSN system. Consequently one can not rely only on the tunneling signature to distinguish the trivial and the topological ZBP.
	Figures~\ref{fig:SSS_A_all} and~\ref{fig:SSS_M_all} demonstrate how ZMP affects trivial and topological ZBP, respectively. Here, the trivial ZBP peak splitting is only moderately noticeable. The peak splits, but before the split in the peak is pronounced, the peak height drops and disappears. On the other hand for the topological ZBP, the conductance peak is fixed to zero bias for a range of $n_0$ before it too vanishes.
	We plot the Majorana component for low energy states belonging to the trivial (Fig.~\ref{fig:SSS_overlap_A}) and topological (Fig.~\ref{fig:SSS_overlap_M}) regimes to determine why this is the case for various $n_0$ values. For trivial zero energy states, the overlap increases with the increase in $n_0$, leading to a rise in the energy split. Additionally, the states drift away from the lead, weakening the coupling between the lead and the states. As a result, the peak vanishes long before the peak split can be seen clearly. Figure~\ref{fig:SSS_overlap_M} for the topological regime with the MBS demonstrates the presence of the exact shifting of Majorana component away from the lead, which accounts for the decreased peak width and the peak's disappearance for higher $n_0$ values as depicted in Fig.~\ref{fig:SSS_M_all}.
	
	\begin{figure}[t!]
		\includegraphics[clip=true,width=\columnwidth]{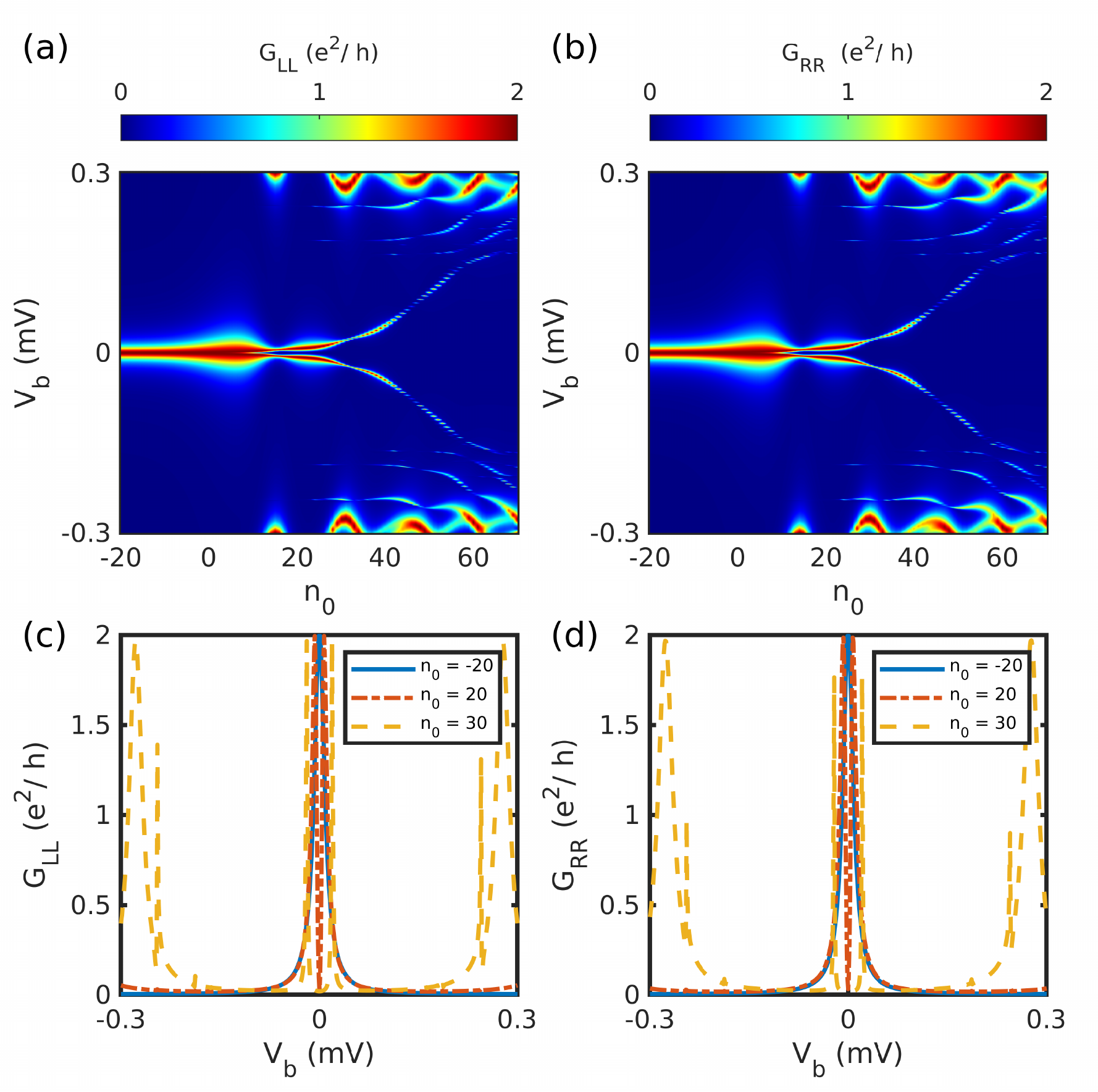}
		\caption{S${}^\prime$SS${}^\prime$ System tunneling conductance signature on the application of PMP to trivial ZBP at $\Gamma = 0.9 meV < \Gamma_c$ with $V_0 = 2$ meV. (a-b) Local tunneling conductance for range of $n_0$ values and, (c-d) vertical line-cut of conductance for specific $n_0$, capturing the splitting signatures.}
		\label{fig:SSS_A_mu}
	\end{figure}
	
	\begin{figure}[t!]
		\includegraphics[clip=true,width=\columnwidth]{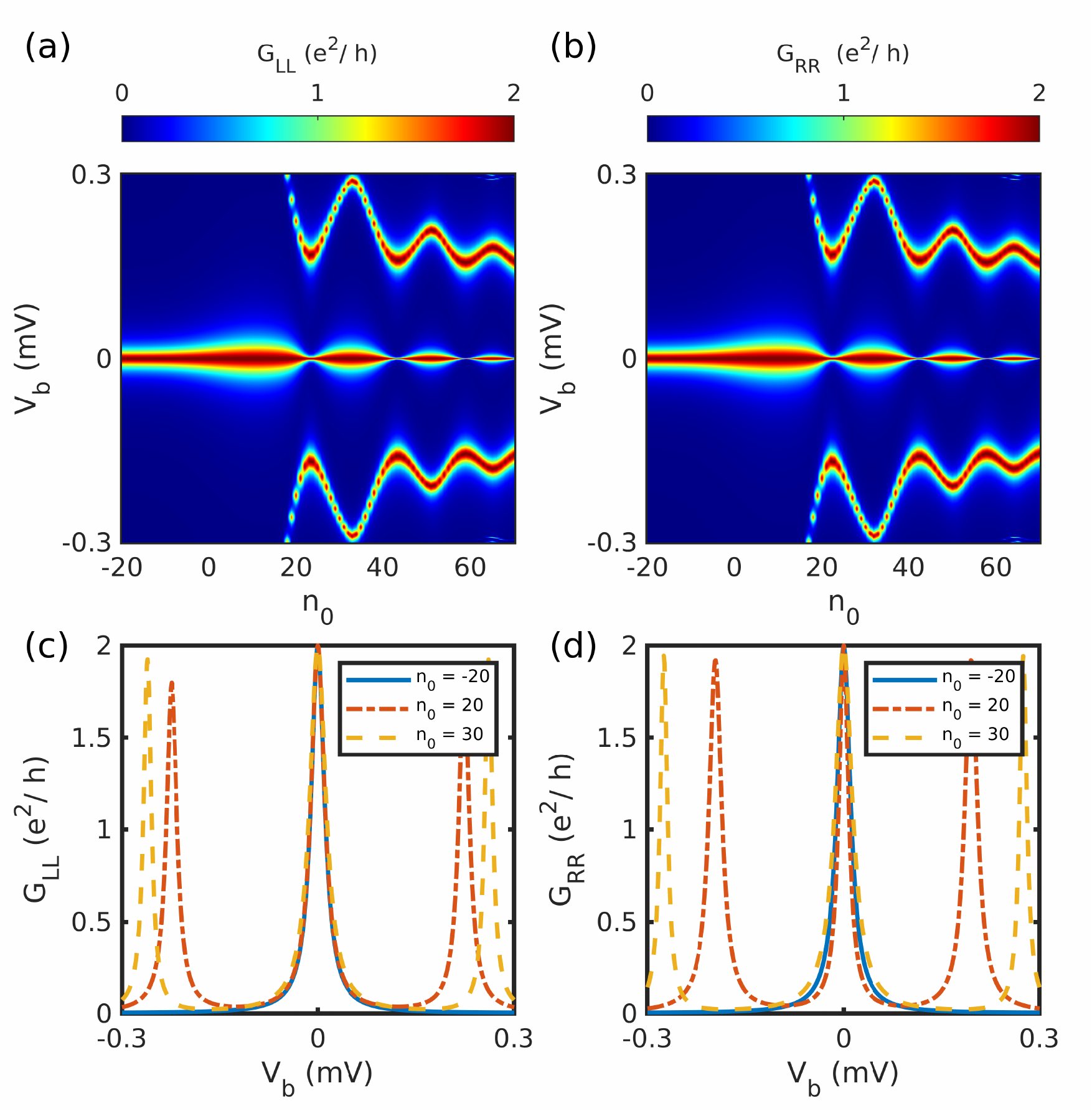}%
		\caption{S${}^\prime$SS${}^\prime$ System tunneling conductance signature on the application of PMP to topological ZBP at $\Gamma = 1.4 meV > \Gamma_c$  with $V_0 = 2$ meV. (a-b) Local tunneling conductance for range of $n_0$ values and, (c-d) vertical linecut of conductance for specific $n_0$, showing robustness of topological ZBP.}
		\label{fig:SSS_M_mu}
	\end{figure}
	
	\begin{figure}[t!]
		\includegraphics[clip=true,width=\columnwidth]{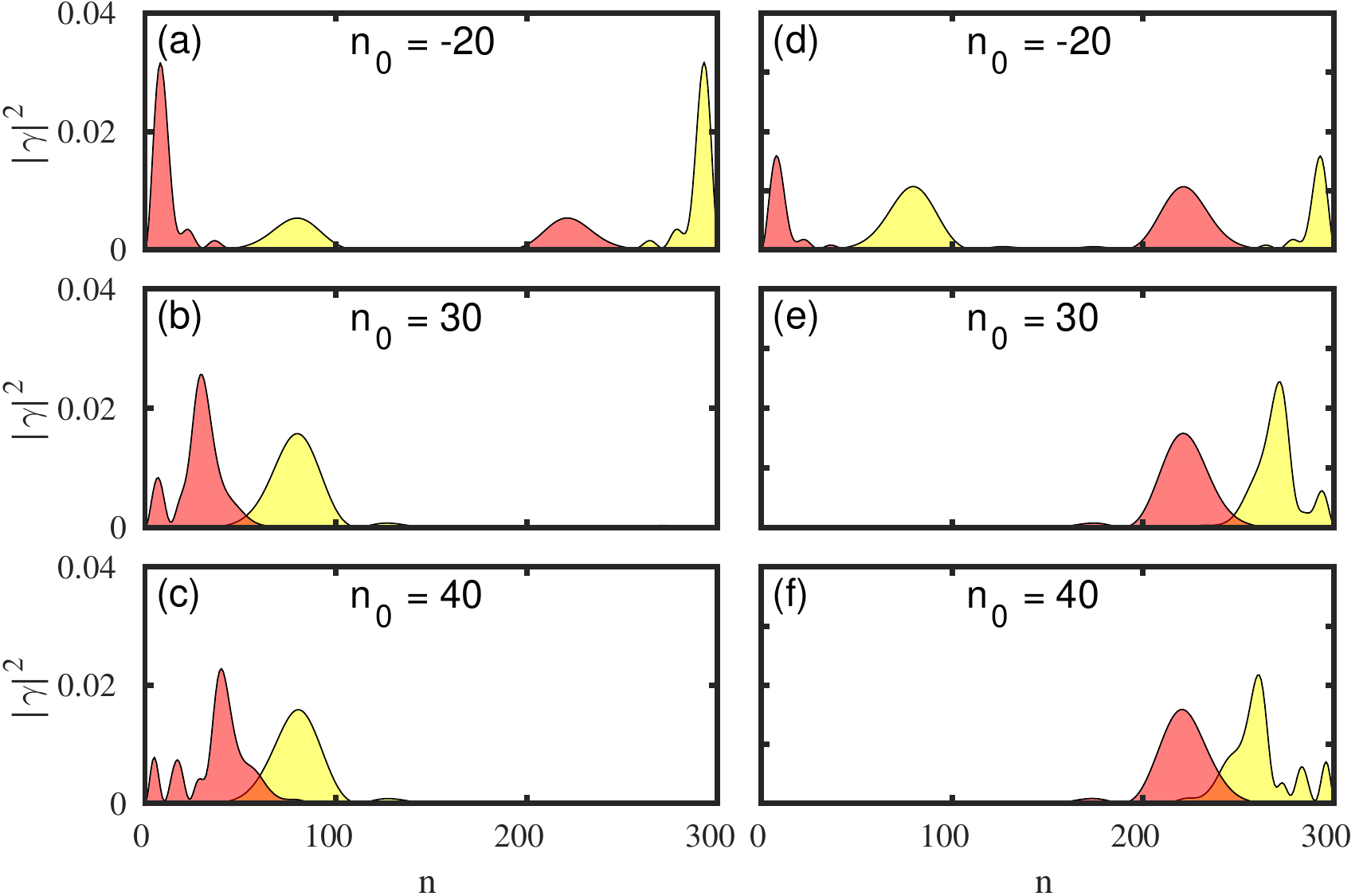}
		\caption{Majorana components of low energy states in S${}^\prime$SS${}^\prime$ system under PMP for $\Gamma = 0.9 meV < \Gamma_c$ with $V_0 = 2$ meV, denotes the overlap increasing as well as states leaking towards the ends.}
		\label{fig:SSS_overlap_A_mu}
	\end{figure}
	
	\begin{figure}[t!]
		\includegraphics[clip=true,width=\columnwidth]{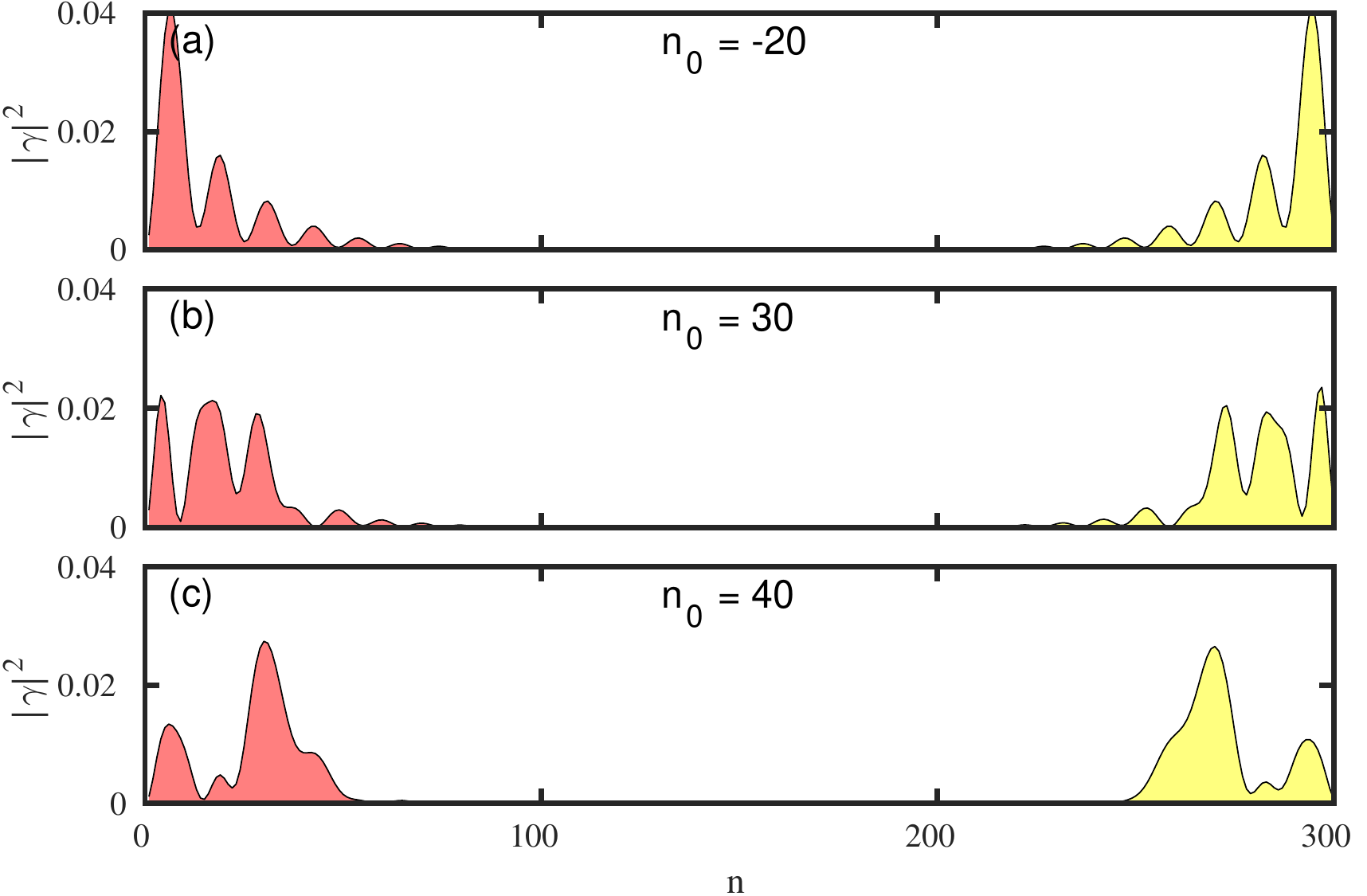}
		\caption{Majorana components of low energy states in S${}^\prime$SS${}^\prime$ system under PMP at $\Gamma = 1.4 meV > \Gamma_c$ $V_0 = 2$ meV, shows separated components.}
		\label{fig:SSS_overlap_M_mu}
	\end{figure}
	
	This shortcoming of ZMP is mainly due to the reliance on the local probes to pick the energy split signature. The protocol moves the states away from the lead, which causes the signature to disappear. As we demonstrate in subsection~\ref{subsec:mu_moving}, the PMP is immune to this effect because of the leakage of states towards the ends of the wire, retaining the coupling to the lead. For trivial and topological ZBP, we show the results corresponding to the PMP on the S${}^\prime$SS${}^\prime$ system in  Figs.~\ref{fig:SSS_A_mu} and~\ref{fig:SSS_M_mu}, respectively. Figure~\ref{fig:SSS_A_mu} clearly illustrates the trivial ZBP splitting during the moving protocol. As shown in Fig.~\ref{fig:SSS_overlap_A_mu}, the states leak towards the ends,thereby improving the split visibility. From tunneling conductance plot in Fig.~\ref{fig:SSS_M_mu} and the analogous Majorana component plot \textcolor{blue}{in} Fig.~\ref{fig:SSS_overlap_M_mu}, it is clear that the resilience of the topological ZBP is also present for this scenario. We want to stress the fact that both moving protocols can induce energy splitting in trivial zero energy states, but as we use local tunneling conductance probe to detect this splitting the PMP is better at capturing the split as well as the robustness in the ZBP.
	
	
	
	

	\section{\label{sec:Conclusion}Conclusion}
	
	In this article, we have reproduced the tunneling conductance signatures from heterostructures and have verified the inability of local and non-local conductance to distinguish ABS from the MBS. Since both the trivial and non-trivial zero energy states produce a quantized ZBP in local conductance, we focus on the effects of the topological length of the nanowire on the ZBP. We show that by applying the moving protocol, the trivial and topological ZBP behaves differently. As the Majorana component of trivial and topological states are fundamentally different, the protocol affects them differently. For MBS, the Majorana components are entirely separated and localized at the edges of the topological region. The Majorana components move away from the wire's edges when the protocol is applied, however they remain separated and are therefore pinned at zero energy, causing the peak in local conductance to remain at zero bias throughout the protocol.
	For trivial zero energy states when the protocol is applied, the overlap between the partially separated Majorana components increases, causing the trivial states to move away from zero energy. As a result, the trivial ZBP splits into two peaks. This effect of moving protocol on topological and trivial ZBP remains the same irrespective of the method of reducing the topological length by putting certain portion of the wire in a trivial regime. We compare the results of ZBP under ZMP and PMP in our numerical simulation. We discuss the shortcoming of the moving protocol in the S${}^\prime$SS${}^\prime$ system, the origin of which is due to the reliance on the local measurements to detect the split in energy. As the states are moved further away from the lead the splitting information is less pronounced in tunneling conductance. However for PMP, the leakage of the states towards the ends preserve the coupling to the leads, making the protocol immune to this vulnerability. Therefore, the PMP which is relatively easier to implement in an experimental setup will be the more suitable protocol for the detection. 
	Our moving protocol combined with existing local tunneling conductance measurement provides a way to differentiate the MBS from the ps-ABS and quasi-MBS. This different response of ZBP under moving protocol clearly distinguishes trivial ZBP from topological ones. 
	
	\begin{acknowledgments}
		S.G. is grateful to the SERB, Government of India, for the
		support via the Core Research Grant No. CRG/2020/002731.
		We also thank J. Carlos Egues and  Poliana H. Penteado for the fruitful discussion.
	\end{acknowledgments}
	\appendix
	\section{Calculation of topological quantum number\label{app:TI}}
	\subsection{\label{scattering}Scattering matrix method}
	To calculate the topological quantum number, we briefly discuss the scattering matrix method here~\cite{Nori}. A scattering matrix characterises a linear relation between the outgoing wave amplitudes and the incoming wave amplitudes at each energy.
	\begin{equation}
		\label{1}
		\begin{pmatrix}
			c^o_{l}\\
			c^o_{r}
		\end{pmatrix} = S(E) \begin{pmatrix}
			c^i_{l}\\
			c^i_{r}
		\end{pmatrix} 
	\end{equation}
	The superscript $i$ $(o)$ is for amplitudes of the waves moving towards (from) the scatterer and the subscript $l$ $(r)$ is for waves on the left (right) of the scatterer. We can express the scattering matrix in terms of transmission and reflection amplitudes in the following form:\\
	\begin{equation}
		\label{2}
		S = \begin{pmatrix}
			r & t' \\
			t & r'
		\end{pmatrix}
	\end{equation}
	Here, $\{t,t'\}$ $(\{r,r'\})$ represent the 4 $\times$ 4 transmission (reflection) matrices.
	The transfer matrix relates the wave amplitudes on the right-hand side to the left-hand side of the sample.
	\begin{equation}
		\label{4}
		\begin{pmatrix}
			c^o_{r}\\
			c^i_{r}
		\end{pmatrix} = M \begin{pmatrix}
			c^o_{l}\\
			c^i_{l}
		\end{pmatrix} 
	\end{equation}
	From Eq.~(\ref{1}) and Eq.~(\ref{4}), one can express the elements of transfer matrix in terms of the elements of scattering matrix as follows:\\
	\begin{equation}
		\label{5}
		M = \begin{pmatrix}
			M_{11} & M_{12} \\
			M_{21} & M_{22} 
		\end{pmatrix} = \begin{pmatrix}
			t-r't'^{-1}r & r't'^{-1} \\
			-t'^{-1}r & t'^{-1}
		\end{pmatrix}
	\end{equation}
	Analogous to Eq.~(\ref{5}), we can also decompose the scattering matrix in terms of the elements of the transfer matrix as follows:
	\begin{equation}
		\label{6}
		S = \begin{pmatrix}
			r & t' \\
			t & r'
		\end{pmatrix} = \begin{pmatrix}
			-M^{-1}_{22} M_{21} & M^{-1}_{22} \\
			M_{11} - M_{12}M^{-1}_{22} M_{21} & M_{12}M^{-1}_{22}
		\end{pmatrix}
	\end{equation}
	The conservation of the probability current also introduces a relationship between the elements of the transfer matrix $M$~\cite{Main_book}:
	\begin{equation}
		\label{M}
		M^\dagger \Sigma_z M = \Sigma_z , \ \ \Sigma_z = \begin{pmatrix}
			\mathbb{I} & 0 \\
			0 & -\mathbb{I}
		\end{pmatrix}  
	\end{equation}
	From the definition of transfer matrix it is clear that it obeys the multiplicative composition law,\\
	\begin{equation}
		\label{7}
		M = M_2 M_1
	\end{equation}
	This composition law for transfer matrices results into a non linear composition of scattering matrices. Using Eq.~(\ref{5}) and Eq.~(\ref{7}), we obtain\\
	\begin{small}
		\begin{equation}
			\begin{split}
				M &= \begin{pmatrix}
					t-r't'^{-1}r & r't'^{-1} \\
					-t'^{-1}r & t'^{-1}
				\end{pmatrix} \\
				&=\begin{pmatrix}
					t_{2}-r_{2}'t_{2}'^{-1}r_{2} & r_{2}'t_{2}'^{-1} \\
					-t_{2}'^{-1}r_{2} & t_{2}'^{-1}
				\end{pmatrix} \begin{pmatrix}
					t_{1}-r_{1}'t_{1}'^{-1}r_{1} & r_{1}'t_{1}'^{-1} \\
					-t_{1}'^{-1}r_{1} & t_{1}'^{-1}
				\end{pmatrix}   
			\end{split}
		\end{equation}
	\end{small}
	This gives the composite $S$-matrix of the system~\cite{Main_book}:\\
	\begin{small}
		\begin{equation}
			\label{8}
			\begin{split}
				S &= \begin{pmatrix}
					r & t' \\
					t & r'
				\end{pmatrix}\\ 
				&= \begin{pmatrix}
					r_1 + t'_1 (\mathbb{I} - r_2 r'_1)^{-1} r_2 t_1 & t'_1 (\mathbb{I} - r_2 r'_1)^{-1} t'_2  \\
					t_2 (\mathbb{I} - r'_1 r_2)^{-1} t_1 & r'_2 + t_2 (\mathbb{I} - r'_1 r_2)^{-1} r'_1 t'_2
				\end{pmatrix}    
			\end{split}
		\end{equation}
	\end{small}
	The composition law Eq.~(\ref{8}) for scattering matrices is denoted by:\\
	\begin{equation}
		S = S_2 \otimes S_1
	\end{equation}
	\subsection{Topological quantum number}
	The Hamiltonian in Eq.~(\ref{eqn:Hamiltonian}) can be rewritten in the Bogoliubov de-Gennes basis $\Psi_n = (c_{n\uparrow},c_{n\downarrow},c^\dagger_{n\downarrow},-c^\dagger_{n\uparrow})^{T}$:\\
	\begin{equation}
		\label{BDG}
		H_{\text{BDG}} = \frac{1}{2} \sum_n \left[ \Psi^{\dagger}_n \hat{h}_n \Psi_n + (\Psi^{\dagger}_n \hat{t}_n \Psi_{n+1} + \textbf{h.c.}) \right],
	\end{equation} 
	Writing the zero-energy Schrödinger’s equation for the BDG Hamiltonian Eq.~(\ref{BDG}) gives us a recursive relation between two sites in terms of the transfer matrix $\mathcal{M}_n$~\cite{Nori,Same_as_Nori}:
	\begin{equation}
		\mathcal{M}_n =  \begin{pmatrix}
			0 & \hat{t}^\dagger_n \\
			-\hat{t}^{-1}_n & -\hat{t}^{-1}_n \hat{h}_n 
		\end{pmatrix}.
	\end{equation}
	Now, $\mathcal{M}_n$ gives the relation between the two nearest-neighbor sites, but it is different from the definition used in Eq.~(\ref{4}). Therefore, we transform to the new basis using a unitary transformation:\\
	\begin{equation}
		M_n =   \mathcal{U}^\dagger \mathcal{M}_n \mathcal{U} ,\ \ \ \mathcal{U} = \frac{1}{\sqrt{2}} \begin{pmatrix}
			\mathbb{I} & \mathbb{I} \\
			i \mathbb{I} & -i \mathbb{I} 
		\end{pmatrix} .
	\end{equation}
	In this basis, the transfer matrix $M_n$ now satisfies Eq.~(\ref{M}). As a result, all of the properties of the transfer matrix outlined in section Eq.~(\ref{scattering}) apply to this transfer matrix. Thus, using Eq.~(\ref{6}), we construct the unitary scattering matrix at each site and then using the composition law Eq.~(\ref{8}), we obtain the composite scattering matrix of the N-dot chain:\\
	\begin{equation}
		S = \begin{pmatrix}
			R & T' \\
			T & R' 
		\end{pmatrix} = S_N \otimes S_{N-1} \otimes ... \otimes S_1   . 
	\end{equation}
	The topological quantum number TI is given by~\cite{TopQuantNumber,FulgaTopologicalQ}
	\begin{equation}
		\label{Q}
		\textnormal{TI} = \text{sgn} [\text{Det}(R)],
	\end{equation}
	where $r$ is the subblock of the total scattering matrix $S$ Eq.~(\ref{2}) of the chain at Fermi level.
	The Majorana bound states exist at the endpoints of the chain if $\textnormal{TI}=-1$. This phase is the topologically nontrivial phase. On the other hand, a value $\textnormal{TI}=+1$ means that the system is in the trivial phase.
	
	\section{\label{app:Tunneling} Tunneling Conductance}
	The expresion of tunneling conductance in terms of scattering matrix is well known in the literature~\cite{disordered_Majorana,Lobos_tunneling_2015} and is given by
	
	\begin{equation}
		\begin{split}
			G_{LL} = \frac{e^2}{h}\int_{-\infty}^{\infty}d\omega {}& \left[-\frac{dn_L(\omega)}{d(eV_L)}\right]\bigg\{ Tr\left[2 r^{LL}_{eh}\left(r^{LL}_{eh}\right)^\dagger \right.\\ &\left.+t^{LR}_{ee}\left(t^{LR}_{ee}\right)^\dagger+ t^{LR}_{eh}\left(t^{LR}_{eh}\right)^\dagger \right] \bigg\}_\omega
		\end{split}
	\end{equation}
	\noindent
	\begin{equation}
		\begin{split}
			G_{LR} = \frac{e^2}{h}\int_{-\infty}^{\infty}d\omega {} \left[\frac{dn_R(\omega)}{d(eV_R)}\right]\bigg\{& \left[t^{LR}_{ee}\left(t^{LR}_{ee}\right)^\dagger\right.\\ 
			&\left. -t^{LR}_{eh}\left(t^{LR}_{eh}\right)^\dagger \right] \bigg\}_\omega
		\end{split}
	\end{equation}
	\noindent
	\begin{equation}
		\begin{split}
			G_{RL} = \frac{e^2}{h}\int_{-\infty}^{\infty}d\omega {} \left[\frac{dn_L(\omega)}{d(eV_L)}\right]\bigg\{&\left[ t^{RL}_{ee}\left(t^{RL}_{ee}\right)^\dagger\right.\\
			&\left. -t^{RL}_{eh}\left(t^{RL}_{eh}\right)^\dagger \right] \bigg\}_\omega
		\end{split}
	\end{equation}
	\noindent
	\begin{equation}
		\begin{split}
			G_{RR} = \frac{e^2}{h}\int_{-\infty}^{\infty}d\omega {}& \left[-\frac{dn_R(\omega)}{d(eV_R)}\right]\bigg\{ Tr\left[2 r^{RR}_{eh}\left(r^{RR}_{eh}\right)^\dagger \right.\\ &\left.+t^{RL}_{ee}\left(t^{RL}_{ee}\right)^\dagger+ t^{RL}_{eh}\left(t^{RL}_{eh}\right)^\dagger \right] \bigg\}_\omega
		\end{split}
	\end{equation}
	\noindent Where $n_\alpha(\omega) = n_F(\omega+eV_\alpha)$ is the fermi function at the $\alpha$ lead with $\alpha$ being L or R. The derivative of fermi function becomes the heavyside fuction at zero kelvin temprature leading to simple equations
	
	\begin{align}
		\begin{split}
			G_{LL} = \frac{e^2}{h}\bigg\{ Tr\left[2 r^{LL}_{eh}\left(r^{LL}_{eh}\right)^\dagger +t^{LR}_{ee}\left(t^{LR}_{ee}\right)^\dagger\right.\\
			\left.+ t^{LR}_{eh}\left(t^{LR}_{eh}\right)^\dagger \right] \bigg\}_{\omega = eV_L}
		\end{split}\\
		\begin{split}
			G_{LR} = -\frac{e^2}{h}\bigg\{ \left[t^{LR}_{ee}\left(t^{LR}_{ee}\right)^\dagger-t^{LR}_{eh}\left(t^{LR}_{eh}\right)^\dagger \right] \bigg\}_{\omega = eV_R}
		\end{split}\\
		\begin{split}
			G_{RL} = -\frac{e^2}{h}\bigg\{\left[ t^{RL}_{ee}\left(t^{RL}_{ee}\right)^\dagger-t^{RL}_{eh}\left(t^{RL}_{eh}\right)^\dagger \right] \bigg\}_{\omega = eV_L}
		\end{split}\\
		\begin{split}
			G_{RR} = \frac{e^2}{h}\bigg\{ Tr\left[2 r^{RR}_{eh}\left(r^{RR}_{eh}\right)^\dagger+t^{RL}_{ee}\left(t^{RL}_{ee}\right)^\dagger\right.\\
			\left.+ t^{RL}_{eh}\left(t^{RL}_{eh}\right)^\dagger \right] \bigg\}_{\omega = eV_R}
		\end{split}
	\end{align}
	
	We use Green function formalism to calculate the scatering matrixs needed for conductance. The retarded green function for the system conneced to two leads is given by
	\begin{equation}
		G^r(\omega) = \frac{1}{\left(\omega+i\eta\right)I - H_{BdG}-\Sigma^r_L-\Sigma^r_R}
	\end{equation}
	with $I$ as identity matrix and $\Sigma^r_\alpha$ denotes the self energy due to $\alpha$ lead. Under the broad band approximation the self energy can be written in terms of level broaderning  matrix as $\Sigma^r_\alpha = -i \Gamma_\alpha/2$. The broaderning matrix is diagonal taking the form $\Gamma_\alpha = \gamma_\alpha I$. The broadering $\gamma_\alpha$ is treated as a parameter in the numerics to calculate the retarded green function which we fix to $\gamma_\alpha = 0.2 t$ for all our calculation. As the BdG hamiltonian is written in Nambu basis the Green fuction has the form
	\begin{equation}
		G_{i,j}^r(\omega) = \begin{pmatrix}
			g_{i,j}^r(\omega) & f_{i,j}^r(\omega)\\
			\bar f_{i,j}^r(\omega) & \bar g_{i,j}^r(\omega)
		\end{pmatrix}
	\end{equation}
	with this perticular form for Green function we could define the scattering matrix elements as
	\begin{align}
		r^{LL}_{ee} &= \gamma_L g^r_{1,1},& r^{RR}_{ee} &= \gamma_R g^r_{N,N}\\
		r^{LL}_{eh} &= \gamma_L f^r_{1,1},& r^{RR}_{ee} &= \gamma_R f^r_{N,N}\\
		t^{LR}_{ee} &= \sqrt{\gamma_L\gamma_R} g^r_{1,N},& t^{LR}_{eh} &= \sqrt{\gamma_L\gamma_R} f^r_{1,N}\\
		t^{RL}_{ee} &= \sqrt{\gamma_L\gamma_R} g^r_{N,1},& t^{RL}_{eh} &= \sqrt{\gamma_L\gamma_R} f^r_{N,1}
	\end{align}
	
	\bibliography{MBS}

\begin{thebibliography}{62}%
\makeatletter
\providecommand \@ifxundefined [1]{%
 \@ifx{#1\undefined}
}%
\providecommand \@ifnum [1]{%
 \ifnum #1\expandafter \@firstoftwo
 \else \expandafter \@secondoftwo
 \fi
}%
\providecommand \@ifx [1]{%
 \ifx #1\expandafter \@firstoftwo
 \else \expandafter \@secondoftwo
 \fi
}%
\providecommand \natexlab [1]{#1}%
\providecommand \enquote  [1]{``#1''}%
\providecommand \bibnamefont  [1]{#1}%
\providecommand \bibfnamefont [1]{#1}%
\providecommand \citenamefont [1]{#1}%
\providecommand \href@noop [0]{\@secondoftwo}%
\providecommand \href [0]{\begingroup \@sanitize@url \@href}%
\providecommand \@href[1]{\@@startlink{#1}\@@href}%
\providecommand \@@href[1]{\endgroup#1\@@endlink}%
\providecommand \@sanitize@url [0]{\catcode `\\12\catcode `\$12\catcode
  `\&12\catcode `\#12\catcode `\^12\catcode `\_12\catcode `\%12\relax}%
\providecommand \@@startlink[1]{}%
\providecommand \@@endlink[0]{}%
\providecommand \url  [0]{\begingroup\@sanitize@url \@url }%
\providecommand \@url [1]{\endgroup\@href {#1}{\urlprefix }}%
\providecommand \urlprefix  [0]{URL }%
\providecommand \Eprint [0]{\href }%
\providecommand \doibase [0]{https://doi.org/}%
\providecommand \selectlanguage [0]{\@gobble}%
\providecommand \bibinfo  [0]{\@secondoftwo}%
\providecommand \bibfield  [0]{\@secondoftwo}%
\providecommand \translation [1]{[#1]}%
\providecommand \BibitemOpen [0]{}%
\providecommand \bibitemStop [0]{}%
\providecommand \bibitemNoStop [0]{.\EOS\space}%
\providecommand \EOS [0]{\spacefactor3000\relax}%
\providecommand \BibitemShut  [1]{\csname bibitem#1\endcsname}%
\let\auto@bib@innerbib\@empty
\bibitem [{\citenamefont {Kitaev}(2001)}]{Kitaev}%
  \BibitemOpen
  \bibfield  {author} {\bibinfo {author} {\bibfnamefont {A.~Y.}\ \bibnamefont
  {Kitaev}},\ }\bibfield  {title} {\bibinfo {title} {Unpaired majorana fermions
  in quantum wires},\ }\href {https://doi.org/10.1070/1063-7869/44/10s/s29} {\
  \textbf {\bibinfo {volume} {44}},\ \bibinfo {pages} {131} (\bibinfo {year}
  {2001})}\BibitemShut {NoStop}%
\bibitem [{\citenamefont {Nayak}\ \emph {et~al.}(2008)\citenamefont {Nayak},
  \citenamefont {Simon}, \citenamefont {Stern}, \citenamefont {Freedman},\ and\
  \citenamefont {Das~Sarma}}]{Chetan_Das_sharma_Non_Abelian_anyons_2008}%
  \BibitemOpen
  \bibfield  {author} {\bibinfo {author} {\bibfnamefont {C.}~\bibnamefont
  {Nayak}}, \bibinfo {author} {\bibfnamefont {S.~H.}\ \bibnamefont {Simon}},
  \bibinfo {author} {\bibfnamefont {A.}~\bibnamefont {Stern}}, \bibinfo
  {author} {\bibfnamefont {M.}~\bibnamefont {Freedman}},\ and\ \bibinfo
  {author} {\bibfnamefont {S.}~\bibnamefont {Das~Sarma}},\ }\bibfield  {title}
  {\bibinfo {title} {Non-abelian anyons and topological quantum computation},\
  }\href {https://doi.org/10.1103/RevModPhys.80.1083} {\bibfield  {journal}
  {\bibinfo  {journal} {Rev. Mod. Phys.}\ }\textbf {\bibinfo {volume} {80}},\
  \bibinfo {pages} {1083} (\bibinfo {year} {2008})}\BibitemShut {NoStop}%
\bibitem [{\citenamefont
  {Alicea}(2012)}]{alicea_New_directions_in_the_pursuit_of_MF}%
  \BibitemOpen
  \bibfield  {author} {\bibinfo {author} {\bibfnamefont {J.}~\bibnamefont
  {Alicea}},\ }\bibfield  {title} {\bibinfo {title} {New directions in the
  pursuit of majorana fermions in solid state systems},\ }\href@noop {}
  {\bibfield  {journal} {\bibinfo  {journal} {Reports on progress in physics}\
  }\textbf {\bibinfo {volume} {75}},\ \bibinfo {pages} {076501} (\bibinfo
  {year} {2012})}\BibitemShut {NoStop}%
\bibitem [{\citenamefont {Sarma}\ \emph
  {et~al.}(2015{\natexlab{a}})\citenamefont {Sarma}, \citenamefont {Freedman},\
  and\ \citenamefont {Nayak}}]{Das_Sharma_Chetan_Majorana_zero_modes_TQC_2015}%
  \BibitemOpen
  \bibfield  {author} {\bibinfo {author} {\bibfnamefont {S.~D.}\ \bibnamefont
  {Sarma}}, \bibinfo {author} {\bibfnamefont {M.}~\bibnamefont {Freedman}},\
  and\ \bibinfo {author} {\bibfnamefont {C.}~\bibnamefont {Nayak}},\ }\bibfield
   {title} {\bibinfo {title} {Majorana zero modes and topological quantum
  computation},\ }\href@noop {} {\bibfield  {journal} {\bibinfo  {journal} {npj
  Quantum Information}\ }\textbf {\bibinfo {volume} {1}},\ \bibinfo {pages} {1}
  (\bibinfo {year} {2015}{\natexlab{a}})}\BibitemShut {NoStop}%
\bibitem [{\citenamefont {Moore}\ and\ \citenamefont
  {Read}(1991)}]{Moore_Nonabelions_FQH_1991}%
  \BibitemOpen
  \bibfield  {author} {\bibinfo {author} {\bibfnamefont {G.}~\bibnamefont
  {Moore}}\ and\ \bibinfo {author} {\bibfnamefont {N.}~\bibnamefont {Read}},\
  }\bibfield  {title} {\bibinfo {title} {Nonabelions in the fractional quantum
  hall effect},\ }\href
  {https://doi.org/https://doi.org/10.1016/0550-3213(91)90407-O} {\bibfield
  {journal} {\bibinfo  {journal} {Nuclear Physics B}\ }\textbf {\bibinfo
  {volume} {360}},\ \bibinfo {pages} {362} (\bibinfo {year}
  {1991})}\BibitemShut {NoStop}%
\bibitem [{\citenamefont {Rice}\ and\ \citenamefont
  {Sigrist}(1995)}]{rice1995sr2ruo4}%
  \BibitemOpen
  \bibfield  {author} {\bibinfo {author} {\bibfnamefont {T.}~\bibnamefont
  {Rice}}\ and\ \bibinfo {author} {\bibfnamefont {M.}~\bibnamefont {Sigrist}},\
  }\bibfield  {title} {\bibinfo {title} {Sr2ruo4: an electronic analogue of
  3he?},\ }\href@noop {} {\bibfield  {journal} {\bibinfo  {journal} {Journal of
  Physics: Condensed Matter}\ }\textbf {\bibinfo {volume} {7}},\ \bibinfo
  {pages} {L643} (\bibinfo {year} {1995})}\BibitemShut {NoStop}%
\bibitem [{\citenamefont {Fu}\ and\ \citenamefont
  {Kane}(2008)}]{Fu_Superconducting_Proximity_Effect_2008}%
  \BibitemOpen
  \bibfield  {author} {\bibinfo {author} {\bibfnamefont {L.}~\bibnamefont
  {Fu}}\ and\ \bibinfo {author} {\bibfnamefont {C.~L.}\ \bibnamefont {Kane}},\
  }\bibfield  {title} {\bibinfo {title} {Superconducting proximity effect and
  majorana fermions at the surface of a topological insulator},\ }\href
  {https://doi.org/10.1103/PhysRevLett.100.096407} {\bibfield  {journal}
  {\bibinfo  {journal} {Phys. Rev. Lett.}\ }\textbf {\bibinfo {volume} {100}},\
  \bibinfo {pages} {096407} (\bibinfo {year} {2008})}\BibitemShut {NoStop}%
\bibitem [{\citenamefont {Sau}\ \emph {et~al.}(2010)\citenamefont {Sau},
  \citenamefont {Lutchyn}, \citenamefont {Tewari},\ and\ \citenamefont
  {Sarma}}]{Sau_Das_sharma_Generic_new_platform_2010}%
  \BibitemOpen
  \bibfield  {author} {\bibinfo {author} {\bibfnamefont {J.~D.}\ \bibnamefont
  {Sau}}, \bibinfo {author} {\bibfnamefont {R.~M.}\ \bibnamefont {Lutchyn}},
  \bibinfo {author} {\bibfnamefont {S.}~\bibnamefont {Tewari}},\ and\ \bibinfo
  {author} {\bibfnamefont {S.~D.}\ \bibnamefont {Sarma}},\ }\bibfield  {title}
  {\bibinfo {title} {Generic new platform for topological quantum computation
  using semiconductor heterostructures},\ }\bibfield  {journal} {\bibinfo
  {journal} {Physical Review Letters}\ }\textbf {\bibinfo {volume} {104}},\
  \href {https://doi.org/10.1103/PHYSREVLETT.104.040502}
  {10.1103/PHYSREVLETT.104.040502} (\bibinfo {year} {2010})\BibitemShut
  {NoStop}%
\bibitem [{\citenamefont {Oreg}\ \emph {et~al.}(2010)\citenamefont {Oreg},
  \citenamefont {Refael},\ and\ \citenamefont {von
  Oppen}}]{Oreg_Von_Helical_Liquids_2010}%
  \BibitemOpen
  \bibfield  {author} {\bibinfo {author} {\bibfnamefont {Y.}~\bibnamefont
  {Oreg}}, \bibinfo {author} {\bibfnamefont {G.}~\bibnamefont {Refael}},\ and\
  \bibinfo {author} {\bibfnamefont {F.}~\bibnamefont {von Oppen}},\ }\bibfield
  {title} {\bibinfo {title} {Helical liquids and majorana bound states in
  quantum wires},\ }\href {https://doi.org/10.1103/PhysRevLett.105.177002}
  {\bibfield  {journal} {\bibinfo  {journal} {Phys. Rev. Lett.}\ }\textbf
  {\bibinfo {volume} {105}},\ \bibinfo {pages} {177002} (\bibinfo {year}
  {2010})}\BibitemShut {NoStop}%
\bibitem [{\citenamefont {Stanescu}\ \emph {et~al.}(2010)\citenamefont
  {Stanescu}, \citenamefont {Sau}, \citenamefont {Lutchyn},\ and\ \citenamefont
  {Sarma}}]{Stanescu_Das_sharma_Proximity_effect_2010}%
  \BibitemOpen
  \bibfield  {author} {\bibinfo {author} {\bibfnamefont {T.~D.}\ \bibnamefont
  {Stanescu}}, \bibinfo {author} {\bibfnamefont {J.~D.}\ \bibnamefont {Sau}},
  \bibinfo {author} {\bibfnamefont {R.~M.}\ \bibnamefont {Lutchyn}},\ and\
  \bibinfo {author} {\bibfnamefont {S.~D.}\ \bibnamefont {Sarma}},\ }\bibfield
  {title} {\bibinfo {title} {Proximity effect at the superconductor-topological
  insulator interface},\ }\bibfield  {journal} {\bibinfo  {journal} {Physical
  Review B - Condensed Matter and Materials Physics}\ }\textbf {\bibinfo
  {volume} {81}},\ \href {https://doi.org/10.1103/PHYSREVB.81.241310}
  {10.1103/PHYSREVB.81.241310} (\bibinfo {year} {2010})\BibitemShut {NoStop}%
\bibitem [{\citenamefont {Albrecht}\ \emph {et~al.}(2016)\citenamefont
  {Albrecht}, \citenamefont {Higginbotham}, \citenamefont {Madsen},
  \citenamefont {Kuemmeth}, \citenamefont {Jespersen}, \citenamefont
  {Nyg{\aa}rd}, \citenamefont {Krogstrup},\ and\ \citenamefont
  {Marcus}}]{Albrecht_Exponential_protection_2016}%
  \BibitemOpen
  \bibfield  {author} {\bibinfo {author} {\bibfnamefont {S.~M.}\ \bibnamefont
  {Albrecht}}, \bibinfo {author} {\bibfnamefont {A.~P.}\ \bibnamefont
  {Higginbotham}}, \bibinfo {author} {\bibfnamefont {M.}~\bibnamefont
  {Madsen}}, \bibinfo {author} {\bibfnamefont {F.}~\bibnamefont {Kuemmeth}},
  \bibinfo {author} {\bibfnamefont {T.~S.}\ \bibnamefont {Jespersen}}, \bibinfo
  {author} {\bibfnamefont {J.}~\bibnamefont {Nyg{\aa}rd}}, \bibinfo {author}
  {\bibfnamefont {P.}~\bibnamefont {Krogstrup}},\ and\ \bibinfo {author}
  {\bibfnamefont {C.}~\bibnamefont {Marcus}},\ }\bibfield  {title} {\bibinfo
  {title} {Exponential protection of zero modes in majorana islands},\
  }\href@noop {} {\bibfield  {journal} {\bibinfo  {journal} {Nature}\ }\textbf
  {\bibinfo {volume} {531}},\ \bibinfo {pages} {206} (\bibinfo {year}
  {2016})}\BibitemShut {NoStop}%
\bibitem [{\citenamefont {Churchill}\ \emph {et~al.}(2013)\citenamefont
  {Churchill}, \citenamefont {Fatemi}, \citenamefont {Grove-Rasmussen},
  \citenamefont {Deng}, \citenamefont {Caroff}, \citenamefont {Xu},\ and\
  \citenamefont {Marcus}}]{Churchill_SC_nanowire_2013}%
  \BibitemOpen
  \bibfield  {author} {\bibinfo {author} {\bibfnamefont {H.~O.~H.}\
  \bibnamefont {Churchill}}, \bibinfo {author} {\bibfnamefont {V.}~\bibnamefont
  {Fatemi}}, \bibinfo {author} {\bibfnamefont {K.}~\bibnamefont
  {Grove-Rasmussen}}, \bibinfo {author} {\bibfnamefont {M.~T.}\ \bibnamefont
  {Deng}}, \bibinfo {author} {\bibfnamefont {P.}~\bibnamefont {Caroff}},
  \bibinfo {author} {\bibfnamefont {H.~Q.}\ \bibnamefont {Xu}},\ and\ \bibinfo
  {author} {\bibfnamefont {C.~M.}\ \bibnamefont {Marcus}},\ }\bibfield  {title}
  {\bibinfo {title} {Superconductor-nanowire devices from tunneling to the
  multichannel regime: Zero-bias oscillations and magnetoconductance
  crossover},\ }\href {https://doi.org/10.1103/PhysRevB.87.241401} {\bibfield
  {journal} {\bibinfo  {journal} {Phys. Rev. B}\ }\textbf {\bibinfo {volume}
  {87}},\ \bibinfo {pages} {241401} (\bibinfo {year} {2013})}\BibitemShut
  {NoStop}%
\bibitem [{\citenamefont {Das}\ \emph {et~al.}(2012)\citenamefont {Das},
  \citenamefont {Ronen}, \citenamefont {Most}, \citenamefont {Oreg},
  \citenamefont {Heiblum},\ and\ \citenamefont
  {Shtrikman}}]{Das_ZBP_InAs_2012}%
  \BibitemOpen
  \bibfield  {author} {\bibinfo {author} {\bibfnamefont {A.}~\bibnamefont
  {Das}}, \bibinfo {author} {\bibfnamefont {Y.}~\bibnamefont {Ronen}}, \bibinfo
  {author} {\bibfnamefont {Y.}~\bibnamefont {Most}}, \bibinfo {author}
  {\bibfnamefont {Y.}~\bibnamefont {Oreg}}, \bibinfo {author} {\bibfnamefont
  {M.}~\bibnamefont {Heiblum}},\ and\ \bibinfo {author} {\bibfnamefont
  {H.}~\bibnamefont {Shtrikman}},\ }\bibfield  {title} {\bibinfo {title}
  {Zero-bias peaks and splitting in an al--inas nanowire topological
  superconductor as a signature of majorana fermions},\ }\href@noop {}
  {\bibfield  {journal} {\bibinfo  {journal} {Nature Physics}\ }\textbf
  {\bibinfo {volume} {8}},\ \bibinfo {pages} {887} (\bibinfo {year}
  {2012})}\BibitemShut {NoStop}%
\bibitem [{\citenamefont {Mourik}\ \emph {et~al.}(2012)\citenamefont {Mourik},
  \citenamefont {Zuo}, \citenamefont {Frolov}, \citenamefont {Plissard},
  \citenamefont {Bakkers},\ and\ \citenamefont
  {Kouwenhoven}}]{mourik_signatures_MBS_InSb_2012}%
  \BibitemOpen
  \bibfield  {author} {\bibinfo {author} {\bibfnamefont {V.}~\bibnamefont
  {Mourik}}, \bibinfo {author} {\bibfnamefont {K.}~\bibnamefont {Zuo}},
  \bibinfo {author} {\bibfnamefont {S.~M.}\ \bibnamefont {Frolov}}, \bibinfo
  {author} {\bibfnamefont {S.}~\bibnamefont {Plissard}}, \bibinfo {author}
  {\bibfnamefont {E.~P.}\ \bibnamefont {Bakkers}},\ and\ \bibinfo {author}
  {\bibfnamefont {L.~P.}\ \bibnamefont {Kouwenhoven}},\ }\bibfield  {title}
  {\bibinfo {title} {Signatures of majorana fermions in hybrid
  superconductor-semiconductor nanowire devices},\ }\href@noop {} {\bibfield
  {journal} {\bibinfo  {journal} {Science}\ }\textbf {\bibinfo {volume}
  {336}},\ \bibinfo {pages} {1003} (\bibinfo {year} {2012})}\BibitemShut
  {NoStop}%
\bibitem [{\citenamefont {Sengupta}\ \emph {et~al.}(2001)\citenamefont
  {Sengupta}, \citenamefont {\ifmmode \check{Z}\else
  \v{Z}\fi{}uti\ifmmode~\acute{c}\else \'{c}\fi{}}, \citenamefont {Kwon},
  \citenamefont {Yakovenko},\ and\ \citenamefont
  {Das~Sarma}}]{Das_Sharma_Sengupta_Midgap_edge_state_2001}%
  \BibitemOpen
  \bibfield  {author} {\bibinfo {author} {\bibfnamefont {K.}~\bibnamefont
  {Sengupta}}, \bibinfo {author} {\bibfnamefont {I.}~\bibnamefont {\ifmmode
  \check{Z}\else \v{Z}\fi{}uti\ifmmode~\acute{c}\else \'{c}\fi{}}}, \bibinfo
  {author} {\bibfnamefont {H.-J.}\ \bibnamefont {Kwon}}, \bibinfo {author}
  {\bibfnamefont {V.~M.}\ \bibnamefont {Yakovenko}},\ and\ \bibinfo {author}
  {\bibfnamefont {S.}~\bibnamefont {Das~Sarma}},\ }\bibfield  {title} {\bibinfo
  {title} {Midgap edge states and pairing symmetry of quasi-one-dimensional
  organic superconductors},\ }\href
  {https://doi.org/10.1103/PhysRevB.63.144531} {\bibfield  {journal} {\bibinfo
  {journal} {Phys. Rev. B}\ }\textbf {\bibinfo {volume} {63}},\ \bibinfo
  {pages} {144531} (\bibinfo {year} {2001})}\BibitemShut {NoStop}%
\bibitem [{\citenamefont {Setiawan}\ \emph
  {et~al.}(2017{\natexlab{a}})\citenamefont {Setiawan}, \citenamefont {Liu},
  \citenamefont {Sau},\ and\ \citenamefont
  {Das~Sarma}}]{Das_sharma_Setiawan_ZBP_Electron_temp_and_tunnel_coupling_dependence}%
  \BibitemOpen
  \bibfield  {author} {\bibinfo {author} {\bibfnamefont {F.}~\bibnamefont
  {Setiawan}}, \bibinfo {author} {\bibfnamefont {C.-X.}\ \bibnamefont {Liu}},
  \bibinfo {author} {\bibfnamefont {J.~D.}\ \bibnamefont {Sau}},\ and\ \bibinfo
  {author} {\bibfnamefont {S.}~\bibnamefont {Das~Sarma}},\ }\bibfield  {title}
  {\bibinfo {title} {Electron temperature and tunnel coupling dependence of
  zero-bias and almost-zero-bias conductance peaks in majorana nanowires},\
  }\href {https://doi.org/10.1103/PhysRevB.96.184520} {\bibfield  {journal}
  {\bibinfo  {journal} {Phys. Rev. B}\ }\textbf {\bibinfo {volume} {96}},\
  \bibinfo {pages} {184520} (\bibinfo {year} {2017}{\natexlab{a}})}\BibitemShut
  {NoStop}%
\bibitem [{\citenamefont {Yu}\ \emph {et~al.}(2021)\citenamefont {Yu},
  \citenamefont {Chen}, \citenamefont {Gomanko}, \citenamefont {Badawy},
  \citenamefont {Bakkers}, \citenamefont {Zuo}, \citenamefont {Mourik},\ and\
  \citenamefont {Frolov}}]{Yu_Forlov_non_MBS_nearly_quantized_2021}%
  \BibitemOpen
  \bibfield  {author} {\bibinfo {author} {\bibfnamefont {P.}~\bibnamefont
  {Yu}}, \bibinfo {author} {\bibfnamefont {J.}~\bibnamefont {Chen}}, \bibinfo
  {author} {\bibfnamefont {M.}~\bibnamefont {Gomanko}}, \bibinfo {author}
  {\bibfnamefont {G.}~\bibnamefont {Badawy}}, \bibinfo {author} {\bibfnamefont
  {E.~P.}\ \bibnamefont {Bakkers}}, \bibinfo {author} {\bibfnamefont
  {K.}~\bibnamefont {Zuo}}, \bibinfo {author} {\bibfnamefont {V.}~\bibnamefont
  {Mourik}},\ and\ \bibinfo {author} {\bibfnamefont {S.~M.}\ \bibnamefont
  {Frolov}},\ }\bibfield  {title} {\bibinfo {title} {Non-majorana states yield
  nearly quantized conductance in proximatized nanowires},\ }\href
  {https://doi.org/10.1038/S41567-020-01107-W} {\bibfield  {journal} {\bibinfo
  {journal} {Nature Physics}\ }\textbf {\bibinfo {volume} {17}},\ \bibinfo
  {pages} {482} (\bibinfo {year} {2021})}\BibitemShut {NoStop}%
\bibitem [{\citenamefont {Pan}\ and\ \citenamefont
  {Sarma}(2020)}]{Pan_Das_sharma_Physical_mechanism_ZBP_2020}%
  \BibitemOpen
  \bibfield  {author} {\bibinfo {author} {\bibfnamefont {H.}~\bibnamefont
  {Pan}}\ and\ \bibinfo {author} {\bibfnamefont {S.~D.}\ \bibnamefont
  {Sarma}},\ }\bibfield  {title} {\bibinfo {title} {Physical mechanisms for
  zero-bias conductance peaks in majorana nanowires},\ }\bibfield  {journal}
  {\bibinfo  {journal} {Physical Review Research}\ }\textbf {\bibinfo {volume}
  {2}},\ \href {https://doi.org/10.1103/PHYSREVRESEARCH.2.013377}
  {10.1103/PHYSREVRESEARCH.2.013377} (\bibinfo {year} {2020})\BibitemShut
  {NoStop}%
\bibitem [{\citenamefont {Vuik}\ \emph
  {et~al.}(2019{\natexlab{a}})\citenamefont {Vuik}, \citenamefont {Nijholt},
  \citenamefont {Akhmerov},\ and\ \citenamefont
  {Wimmer}}]{Vuik_Akmerov_Reproducing_topological_2019}%
  \BibitemOpen
  \bibfield  {author} {\bibinfo {author} {\bibfnamefont {A.}~\bibnamefont
  {Vuik}}, \bibinfo {author} {\bibfnamefont {B.}~\bibnamefont {Nijholt}},
  \bibinfo {author} {\bibfnamefont {A.~R.}\ \bibnamefont {Akhmerov}},\ and\
  \bibinfo {author} {\bibfnamefont {M.}~\bibnamefont {Wimmer}},\ }\bibfield
  {title} {\bibinfo {title} {Reproducing topological properties with
  quasi-majorana states},\ }\bibfield  {journal} {\bibinfo  {journal} {SciPost
  Physics}\ }\textbf {\bibinfo {volume} {7}},\ \href
  {https://doi.org/10.21468/SCIPOSTPHYS.7.5.061/PDF}
  {10.21468/SCIPOSTPHYS.7.5.061/PDF} (\bibinfo {year}
  {2019}{\natexlab{a}})\BibitemShut {NoStop}%
\bibitem [{\citenamefont {Pan}\ \emph {et~al.}(2020)\citenamefont {Pan},
  \citenamefont {Cole}, \citenamefont {Sau},\ and\ \citenamefont
  {Sarma}}]{Pan_Das_Sharma_Generic_quantized_ZBP_2020}%
  \BibitemOpen
  \bibfield  {author} {\bibinfo {author} {\bibfnamefont {H.}~\bibnamefont
  {Pan}}, \bibinfo {author} {\bibfnamefont {W.~S.}\ \bibnamefont {Cole}},
  \bibinfo {author} {\bibfnamefont {J.~D.}\ \bibnamefont {Sau}},\ and\ \bibinfo
  {author} {\bibfnamefont {S.~D.}\ \bibnamefont {Sarma}},\ }\bibfield  {title}
  {\bibinfo {title} {Generic quantized zero-bias conductance peaks in
  superconductor-semiconductor hybrid structures},\ }\bibfield  {journal}
  {\bibinfo  {journal} {Physical Review B}\ }\textbf {\bibinfo {volume}
  {101}},\ \href {https://doi.org/10.1103/PHYSREVB.101.024506}
  {10.1103/PHYSREVB.101.024506} (\bibinfo {year} {2020})\BibitemShut {NoStop}%
\bibitem [{\citenamefont {Chen}\ \emph {et~al.}(2019)\citenamefont {Chen},
  \citenamefont {Woods}, \citenamefont {Yu}, \citenamefont {Hocevar},
  \citenamefont {Car}, \citenamefont {Plissard}, \citenamefont {Bakkers},
  \citenamefont {Stanescu},\ and\ \citenamefont
  {Frolov}}]{Chen_Akmerov_Ubiquitous_non_Majorana_ZBP_2019}%
  \BibitemOpen
  \bibfield  {author} {\bibinfo {author} {\bibfnamefont {J.}~\bibnamefont
  {Chen}}, \bibinfo {author} {\bibfnamefont {B.~D.}\ \bibnamefont {Woods}},
  \bibinfo {author} {\bibfnamefont {P.}~\bibnamefont {Yu}}, \bibinfo {author}
  {\bibfnamefont {M.}~\bibnamefont {Hocevar}}, \bibinfo {author} {\bibfnamefont
  {D.}~\bibnamefont {Car}}, \bibinfo {author} {\bibfnamefont {S.~R.}\
  \bibnamefont {Plissard}}, \bibinfo {author} {\bibfnamefont {E.~P.}\
  \bibnamefont {Bakkers}}, \bibinfo {author} {\bibfnamefont {T.~D.}\
  \bibnamefont {Stanescu}},\ and\ \bibinfo {author} {\bibfnamefont {S.~M.}\
  \bibnamefont {Frolov}},\ }\bibfield  {title} {\bibinfo {title} {Ubiquitous
  non-majorana zero-bias conductance peaks in nanowire devices},\ }\bibfield
  {journal} {\bibinfo  {journal} {Physical Review Letters}\ }\textbf {\bibinfo
  {volume} {123}},\ \href {https://doi.org/10.1103/PHYSREVLETT.123.107703}
  {10.1103/PHYSREVLETT.123.107703} (\bibinfo {year} {2019})\BibitemShut
  {NoStop}%
\bibitem [{\citenamefont {Woods}\ \emph {et~al.}(2019)\citenamefont {Woods},
  \citenamefont {Chen}, \citenamefont {Frolov},\ and\ \citenamefont
  {Stanescu}}]{Woods_Stanescu_Zero_energy_pinning_trivial_BS_2019}%
  \BibitemOpen
  \bibfield  {author} {\bibinfo {author} {\bibfnamefont {B.~D.}\ \bibnamefont
  {Woods}}, \bibinfo {author} {\bibfnamefont {J.}~\bibnamefont {Chen}},
  \bibinfo {author} {\bibfnamefont {S.~M.}\ \bibnamefont {Frolov}},\ and\
  \bibinfo {author} {\bibfnamefont {T.~D.}\ \bibnamefont {Stanescu}},\
  }\bibfield  {title} {\bibinfo {title} {Zero-energy pinning of topologically
  trivial bound states in multiband semiconductor-superconductor nanowires},\
  }\bibfield  {journal} {\bibinfo  {journal} {Physical Review B}\ }\textbf
  {\bibinfo {volume} {100}},\ \href
  {https://doi.org/10.1103/PHYSREVB.100.125407} {10.1103/PHYSREVB.100.125407}
  (\bibinfo {year} {2019})\BibitemShut {NoStop}%
\bibitem [{\citenamefont {Vernek}\ \emph {et~al.}(2014)\citenamefont {Vernek},
  \citenamefont {Penteado}, \citenamefont {Seridonio},\ and\ \citenamefont
  {Egues}}]{Vernek_Subtle_leakage_of_Majorana_2014}%
  \BibitemOpen
  \bibfield  {author} {\bibinfo {author} {\bibfnamefont {E.}~\bibnamefont
  {Vernek}}, \bibinfo {author} {\bibfnamefont {P.~H.}\ \bibnamefont
  {Penteado}}, \bibinfo {author} {\bibfnamefont {A.~C.}\ \bibnamefont
  {Seridonio}},\ and\ \bibinfo {author} {\bibfnamefont {J.~C.}\ \bibnamefont
  {Egues}},\ }\bibfield  {title} {\bibinfo {title} {Subtle leakage of a
  majorana mode into a quantum dot},\ }\href
  {https://doi.org/10.1103/PhysRevB.89.165314} {\bibfield  {journal} {\bibinfo
  {journal} {Phys. Rev. B}\ }\textbf {\bibinfo {volume} {89}},\ \bibinfo
  {pages} {165314} (\bibinfo {year} {2014})}\BibitemShut {NoStop}%
\bibitem [{\citenamefont {Grivnin}\ \emph {et~al.}(2019)\citenamefont
  {Grivnin}, \citenamefont {Bor}, \citenamefont {Heiblum}, \citenamefont
  {Oreg},\ and\ \citenamefont {Shtrikman}}]{Grivnin_MZM_2019}%
  \BibitemOpen
  \bibfield  {author} {\bibinfo {author} {\bibfnamefont {A.}~\bibnamefont
  {Grivnin}}, \bibinfo {author} {\bibfnamefont {E.}~\bibnamefont {Bor}},
  \bibinfo {author} {\bibfnamefont {M.}~\bibnamefont {Heiblum}}, \bibinfo
  {author} {\bibfnamefont {Y.}~\bibnamefont {Oreg}},\ and\ \bibinfo {author}
  {\bibfnamefont {H.}~\bibnamefont {Shtrikman}},\ }\bibfield  {title} {\bibinfo
  {title} {Concomitant opening of a bulk-gap with an emerging possible majorana
  zero mode},\ }\bibfield  {journal} {\bibinfo  {journal} {Nature
  Communications}\ }\textbf {\bibinfo {volume} {10}},\ \href
  {https://doi.org/10.1038/S41467-019-09771-0} {10.1038/S41467-019-09771-0}
  (\bibinfo {year} {2019})\BibitemShut {NoStop}%
\bibitem [{\citenamefont {Moore}\ \emph
  {et~al.}(2018{\natexlab{a}})\citenamefont {Moore}, \citenamefont {Zeng},
  \citenamefont {Stanescu},\ and\ \citenamefont
  {Tewari}}]{Moore_Tiwari_Quantized_ZBP_trivial_2018}%
  \BibitemOpen
  \bibfield  {author} {\bibinfo {author} {\bibfnamefont {C.}~\bibnamefont
  {Moore}}, \bibinfo {author} {\bibfnamefont {C.}~\bibnamefont {Zeng}},
  \bibinfo {author} {\bibfnamefont {T.~D.}\ \bibnamefont {Stanescu}},\ and\
  \bibinfo {author} {\bibfnamefont {S.}~\bibnamefont {Tewari}},\ }\bibfield
  {title} {\bibinfo {title} {Quantized zero-bias conductance plateau in
  semiconductor-superconductor heterostructures without topological majorana
  zero modes},\ }\bibfield  {journal} {\bibinfo  {journal} {Physical Review B}\
  }\textbf {\bibinfo {volume} {94}},\ \href
  {https://doi.org/10.1103/PHYSREVB.98.155314} {10.1103/PHYSREVB.98.155314}
  (\bibinfo {year} {2018}{\natexlab{a}})\BibitemShut {NoStop}%
\bibitem [{\citenamefont {Huang}\ \emph {et~al.}(2018)\citenamefont {Huang},
  \citenamefont {Pan}, \citenamefont {Liu}, \citenamefont {Sau}, \citenamefont
  {Stanescu},\ and\ \citenamefont
  {Sarma}}]{Huang_Sau_Das_sharma_Metamorphosis_ABS_to_MBS_2018}%
  \BibitemOpen
  \bibfield  {author} {\bibinfo {author} {\bibfnamefont {Y.}~\bibnamefont
  {Huang}}, \bibinfo {author} {\bibfnamefont {H.}~\bibnamefont {Pan}}, \bibinfo
  {author} {\bibfnamefont {C.~X.}\ \bibnamefont {Liu}}, \bibinfo {author}
  {\bibfnamefont {J.~D.}\ \bibnamefont {Sau}}, \bibinfo {author} {\bibfnamefont
  {T.~D.}\ \bibnamefont {Stanescu}},\ and\ \bibinfo {author} {\bibfnamefont
  {S.~D.}\ \bibnamefont {Sarma}},\ }\bibfield  {title} {\bibinfo {title}
  {Metamorphosis of andreev bound states into majorana bound states in pristine
  nanowires},\ }\bibfield  {journal} {\bibinfo  {journal} {Physical Review B}\
  }\textbf {\bibinfo {volume} {98}},\ \href
  {https://doi.org/10.1103/PHYSREVB.98.144511} {10.1103/PHYSREVB.98.144511}
  (\bibinfo {year} {2018})\BibitemShut {NoStop}%
\bibitem [{\citenamefont {Setiawan}\ \emph
  {et~al.}(2017{\natexlab{b}})\citenamefont {Setiawan}, \citenamefont {Liu},
  \citenamefont {Sau},\ and\ \citenamefont
  {Sarma}}]{Setiawan_Das_Sharma_Electron_temp_dependance_ZBP_2017}%
  \BibitemOpen
  \bibfield  {author} {\bibinfo {author} {\bibfnamefont {F.}~\bibnamefont
  {Setiawan}}, \bibinfo {author} {\bibfnamefont {C.~X.}\ \bibnamefont {Liu}},
  \bibinfo {author} {\bibfnamefont {J.~D.}\ \bibnamefont {Sau}},\ and\ \bibinfo
  {author} {\bibfnamefont {S.~D.}\ \bibnamefont {Sarma}},\ }\bibfield  {title}
  {\bibinfo {title} {Electron temperature and tunnel coupling dependence of
  zero-bias and almost-zero-bias conductance peaks in majorana nanowires},\
  }\bibfield  {journal} {\bibinfo  {journal} {Physical Review B}\ }\textbf
  {\bibinfo {volume} {96}},\ \href {https://doi.org/10.1103/PHYSREVB.96.184520}
  {10.1103/PHYSREVB.96.184520} (\bibinfo {year}
  {2017}{\natexlab{b}})\BibitemShut {NoStop}%
\bibitem [{\citenamefont {Nichele}\ \emph {et~al.}(2017)\citenamefont
  {Nichele}, \citenamefont {Drachmann}, \citenamefont {Whiticar}, \citenamefont
  {O'Farrell}, \citenamefont {Suominen}, \citenamefont {Fornieri},
  \citenamefont {Wang}, \citenamefont {Gardner}, \citenamefont {Thomas},
  \citenamefont {Hatke}, \citenamefont {Krogstrup}, \citenamefont {Manfra},
  \citenamefont {Flensberg},\ and\ \citenamefont
  {Marcus}}]{Nichele_Flensberg_Scaling_Majorana_ZBP_2017}%
  \BibitemOpen
  \bibfield  {author} {\bibinfo {author} {\bibfnamefont {F.}~\bibnamefont
  {Nichele}}, \bibinfo {author} {\bibfnamefont {A.~C.}\ \bibnamefont
  {Drachmann}}, \bibinfo {author} {\bibfnamefont {A.~M.}\ \bibnamefont
  {Whiticar}}, \bibinfo {author} {\bibfnamefont {E.~C.}\ \bibnamefont
  {O'Farrell}}, \bibinfo {author} {\bibfnamefont {H.~J.}\ \bibnamefont
  {Suominen}}, \bibinfo {author} {\bibfnamefont {A.}~\bibnamefont {Fornieri}},
  \bibinfo {author} {\bibfnamefont {T.}~\bibnamefont {Wang}}, \bibinfo {author}
  {\bibfnamefont {G.~C.}\ \bibnamefont {Gardner}}, \bibinfo {author}
  {\bibfnamefont {C.}~\bibnamefont {Thomas}}, \bibinfo {author} {\bibfnamefont
  {A.~T.}\ \bibnamefont {Hatke}}, \bibinfo {author} {\bibfnamefont
  {P.}~\bibnamefont {Krogstrup}}, \bibinfo {author} {\bibfnamefont {M.~J.}\
  \bibnamefont {Manfra}}, \bibinfo {author} {\bibfnamefont {K.}~\bibnamefont
  {Flensberg}},\ and\ \bibinfo {author} {\bibfnamefont {C.~M.}\ \bibnamefont
  {Marcus}},\ }\bibfield  {title} {\bibinfo {title} {Scaling of majorana
  zero-bias conductance peaks},\ }\bibfield  {journal} {\bibinfo  {journal}
  {Physical Review Letters}\ }\textbf {\bibinfo {volume} {119}},\ \href
  {https://doi.org/10.1103/PHYSREVLETT.119.136803}
  {10.1103/PHYSREVLETT.119.136803} (\bibinfo {year} {2017})\BibitemShut
  {NoStop}%
\bibitem [{\citenamefont {Kammhuber}\ \emph {et~al.}(2017)\citenamefont
  {Kammhuber}, \citenamefont {Cassidy}, \citenamefont {Pei}, \citenamefont
  {Nowak}, \citenamefont {Vuik}, \citenamefont {Gül}, \citenamefont {Car},
  \citenamefont {Plissard}, \citenamefont {Bakkers}, \citenamefont {Wimmer},\
  and\ \citenamefont
  {Kouwenhoven}}]{Kammhuber_Wimmer_Conductance_halical_states_2017}%
  \BibitemOpen
  \bibfield  {author} {\bibinfo {author} {\bibfnamefont {J.}~\bibnamefont
  {Kammhuber}}, \bibinfo {author} {\bibfnamefont {M.~C.}\ \bibnamefont
  {Cassidy}}, \bibinfo {author} {\bibfnamefont {F.}~\bibnamefont {Pei}},
  \bibinfo {author} {\bibfnamefont {M.~P.}\ \bibnamefont {Nowak}}, \bibinfo
  {author} {\bibfnamefont {A.}~\bibnamefont {Vuik}}, \bibinfo {author}
  {\bibfnamefont {O.}~\bibnamefont {Gül}}, \bibinfo {author} {\bibfnamefont
  {D.}~\bibnamefont {Car}}, \bibinfo {author} {\bibfnamefont {S.~R.}\
  \bibnamefont {Plissard}}, \bibinfo {author} {\bibfnamefont {E.~P.}\
  \bibnamefont {Bakkers}}, \bibinfo {author} {\bibfnamefont {M.}~\bibnamefont
  {Wimmer}},\ and\ \bibinfo {author} {\bibfnamefont {L.~P.}\ \bibnamefont
  {Kouwenhoven}},\ }\bibfield  {title} {\bibinfo {title} {Conductance through a
  helical state in an indium antimonide nanowire},\ }\bibfield  {journal}
  {\bibinfo  {journal} {Nature Communications}\ }\textbf {\bibinfo {volume}
  {8}},\ \href {https://doi.org/10.1038/S41467-017-00315-Y}
  {10.1038/S41467-017-00315-Y} (\bibinfo {year} {2017})\BibitemShut {NoStop}%
\bibitem [{\citenamefont {Reeg}\ \emph {et~al.}(2018)\citenamefont {Reeg},
  \citenamefont {Dmytruk}, \citenamefont {Chevallier}, \citenamefont {Loss},\
  and\ \citenamefont {Klinovaja}}]{Reeg_Zero_energy_ABS_2018}%
  \BibitemOpen
  \bibfield  {author} {\bibinfo {author} {\bibfnamefont {C.}~\bibnamefont
  {Reeg}}, \bibinfo {author} {\bibfnamefont {O.}~\bibnamefont {Dmytruk}},
  \bibinfo {author} {\bibfnamefont {D.}~\bibnamefont {Chevallier}}, \bibinfo
  {author} {\bibfnamefont {D.}~\bibnamefont {Loss}},\ and\ \bibinfo {author}
  {\bibfnamefont {J.}~\bibnamefont {Klinovaja}},\ }\bibfield  {title} {\bibinfo
  {title} {Zero-energy andreev bound states from quantum dots in proximitized
  rashba nanowires},\ }\href {https://doi.org/10.1103/PhysRevB.98.245407}
  {\bibfield  {journal} {\bibinfo  {journal} {Phys. Rev. B}\ }\textbf {\bibinfo
  {volume} {98}},\ \bibinfo {pages} {245407} (\bibinfo {year}
  {2018})}\BibitemShut {NoStop}%
\bibitem [{\citenamefont {Liu}\ \emph {et~al.}(2017{\natexlab{a}})\citenamefont
  {Liu}, \citenamefont {Sau}, \citenamefont {Stanescu},\ and\ \citenamefont
  {Sarma}}]{Liu_Sau_Das_Sharma_ABS_vs_MBS_ZBP_2017}%
  \BibitemOpen
  \bibfield  {author} {\bibinfo {author} {\bibfnamefont {C.~X.}\ \bibnamefont
  {Liu}}, \bibinfo {author} {\bibfnamefont {J.~D.}\ \bibnamefont {Sau}},
  \bibinfo {author} {\bibfnamefont {T.~D.}\ \bibnamefont {Stanescu}},\ and\
  \bibinfo {author} {\bibfnamefont {S.~D.}\ \bibnamefont {Sarma}},\ }\bibfield
  {title} {\bibinfo {title} {Andreev bound states versus majorana bound states
  in quantum dot-nanowire-superconductor hybrid structures: Trivial versus
  topological zero-bias conductance peaks},\ }\bibfield  {journal} {\bibinfo
  {journal} {Physical Review B}\ }\textbf {\bibinfo {volume} {96}},\ \href
  {https://doi.org/10.1103/PHYSREVB.96.075161} {10.1103/PHYSREVB.96.075161}
  (\bibinfo {year} {2017}{\natexlab{a}})\BibitemShut {NoStop}%
\bibitem [{\citenamefont {Cayao}\ \emph {et~al.}(2015)\citenamefont {Cayao},
  \citenamefont {Prada}, \citenamefont {San-Jose},\ and\ \citenamefont
  {Aguado}}]{Cayao_SNS_2015}%
  \BibitemOpen
  \bibfield  {author} {\bibinfo {author} {\bibfnamefont {J.}~\bibnamefont
  {Cayao}}, \bibinfo {author} {\bibfnamefont {E.}~\bibnamefont {Prada}},
  \bibinfo {author} {\bibfnamefont {P.}~\bibnamefont {San-Jose}},\ and\
  \bibinfo {author} {\bibfnamefont {R.}~\bibnamefont {Aguado}},\ }\bibfield
  {title} {\bibinfo {title} {Sns junctions in nanowires with spin-orbit
  coupling: Role of confinement and helicity on the subgap spectrum},\ }\href
  {https://doi.org/10.1103/PhysRevB.91.024514} {\bibfield  {journal} {\bibinfo
  {journal} {Phys. Rev. B}\ }\textbf {\bibinfo {volume} {91}},\ \bibinfo
  {pages} {024514} (\bibinfo {year} {2015})}\BibitemShut {NoStop}%
\bibitem [{\citenamefont {Zhang}\ \emph {et~al.}(2017)\citenamefont {Zhang},
  \citenamefont {Önder Gül}, \citenamefont {Conesa-Boj}, \citenamefont
  {Nowak}, \citenamefont {Wimmer}, \citenamefont {Zuo}, \citenamefont {Mourik},
  \citenamefont {Vries}, \citenamefont {Veen}, \citenamefont {Moor},
  \citenamefont {Bommer}, \citenamefont {Woerkom}, \citenamefont {Car},
  \citenamefont {Plissard}, \citenamefont {Bakkers}, \citenamefont
  {Quintero-Pérez}, \citenamefont {Cassidy}, \citenamefont {Koelling},
  \citenamefont {Goswami}, \citenamefont {Watanabe}, \citenamefont
  {Taniguchi},\ and\ \citenamefont
  {Kouwenhoven}}]{Zhang_Balastic_SC_SM_nanowire_2017}%
  \BibitemOpen
  \bibfield  {author} {\bibinfo {author} {\bibfnamefont {H.}~\bibnamefont
  {Zhang}}, \bibinfo {author} {\bibnamefont {Önder Gül}}, \bibinfo {author}
  {\bibfnamefont {S.}~\bibnamefont {Conesa-Boj}}, \bibinfo {author}
  {\bibfnamefont {M.~P.}\ \bibnamefont {Nowak}}, \bibinfo {author}
  {\bibfnamefont {M.}~\bibnamefont {Wimmer}}, \bibinfo {author} {\bibfnamefont
  {K.}~\bibnamefont {Zuo}}, \bibinfo {author} {\bibfnamefont {V.}~\bibnamefont
  {Mourik}}, \bibinfo {author} {\bibfnamefont {F.~K.~D.}\ \bibnamefont
  {Vries}}, \bibinfo {author} {\bibfnamefont {J.~V.}\ \bibnamefont {Veen}},
  \bibinfo {author} {\bibfnamefont {M.~W.~D.}\ \bibnamefont {Moor}}, \bibinfo
  {author} {\bibfnamefont {J.~D.}\ \bibnamefont {Bommer}}, \bibinfo {author}
  {\bibfnamefont {D.~J.~V.}\ \bibnamefont {Woerkom}}, \bibinfo {author}
  {\bibfnamefont {D.}~\bibnamefont {Car}}, \bibinfo {author} {\bibfnamefont
  {S.~R.}\ \bibnamefont {Plissard}}, \bibinfo {author} {\bibfnamefont {E.~P.}\
  \bibnamefont {Bakkers}}, \bibinfo {author} {\bibfnamefont {M.}~\bibnamefont
  {Quintero-Pérez}}, \bibinfo {author} {\bibfnamefont {M.~C.}\ \bibnamefont
  {Cassidy}}, \bibinfo {author} {\bibfnamefont {S.}~\bibnamefont {Koelling}},
  \bibinfo {author} {\bibfnamefont {S.}~\bibnamefont {Goswami}}, \bibinfo
  {author} {\bibfnamefont {K.}~\bibnamefont {Watanabe}}, \bibinfo {author}
  {\bibfnamefont {T.}~\bibnamefont {Taniguchi}},\ and\ \bibinfo {author}
  {\bibfnamefont {L.~P.}\ \bibnamefont {Kouwenhoven}},\ }\bibfield  {title}
  {\bibinfo {title} {Ballistic superconductivity in semiconductor nanowires},\
  }\bibfield  {journal} {\bibinfo  {journal} {Nature Communications}\ }\textbf
  {\bibinfo {volume} {8}},\ \href {https://doi.org/10.1038/NCOMMS16025}
  {10.1038/NCOMMS16025} (\bibinfo {year} {2017})\BibitemShut {NoStop}%
\bibitem [{\citenamefont {Higginbotham}\ \emph {et~al.}(2015)\citenamefont
  {Higginbotham}, \citenamefont {Albrecht}, \citenamefont {Kir{\v{s}}anskas},
  \citenamefont {Chang}, \citenamefont {Kuemmeth}, \citenamefont {Krogstrup},
  \citenamefont {Jespersen}, \citenamefont {Nyg{\aa}rd}, \citenamefont
  {Flensberg},\ and\ \citenamefont {Marcus}}]{higginbotham_parity_2015}%
  \BibitemOpen
  \bibfield  {author} {\bibinfo {author} {\bibfnamefont {A.~P.}\ \bibnamefont
  {Higginbotham}}, \bibinfo {author} {\bibfnamefont {S.~M.}\ \bibnamefont
  {Albrecht}}, \bibinfo {author} {\bibfnamefont {G.}~\bibnamefont
  {Kir{\v{s}}anskas}}, \bibinfo {author} {\bibfnamefont {W.}~\bibnamefont
  {Chang}}, \bibinfo {author} {\bibfnamefont {F.}~\bibnamefont {Kuemmeth}},
  \bibinfo {author} {\bibfnamefont {P.}~\bibnamefont {Krogstrup}}, \bibinfo
  {author} {\bibfnamefont {T.~S.}\ \bibnamefont {Jespersen}}, \bibinfo {author}
  {\bibfnamefont {J.}~\bibnamefont {Nyg{\aa}rd}}, \bibinfo {author}
  {\bibfnamefont {K.}~\bibnamefont {Flensberg}},\ and\ \bibinfo {author}
  {\bibfnamefont {C.~M.}\ \bibnamefont {Marcus}},\ }\bibfield  {title}
  {\bibinfo {title} {Parity lifetime of bound states in a proximitized
  semiconductor nanowire},\ }\href@noop {} {\bibfield  {journal} {\bibinfo
  {journal} {Nature Physics}\ }\textbf {\bibinfo {volume} {11}},\ \bibinfo
  {pages} {1017} (\bibinfo {year} {2015})}\BibitemShut {NoStop}%
\bibitem [{\citenamefont {Dmytruk}\ \emph {et~al.}(2020)\citenamefont
  {Dmytruk}, \citenamefont {Loss},\ and\ \citenamefont
  {Klinovaja}}]{Dmytruk_Pinning_2020}%
  \BibitemOpen
  \bibfield  {author} {\bibinfo {author} {\bibfnamefont {O.}~\bibnamefont
  {Dmytruk}}, \bibinfo {author} {\bibfnamefont {D.}~\bibnamefont {Loss}},\ and\
  \bibinfo {author} {\bibfnamefont {J.}~\bibnamefont {Klinovaja}},\ }\bibfield
  {title} {\bibinfo {title} {Pinning of andreev bound states to zero energy in
  two-dimensional superconductor- semiconductor rashba heterostructures},\
  }\href {https://doi.org/10.1103/PhysRevB.102.245431} {\bibfield  {journal}
  {\bibinfo  {journal} {Phys. Rev. B}\ }\textbf {\bibinfo {volume} {102}},\
  \bibinfo {pages} {245431} (\bibinfo {year} {2020})}\BibitemShut {NoStop}%
\bibitem [{\citenamefont {Liu}\ \emph {et~al.}(2017{\natexlab{b}})\citenamefont
  {Liu}, \citenamefont {Sau},\ and\ \citenamefont
  {Sarma}}]{Liu_Sau_Das_sharma_Role_of_dissipation_MBS_2017}%
  \BibitemOpen
  \bibfield  {author} {\bibinfo {author} {\bibfnamefont {C.~X.}\ \bibnamefont
  {Liu}}, \bibinfo {author} {\bibfnamefont {J.~D.}\ \bibnamefont {Sau}},\ and\
  \bibinfo {author} {\bibfnamefont {S.~D.}\ \bibnamefont {Sarma}},\ }\bibfield
  {title} {\bibinfo {title} {Role of dissipation in realistic majorana
  nanowires},\ }\bibfield  {journal} {\bibinfo  {journal} {Physical Review B}\
  }\textbf {\bibinfo {volume} {95}},\ \href
  {https://doi.org/10.1103/PHYSREVB.95.054502} {10.1103/PHYSREVB.95.054502}
  (\bibinfo {year} {2017}{\natexlab{b}})\BibitemShut {NoStop}%
\bibitem [{\citenamefont {Prada}\ \emph {et~al.}(2012)\citenamefont {Prada},
  \citenamefont {San-Jose},\ and\ \citenamefont
  {Aguado}}]{Prada_Transport_2012}%
  \BibitemOpen
  \bibfield  {author} {\bibinfo {author} {\bibfnamefont {E.}~\bibnamefont
  {Prada}}, \bibinfo {author} {\bibfnamefont {P.}~\bibnamefont {San-Jose}},\
  and\ \bibinfo {author} {\bibfnamefont {R.}~\bibnamefont {Aguado}},\
  }\bibfield  {title} {\bibinfo {title} {Transport spectroscopy of $ns$
  nanowire junctions with majorana fermions},\ }\href
  {https://doi.org/10.1103/PhysRevB.86.180503} {\bibfield  {journal} {\bibinfo
  {journal} {Phys. Rev. B}\ }\textbf {\bibinfo {volume} {86}},\ \bibinfo
  {pages} {180503} (\bibinfo {year} {2012})}\BibitemShut {NoStop}%
\bibitem [{\citenamefont {Lee}\ \emph {et~al.}(2012)\citenamefont {Lee},
  \citenamefont {Jiang}, \citenamefont {Aguado}, \citenamefont {Katsaros},
  \citenamefont {Lieber},\ and\ \citenamefont {De~Franceschi}}]{Lee_ZBA_2012}%
  \BibitemOpen
  \bibfield  {author} {\bibinfo {author} {\bibfnamefont {E.~J.~H.}\
  \bibnamefont {Lee}}, \bibinfo {author} {\bibfnamefont {X.}~\bibnamefont
  {Jiang}}, \bibinfo {author} {\bibfnamefont {R.}~\bibnamefont {Aguado}},
  \bibinfo {author} {\bibfnamefont {G.}~\bibnamefont {Katsaros}}, \bibinfo
  {author} {\bibfnamefont {C.~M.}\ \bibnamefont {Lieber}},\ and\ \bibinfo
  {author} {\bibfnamefont {S.}~\bibnamefont {De~Franceschi}},\ }\bibfield
  {title} {\bibinfo {title} {Zero-bias anomaly in a nanowire quantum dot
  coupled to superconductors},\ }\href
  {https://doi.org/10.1103/PhysRevLett.109.186802} {\bibfield  {journal}
  {\bibinfo  {journal} {Phys. Rev. Lett.}\ }\textbf {\bibinfo {volume} {109}},\
  \bibinfo {pages} {186802} (\bibinfo {year} {2012})}\BibitemShut {NoStop}%
\bibitem [{\citenamefont {Prada}\ \emph {et~al.}(2020)\citenamefont {Prada},
  \citenamefont {San-Jose}, \citenamefont {de~Moor}, \citenamefont {Geresdi},
  \citenamefont {Lee}, \citenamefont {Klinovaja}, \citenamefont {Loss},
  \citenamefont {Nyg{\aa}rd}, \citenamefont {Aguado},\ and\ \citenamefont
  {Kouwenhoven}}]{prada_andreev_2020}%
  \BibitemOpen
  \bibfield  {author} {\bibinfo {author} {\bibfnamefont {E.}~\bibnamefont
  {Prada}}, \bibinfo {author} {\bibfnamefont {P.}~\bibnamefont {San-Jose}},
  \bibinfo {author} {\bibfnamefont {M.~W.}\ \bibnamefont {de~Moor}}, \bibinfo
  {author} {\bibfnamefont {A.}~\bibnamefont {Geresdi}}, \bibinfo {author}
  {\bibfnamefont {E.~J.}\ \bibnamefont {Lee}}, \bibinfo {author} {\bibfnamefont
  {J.}~\bibnamefont {Klinovaja}}, \bibinfo {author} {\bibfnamefont
  {D.}~\bibnamefont {Loss}}, \bibinfo {author} {\bibfnamefont {J.}~\bibnamefont
  {Nyg{\aa}rd}}, \bibinfo {author} {\bibfnamefont {R.}~\bibnamefont {Aguado}},\
  and\ \bibinfo {author} {\bibfnamefont {L.~P.}\ \bibnamefont {Kouwenhoven}},\
  }\bibfield  {title} {\bibinfo {title} {From andreev to majorana bound states
  in hybrid superconductor--semiconductor nanowires},\ }\href@noop {}
  {\bibfield  {journal} {\bibinfo  {journal} {Nature Reviews Physics}\ }\textbf
  {\bibinfo {volume} {2}},\ \bibinfo {pages} {575} (\bibinfo {year}
  {2020})}\BibitemShut {NoStop}%
\bibitem [{\citenamefont {San-Jose}\ \emph {et~al.}(2016)\citenamefont
  {San-Jose}, \citenamefont {Cayao}, \citenamefont {Prada},\ and\ \citenamefont
  {Aguado}}]{san_majorana_2016}%
  \BibitemOpen
  \bibfield  {author} {\bibinfo {author} {\bibfnamefont {P.}~\bibnamefont
  {San-Jose}}, \bibinfo {author} {\bibfnamefont {J.}~\bibnamefont {Cayao}},
  \bibinfo {author} {\bibfnamefont {E.}~\bibnamefont {Prada}},\ and\ \bibinfo
  {author} {\bibfnamefont {R.}~\bibnamefont {Aguado}},\ }\bibfield  {title}
  {\bibinfo {title} {Majorana bound states from exceptional points in
  non-topological superconductors},\ }\href@noop {} {\bibfield  {journal}
  {\bibinfo  {journal} {Scientific reports}\ }\textbf {\bibinfo {volume} {6}},\
  \bibinfo {pages} {1} (\bibinfo {year} {2016})}\BibitemShut {NoStop}%
\bibitem [{\citenamefont {Deng}\ \emph {et~al.}(2016)\citenamefont {Deng},
  \citenamefont {Vaitiekenas}, \citenamefont {Hansen}, \citenamefont {Danon},
  \citenamefont {Leijnse}, \citenamefont {Flensberg}, \citenamefont {Nygård},
  \citenamefont {Krogstrup},\ and\ \citenamefont
  {Marcus}}]{Deng_Flensberg_MBS_in_QD_hybrid_nanowire_2016}%
  \BibitemOpen
  \bibfield  {author} {\bibinfo {author} {\bibfnamefont {M.~T.}\ \bibnamefont
  {Deng}}, \bibinfo {author} {\bibfnamefont {S.}~\bibnamefont {Vaitiekenas}},
  \bibinfo {author} {\bibfnamefont {E.~B.}\ \bibnamefont {Hansen}}, \bibinfo
  {author} {\bibfnamefont {J.}~\bibnamefont {Danon}}, \bibinfo {author}
  {\bibfnamefont {M.}~\bibnamefont {Leijnse}}, \bibinfo {author} {\bibfnamefont
  {K.}~\bibnamefont {Flensberg}}, \bibinfo {author} {\bibfnamefont
  {J.}~\bibnamefont {Nygård}}, \bibinfo {author} {\bibfnamefont
  {P.}~\bibnamefont {Krogstrup}},\ and\ \bibinfo {author} {\bibfnamefont
  {C.~M.}\ \bibnamefont {Marcus}},\ }\bibfield  {title} {\bibinfo {title}
  {Majorana bound state in a coupled quantum-dot hybrid-nanowire system},\
  }\href {https://doi.org/10.1126/SCIENCE.AAF3961} {\bibfield  {journal}
  {\bibinfo  {journal} {Science}\ }\textbf {\bibinfo {volume} {354}},\ \bibinfo
  {pages} {1557} (\bibinfo {year} {2016})}\BibitemShut {NoStop}%
\bibitem [{\citenamefont {Song}\ \emph {et~al.}(2022)\citenamefont {Song},
  \citenamefont {Zhang}, \citenamefont {Pan}, \citenamefont {Liu},
  \citenamefont {Wang}, \citenamefont {Cao}, \citenamefont {Liu}, \citenamefont
  {Wen}, \citenamefont {Liao}, \citenamefont {Zhuo}, \citenamefont {Liu},
  \citenamefont {Shang}, \citenamefont {Zhao},\ and\ \citenamefont
  {Zhang}}]{Song_Large_ZBP_2022}%
  \BibitemOpen
  \bibfield  {author} {\bibinfo {author} {\bibfnamefont {H.}~\bibnamefont
  {Song}}, \bibinfo {author} {\bibfnamefont {Z.}~\bibnamefont {Zhang}},
  \bibinfo {author} {\bibfnamefont {D.}~\bibnamefont {Pan}}, \bibinfo {author}
  {\bibfnamefont {D.}~\bibnamefont {Liu}}, \bibinfo {author} {\bibfnamefont
  {Z.}~\bibnamefont {Wang}}, \bibinfo {author} {\bibfnamefont {Z.}~\bibnamefont
  {Cao}}, \bibinfo {author} {\bibfnamefont {L.}~\bibnamefont {Liu}}, \bibinfo
  {author} {\bibfnamefont {L.}~\bibnamefont {Wen}}, \bibinfo {author}
  {\bibfnamefont {D.}~\bibnamefont {Liao}}, \bibinfo {author} {\bibfnamefont
  {R.}~\bibnamefont {Zhuo}}, \bibinfo {author} {\bibfnamefont {D.~E.}\
  \bibnamefont {Liu}}, \bibinfo {author} {\bibfnamefont {R.}~\bibnamefont
  {Shang}}, \bibinfo {author} {\bibfnamefont {J.}~\bibnamefont {Zhao}},\ and\
  \bibinfo {author} {\bibfnamefont {H.}~\bibnamefont {Zhang}},\ }\bibfield
  {title} {\bibinfo {title} {Large zero bias peaks and dips in a four-terminal
  thin inas-al nanowire device},\ }\href
  {https://doi.org/10.1103/PhysRevResearch.4.033235} {\bibfield  {journal}
  {\bibinfo  {journal} {Phys. Rev. Research}\ }\textbf {\bibinfo {volume}
  {4}},\ \bibinfo {pages} {033235} (\bibinfo {year} {2022})}\BibitemShut
  {NoStop}%
\bibitem [{\citenamefont {Avila}\ \emph {et~al.}(2019)\citenamefont {Avila},
  \citenamefont {Pe{\~n}aranda}, \citenamefont {Prada}, \citenamefont
  {San-Jose},\ and\ \citenamefont {Aguado}}]{avila_non_hermitian_2019}%
  \BibitemOpen
  \bibfield  {author} {\bibinfo {author} {\bibfnamefont {J.}~\bibnamefont
  {Avila}}, \bibinfo {author} {\bibfnamefont {F.}~\bibnamefont
  {Pe{\~n}aranda}}, \bibinfo {author} {\bibfnamefont {E.}~\bibnamefont
  {Prada}}, \bibinfo {author} {\bibfnamefont {P.}~\bibnamefont {San-Jose}},\
  and\ \bibinfo {author} {\bibfnamefont {R.}~\bibnamefont {Aguado}},\
  }\bibfield  {title} {\bibinfo {title} {Non-hermitian topology as a unifying
  framework for the andreev versus majorana states controversy},\ }\href@noop
  {} {\bibfield  {journal} {\bibinfo  {journal} {Communications Physics}\
  }\textbf {\bibinfo {volume} {2}},\ \bibinfo {pages} {1} (\bibinfo {year}
  {2019})}\BibitemShut {NoStop}%
\bibitem [{\citenamefont {Moore}\ \emph
  {et~al.}(2018{\natexlab{b}})\citenamefont {Moore}, \citenamefont {Stanescu},\
  and\ \citenamefont {Tewari}}]{Tiwari_Two-terminal_2018}%
  \BibitemOpen
  \bibfield  {author} {\bibinfo {author} {\bibfnamefont {C.}~\bibnamefont
  {Moore}}, \bibinfo {author} {\bibfnamefont {T.~D.}\ \bibnamefont
  {Stanescu}},\ and\ \bibinfo {author} {\bibfnamefont {S.}~\bibnamefont
  {Tewari}},\ }\bibfield  {title} {\bibinfo {title} {Two-terminal charge
  tunneling: Disentangling majorana zero modes from partially separated andreev
  bound states in semiconductor-superconductor heterostructures},\ }\href
  {https://doi.org/10.1103/PhysRevB.97.165302} {\bibfield  {journal} {\bibinfo
  {journal} {Phys. Rev. B}\ }\textbf {\bibinfo {volume} {97}},\ \bibinfo
  {pages} {165302} (\bibinfo {year} {2018}{\natexlab{b}})}\BibitemShut
  {NoStop}%
\bibitem [{\citenamefont {Stanescu}\ and\ \citenamefont
  {Tewari}(2019)}]{Tiwari_Robust_low-energy_ABS_2019}%
  \BibitemOpen
  \bibfield  {author} {\bibinfo {author} {\bibfnamefont {T.~D.}\ \bibnamefont
  {Stanescu}}\ and\ \bibinfo {author} {\bibfnamefont {S.}~\bibnamefont
  {Tewari}},\ }\bibfield  {title} {\bibinfo {title} {Robust low-energy andreev
  bound states in semiconductor-superconductor structures: Importance of
  partial separation of component majorana bound states},\ }\href
  {https://doi.org/10.1103/PhysRevB.100.155429} {\bibfield  {journal} {\bibinfo
   {journal} {Phys. Rev. B}\ }\textbf {\bibinfo {volume} {100}},\ \bibinfo
  {pages} {155429} (\bibinfo {year} {2019})}\BibitemShut {NoStop}%
\bibitem [{\citenamefont {Pan}\ \emph {et~al.}(2021)\citenamefont {Pan},
  \citenamefont {Sau},\ and\ \citenamefont
  {Das~Sarma}}]{Das_sharma_Three_terminal_ZBP_2021}%
  \BibitemOpen
  \bibfield  {author} {\bibinfo {author} {\bibfnamefont {H.}~\bibnamefont
  {Pan}}, \bibinfo {author} {\bibfnamefont {J.~D.}\ \bibnamefont {Sau}},\ and\
  \bibinfo {author} {\bibfnamefont {S.}~\bibnamefont {Das~Sarma}},\ }\bibfield
  {title} {\bibinfo {title} {Three-terminal nonlocal conductance in majorana
  nanowires: Distinguishing topological and trivial in realistic systems with
  disorder and inhomogeneous potential},\ }\href
  {https://doi.org/10.1103/PhysRevB.103.014513} {\bibfield  {journal} {\bibinfo
   {journal} {Phys. Rev. B}\ }\textbf {\bibinfo {volume} {103}},\ \bibinfo
  {pages} {014513} (\bibinfo {year} {2021})}\BibitemShut {NoStop}%
\bibitem [{\citenamefont {Hess}\ \emph {et~al.}(2021)\citenamefont {Hess},
  \citenamefont {Legg}, \citenamefont {Loss},\ and\ \citenamefont
  {Klinovaja}}]{Loss_Local_and_nonlocal_2021}%
  \BibitemOpen
  \bibfield  {author} {\bibinfo {author} {\bibfnamefont {R.}~\bibnamefont
  {Hess}}, \bibinfo {author} {\bibfnamefont {H.~F.}\ \bibnamefont {Legg}},
  \bibinfo {author} {\bibfnamefont {D.}~\bibnamefont {Loss}},\ and\ \bibinfo
  {author} {\bibfnamefont {J.}~\bibnamefont {Klinovaja}},\ }\bibfield  {title}
  {\bibinfo {title} {Local and nonlocal quantum transport due to andreev bound
  states in finite rashba nanowires with superconducting and normal sections},\
  }\href {https://doi.org/10.1103/PhysRevB.104.075405} {\bibfield  {journal}
  {\bibinfo  {journal} {Phys. Rev. B}\ }\textbf {\bibinfo {volume} {104}},\
  \bibinfo {pages} {075405} (\bibinfo {year} {2021})}\BibitemShut {NoStop}%
\bibitem [{\citenamefont {Mishmash}\ \emph {et~al.}(2020)\citenamefont
  {Mishmash}, \citenamefont {Bauer}, \citenamefont {von Oppen},\ and\
  \citenamefont {Alicea}}]{Mishmash_Dephasing_leakage_2020}%
  \BibitemOpen
  \bibfield  {author} {\bibinfo {author} {\bibfnamefont {R.~V.}\ \bibnamefont
  {Mishmash}}, \bibinfo {author} {\bibfnamefont {B.}~\bibnamefont {Bauer}},
  \bibinfo {author} {\bibfnamefont {F.}~\bibnamefont {von Oppen}},\ and\
  \bibinfo {author} {\bibfnamefont {J.}~\bibnamefont {Alicea}},\ }\bibfield
  {title} {\bibinfo {title} {Dephasing and leakage dynamics of noisy
  majorana-based qubits: Topological versus andreev},\ }\href
  {https://doi.org/10.1103/PhysRevB.101.075404} {\bibfield  {journal} {\bibinfo
   {journal} {Phys. Rev. B}\ }\textbf {\bibinfo {volume} {101}},\ \bibinfo
  {pages} {075404} (\bibinfo {year} {2020})}\BibitemShut {NoStop}%
\bibitem [{\citenamefont {Lai}\ \emph {et~al.}(2021)\citenamefont {Lai},
  \citenamefont {Sarma},\ and\ \citenamefont
  {Sau}}]{Das_sharma_Quality_ZBP_2021}%
  \BibitemOpen
  \bibfield  {author} {\bibinfo {author} {\bibfnamefont {Y.-H.}\ \bibnamefont
  {Lai}}, \bibinfo {author} {\bibfnamefont {S.~D.}\ \bibnamefont {Sarma}},\
  and\ \bibinfo {author} {\bibfnamefont {J.~D.}\ \bibnamefont {Sau}},\ }\href
  {https://doi.org/10.48550/ARXIV.2111.01178} {\bibinfo {title} {Quality factor
  for zero-bias conductance peaks in majorana nanowire}} (\bibinfo {year}
  {2021})\BibitemShut {NoStop}%
\bibitem [{\citenamefont {Alicea}\ \emph {et~al.}(2011)\citenamefont {Alicea},
  \citenamefont {Oreg}, \citenamefont {Refael}, \citenamefont {Von~Oppen},\
  and\ \citenamefont {Fisher}}]{alicea_non_Abelian_2011}%
  \BibitemOpen
  \bibfield  {author} {\bibinfo {author} {\bibfnamefont {J.}~\bibnamefont
  {Alicea}}, \bibinfo {author} {\bibfnamefont {Y.}~\bibnamefont {Oreg}},
  \bibinfo {author} {\bibfnamefont {G.}~\bibnamefont {Refael}}, \bibinfo
  {author} {\bibfnamefont {F.}~\bibnamefont {Von~Oppen}},\ and\ \bibinfo
  {author} {\bibfnamefont {M.}~\bibnamefont {Fisher}},\ }\bibfield  {title}
  {\bibinfo {title} {Non-abelian statistics and topological quantum information
  processing in 1d wire networks},\ }\href@noop {} {\bibfield  {journal}
  {\bibinfo  {journal} {Nature Physics}\ }\textbf {\bibinfo {volume} {7}},\
  \bibinfo {pages} {412} (\bibinfo {year} {2011})}\BibitemShut {NoStop}%
\bibitem [{\citenamefont {Sarma}\ \emph
  {et~al.}(2015{\natexlab{b}})\citenamefont {Sarma}, \citenamefont {Freedman},\
  and\ \citenamefont {Nayak}}]{Das_Sharma_MZM_QC_2015}%
  \BibitemOpen
  \bibfield  {author} {\bibinfo {author} {\bibfnamefont {S.~D.}\ \bibnamefont
  {Sarma}}, \bibinfo {author} {\bibfnamefont {M.}~\bibnamefont {Freedman}},\
  and\ \bibinfo {author} {\bibfnamefont {C.}~\bibnamefont {Nayak}},\ }\bibfield
   {title} {\bibinfo {title} {Majorana zero modes and topological quantum
  computation},\ }\href@noop {} {\bibfield  {journal} {\bibinfo  {journal} {npj
  Quantum Information}\ }\textbf {\bibinfo {volume} {1}},\ \bibinfo {pages} {1}
  (\bibinfo {year} {2015}{\natexlab{b}})}\BibitemShut {NoStop}%
\bibitem [{\citenamefont {Bauer}\ \emph {et~al.}(2018)\citenamefont {Bauer},
  \citenamefont {Karzig}, \citenamefont {Mishmash}, \citenamefont {Antipov},\
  and\ \citenamefont {Alicea}}]{Alicea_Antipov_Dynamic_of_Majorana_qubit_2018}%
  \BibitemOpen
  \bibfield  {author} {\bibinfo {author} {\bibfnamefont {B.}~\bibnamefont
  {Bauer}}, \bibinfo {author} {\bibfnamefont {T.}~\bibnamefont {Karzig}},
  \bibinfo {author} {\bibfnamefont {R.~V.}\ \bibnamefont {Mishmash}}, \bibinfo
  {author} {\bibfnamefont {A.~E.}\ \bibnamefont {Antipov}},\ and\ \bibinfo
  {author} {\bibfnamefont {J.}~\bibnamefont {Alicea}},\ }\bibfield  {title}
  {\bibinfo {title} {{Dynamics of Majorana-based qubits operated with an array
  of tunable gates}},\ }\href {https://doi.org/10.21468/SciPostPhys.5.1.004}
  {\bibfield  {journal} {\bibinfo  {journal} {SciPost Phys.}\ }\textbf
  {\bibinfo {volume} {5}},\ \bibinfo {pages} {4} (\bibinfo {year}
  {2018})}\BibitemShut {NoStop}%
\bibitem [{\citenamefont {Hu}\ \emph {et~al.}(2015)\citenamefont {Hu},
  \citenamefont {Cai}, \citenamefont {Baranov},\ and\ \citenamefont
  {Zoller}}]{Zolle_MBS_noisy_kitaev_2015}%
  \BibitemOpen
  \bibfield  {author} {\bibinfo {author} {\bibfnamefont {Y.}~\bibnamefont
  {Hu}}, \bibinfo {author} {\bibfnamefont {Z.}~\bibnamefont {Cai}}, \bibinfo
  {author} {\bibfnamefont {M.~A.}\ \bibnamefont {Baranov}},\ and\ \bibinfo
  {author} {\bibfnamefont {P.}~\bibnamefont {Zoller}},\ }\bibfield  {title}
  {\bibinfo {title} {Majorana fermions in noisy kitaev wires},\ }\href
  {https://doi.org/10.1103/PhysRevB.92.165118} {\bibfield  {journal} {\bibinfo
  {journal} {Phys. Rev. B}\ }\textbf {\bibinfo {volume} {92}},\ \bibinfo
  {pages} {165118} (\bibinfo {year} {2015})}\BibitemShut {NoStop}%
\bibitem [{\citenamefont {Zhang}\ and\ \citenamefont {Nori}(2016)}]{Nori}%
  \BibitemOpen
  \bibfield  {author} {\bibinfo {author} {\bibfnamefont {P.}~\bibnamefont
  {Zhang}}\ and\ \bibinfo {author} {\bibfnamefont {F.}~\bibnamefont {Nori}},\
  }\bibfield  {title} {\bibinfo {title} {Majorana bound states in a disordered
  quantum dot chain},\ }\href {https://doi.org/10.1088/1367-2630/18/4/043033}
  {\bibfield  {journal} {\bibinfo  {journal} {New Journal of Physics}\ }\textbf
  {\bibinfo {volume} {18}},\ \bibinfo {pages} {043033} (\bibinfo {year}
  {2016})}\BibitemShut {NoStop}%
\bibitem [{\citenamefont {Liu}\ \emph {et~al.}(2017{\natexlab{c}})\citenamefont
  {Liu}, \citenamefont {Sau}, \citenamefont {Stanescu},\ and\ \citenamefont
  {Das~Sarma}}]{Das_sharma_2017_ABS_ZBP_split}%
  \BibitemOpen
  \bibfield  {author} {\bibinfo {author} {\bibfnamefont {C.-X.}\ \bibnamefont
  {Liu}}, \bibinfo {author} {\bibfnamefont {J.~D.}\ \bibnamefont {Sau}},
  \bibinfo {author} {\bibfnamefont {T.~D.}\ \bibnamefont {Stanescu}},\ and\
  \bibinfo {author} {\bibfnamefont {S.}~\bibnamefont {Das~Sarma}},\ }\bibfield
  {title} {\bibinfo {title} {Andreev bound states versus majorana bound states
  in quantum dot-nanowire-superconductor hybrid structures: Trivial versus
  topological zero-bias conductance peaks},\ }\href
  {https://doi.org/10.1103/PhysRevB.96.075161} {\bibfield  {journal} {\bibinfo
  {journal} {Phys. Rev. B}\ }\textbf {\bibinfo {volume} {96}},\ \bibinfo
  {pages} {075161} (\bibinfo {year} {2017}{\natexlab{c}})}\BibitemShut
  {NoStop}%
\bibitem [{\citenamefont {Vuik}\ \emph
  {et~al.}(2019{\natexlab{b}})\citenamefont {Vuik}, \citenamefont {Nijholt},
  \citenamefont {Akhmerov},\ and\ \citenamefont
  {Wimmer}}]{akmerov_reproducing_2019}%
  \BibitemOpen
  \bibfield  {author} {\bibinfo {author} {\bibfnamefont {A.}~\bibnamefont
  {Vuik}}, \bibinfo {author} {\bibfnamefont {B.}~\bibnamefont {Nijholt}},
  \bibinfo {author} {\bibfnamefont {A.}~\bibnamefont {Akhmerov}},\ and\
  \bibinfo {author} {\bibfnamefont {M.}~\bibnamefont {Wimmer}},\ }\bibfield
  {title} {\bibinfo {title} {Reproducing topological properties with
  quasi-majorana states},\ }\href@noop {} {\bibfield  {journal} {\bibinfo
  {journal} {SciPost Physics}\ }\textbf {\bibinfo {volume} {7}},\ \bibinfo
  {pages} {061} (\bibinfo {year} {2019}{\natexlab{b}})}\BibitemShut {NoStop}%
\bibitem [{\citenamefont {Markos}\ and\ \citenamefont
  {Soukoulis}(2008)}]{Main_book}%
  \BibitemOpen
  \bibfield  {author} {\bibinfo {author} {\bibfnamefont {P.}~\bibnamefont
  {Markos}}\ and\ \bibinfo {author} {\bibfnamefont {C.~M.}\ \bibnamefont
  {Soukoulis}},\ }\href {https://doi.org/doi:10.1515/9781400835676} {\emph
  {\bibinfo {title} {Wave Propagation: From Electrons to Photonic Crystals and
  Left-Handed Materials}}}\ (\bibinfo  {publisher} {Princeton University
  Press},\ \bibinfo {year} {2008})\BibitemShut {NoStop}%
\bibitem [{\citenamefont {Choy}\ \emph {et~al.}(2011)\citenamefont {Choy},
  \citenamefont {Edge}, \citenamefont {Akhmerov},\ and\ \citenamefont
  {Beenakker}}]{Same_as_Nori}%
  \BibitemOpen
  \bibfield  {author} {\bibinfo {author} {\bibfnamefont {T.-P.}\ \bibnamefont
  {Choy}}, \bibinfo {author} {\bibfnamefont {J.~M.}\ \bibnamefont {Edge}},
  \bibinfo {author} {\bibfnamefont {A.~R.}\ \bibnamefont {Akhmerov}},\ and\
  \bibinfo {author} {\bibfnamefont {C.~W.~J.}\ \bibnamefont {Beenakker}},\
  }\bibfield  {title} {\bibinfo {title} {Majorana fermions emerging from
  magnetic nanoparticles on a superconductor without spin-orbit coupling},\
  }\href {https://doi.org/10.1103/PhysRevB.84.195442} {\bibfield  {journal}
  {\bibinfo  {journal} {Phys. Rev. B}\ }\textbf {\bibinfo {volume} {84}},\
  \bibinfo {pages} {195442} (\bibinfo {year} {2011})}\BibitemShut {NoStop}%
\bibitem [{\citenamefont {Akhmerov}\ \emph {et~al.}(2011)\citenamefont
  {Akhmerov}, \citenamefont {Dahlhaus}, \citenamefont {Hassler}, \citenamefont
  {Wimmer},\ and\ \citenamefont {Beenakker}}]{TopQuantNumber}%
  \BibitemOpen
  \bibfield  {author} {\bibinfo {author} {\bibfnamefont {A.~R.}\ \bibnamefont
  {Akhmerov}}, \bibinfo {author} {\bibfnamefont {J.~P.}\ \bibnamefont
  {Dahlhaus}}, \bibinfo {author} {\bibfnamefont {F.}~\bibnamefont {Hassler}},
  \bibinfo {author} {\bibfnamefont {M.}~\bibnamefont {Wimmer}},\ and\ \bibinfo
  {author} {\bibfnamefont {C.~W.~J.}\ \bibnamefont {Beenakker}},\ }\bibfield
  {title} {\bibinfo {title} {Quantized conductance at the majorana phase
  transition in a disordered superconducting wire},\ }\href
  {https://doi.org/10.1103/PhysRevLett.106.057001} {\bibfield  {journal}
  {\bibinfo  {journal} {Phys. Rev. Lett.}\ }\textbf {\bibinfo {volume} {106}},\
  \bibinfo {pages} {057001} (\bibinfo {year} {2011})}\BibitemShut {NoStop}%
\bibitem [{\citenamefont {Fulga}\ \emph {et~al.}(2011)\citenamefont {Fulga},
  \citenamefont {Hassler}, \citenamefont {Akhmerov},\ and\ \citenamefont
  {Beenakker}}]{FulgaTopologicalQ}%
  \BibitemOpen
  \bibfield  {author} {\bibinfo {author} {\bibfnamefont {I.~C.}\ \bibnamefont
  {Fulga}}, \bibinfo {author} {\bibfnamefont {F.}~\bibnamefont {Hassler}},
  \bibinfo {author} {\bibfnamefont {A.~R.}\ \bibnamefont {Akhmerov}},\ and\
  \bibinfo {author} {\bibfnamefont {C.~W.~J.}\ \bibnamefont {Beenakker}},\
  }\bibfield  {title} {\bibinfo {title} {Scattering formula for the topological
  quantum number of a disordered multimode wire},\ }\href
  {https://doi.org/10.1103/PhysRevB.83.155429} {\bibfield  {journal} {\bibinfo
  {journal} {Phys. Rev. B}\ }\textbf {\bibinfo {volume} {83}},\ \bibinfo
  {pages} {155429} (\bibinfo {year} {2011})}\BibitemShut {NoStop}%
\bibitem [{\citenamefont {Fregoso}\ \emph {et~al.}(2013)\citenamefont
  {Fregoso}, \citenamefont {Lobos},\ and\ \citenamefont
  {Das~Sarma}}]{disordered_Majorana}%
  \BibitemOpen
  \bibfield  {author} {\bibinfo {author} {\bibfnamefont {B.~M.}\ \bibnamefont
  {Fregoso}}, \bibinfo {author} {\bibfnamefont {A.~M.}\ \bibnamefont {Lobos}},\
  and\ \bibinfo {author} {\bibfnamefont {S.}~\bibnamefont {Das~Sarma}},\
  }\bibfield  {title} {\bibinfo {title} {Electrical detection of topological
  quantum phase transitions in disordered majorana nanowires},\ }\href
  {https://doi.org/10.1103/PhysRevB.88.180507} {\bibfield  {journal} {\bibinfo
  {journal} {Phys. Rev. B}\ }\textbf {\bibinfo {volume} {88}},\ \bibinfo
  {pages} {180507} (\bibinfo {year} {2013})}\BibitemShut {NoStop}%
\bibitem [{\citenamefont {Lobos}\ and\ \citenamefont
  {Sarma}(2015)}]{Lobos_tunneling_2015}%
  \BibitemOpen
  \bibfield  {author} {\bibinfo {author} {\bibfnamefont {A.~M.}\ \bibnamefont
  {Lobos}}\ and\ \bibinfo {author} {\bibfnamefont {S.~D.}\ \bibnamefont
  {Sarma}},\ }\bibfield  {title} {\bibinfo {title} {Tunneling transport in nsn
  majorana junctions across the topological quantum phase transition},\
  }\href@noop {} {\bibfield  {journal} {\bibinfo  {journal} {New Journal of
  Physics}\ }\textbf {\bibinfo {volume} {17}},\ \bibinfo {pages} {065010}
  (\bibinfo {year} {2015})}\BibitemShut {NoStop}%
\end{thebibliography}%
\end{document}